\RequirePackage{rotating}
\documentclass[12pt]{article}%
\usepackage{subcaption}
\usepackage{adjustbox}
\usepackage{bm}
\usepackage{amsfonts}
\usepackage[colorlinks,citecolor=blue,urlcolor=blue,hyperfootnotes=false]
{hyperref}
\usepackage{amssymb}
\usepackage[bottom]{footmisc}
\usepackage[hyperref]{ntheorem}
\usepackage{natbib}
\usepackage{graphicx}
\usepackage{float}
\usepackage[doublespacing,nodisplayskipstretch]{setspace}
\usepackage[margin=1in, a4paper]{geometry}
\usepackage{threeparttable}
\usepackage{amsmath}
\usepackage{lscape}
\usepackage{bbm}
\usepackage{scrextend}
\usepackage{relsize}
\usepackage{multicol}
\usepackage{chngcntr}
\usepackage{etoolbox}
\usepackage{tabularx}
\usepackage[english]{babel}
\usepackage{multirow}
\usepackage{booktabs}
\usepackage{bookmark}
\usepackage[labelfont=bf]{caption}
\usepackage{titlesec}
\usepackage{array}
\usepackage{color}
\usepackage{afterpage}
\usepackage{framed}
\usepackage{mathrsfs}
\usepackage{xcolor}
\usepackage{longtable}
\usepackage{rotating}
\usepackage{makecell}
\usepackage{afterpage}%
\providecommand{\U}[1]{\protect\rule{.1in}{.1in}}
\allowdisplaybreaks

\theoremstyle{plain}
\newtheorem{theorem}{Theorem}

\newtheorem{remark}{Remark}
\newtheorem{corollary}{Corollary}

\newtheorem{assumption}{Assumption}

\theorembodyfont{\normalfont}
\deffootnote{1em}{1.6em}{\thefootnotemark \enskip}

\numberwithin{equation}{section}
\numberwithin{assumption}{section}

\providecommand{\keywords}[1]
{
  {
  \small
  \textbf{Keywords:} #1
  }
}
\definecolor{lightgray}{gray}{0.9}
\newcommand{\var}{\operatorname{Var}}

\begin{document}

\title{Data-driven Smooth Tests for Normality in ANOVA When the Number of Groups is Large}
\author{Peiwen Jia\thanks{Department of Business Statistics and Econometrics,
Guanghua School of Management, Peking University, Beijing, 100871, China. E-mail: \texttt{2201111026@stu.pku.edu.cn}.}\\Peking University
\and Xiaojun Song\thanks{Corresponding author: Department of Business Statistics and Econometrics, Guanghua School of Management, Peking University, Beijing, 100871, China. E-mail: \texttt{sxj@gsm.pku.edu.cn}. This work was supported by the National Natural Science Foundation of China [Grant Numbers 72373007 and 72333001]. The author also gratefully acknowledges the research support from the Center for Statistical Science of Peking University, China, and the Key Laboratory of Mathematical Economics and Quantitative Finance (Peking University) of the Ministry of Education, China.}
\\Peking University
\and Haoyu Wei\thanks{Department of Economics, University of California, San Diego, La Jolla, 92092, USA. E-mail: \texttt{h8wei@ucsd.edu}.} \\University of California, San Diego}

\maketitle

\begin{abstract}
The normality assumption for random errors is fundamental in the analysis of variance (ANOVA) models. However, it is rarely subjected to formal testing in practice, and theoretically justified procedures are largely unavailable, especially when the number of groups diverges. In this paper, we develop Neyman's smooth tests for assessing normality in a broad class of ANOVA models, allowing the number of groups to diverge. The proposed test statistics are constructed via the Gaussian probability integral transformation of ANOVA residuals. We show that using residuals induces non-negligible parameter estimation effects, whose structure depends on the underlying ANOVA model and plays a crucial role in shaping the form of the test statistics and their asymptotic behavior. Under the null hypothesis of normality, the resulting statistics follow an asymptotic Chi-square distribution, with degrees of freedom determined by the order of the smooth test (i.e., the number of components included in the smooth test). We further propose a modified Schwarz's selection rule to automatically determine the order, thereby yielding fully data-driven smooth tests that require no additional tuning parameters. 
Simulation studies and a real-data example indicate that the proposed tests perform well in practice and are readily applicable.
\end{abstract}


\section{Introduction}
Analysis of Variance (ANOVA) is a fundamental and widely used tool in both exploratory and confirmatory data analysis \citep{gelman2005analysis}, particularly for comparing group means and assessing the significance of factors in experimental designs. The theory of ANOVA has been well established in the literature; see, for example, \cite{scheffe1959anova}, \cite{miller1997beyond}, \cite{wickens2004design}, \cite{dean2017design}, \cite{hirotsu2017advanced}, and \cite{montgomery2017design}. A standard assumption in ANOVA is that the random errors are normally distributed. Violations of this assumption may invalidate normal-based inference. From an estimation perspective, departures from normality undermine the validity of variance component estimators, since the classical variance formulae rely on mean squares of random effects being constant multiples of Chi-square variables \citep{scheffe1959anova}. From a testing perspective, the $F$-test in ANOVA is sensitive to nonnormality, especially when group sample sizes are unbalanced \citep{ali1996robustness} or the number of groups is large \citep{akritas2000asymptotics, akritas2004heteroscedastic, wang2006two}. The impact of nonnormality on the size and power of the $F$-test has been extensively studied \citep{pearson1931analysis, gayen1950distribution, david1951method, david1951effect, srivastava1959effect, atiqullah1962estimation, tiku1964approximating, donaldson1968robustness, tiku1971power, akritas2000asymptotics, janic2000data}. These considerations highlight the importance of rigorously evaluating the normality assumption in ANOVA models. 

Although a variety of normality tests have been proposed for observed data and linear models (see, for example, \cite{dagostino1986goodness} and \cite{bonett2002test}), relatively little attention has been paid to the systematic assessment of normality in ANOVA. A straightforward diagnostic approach is to use the normal probability plots of ANOVA residuals \citep{dean2017design, hirotsu2017advanced}, which is intuitive but lacks theoretical justification. To our knowledge, formal normality testing procedures explicitly designed for ANOVA are scarce in the literature. Most existing methods instead apply classical normality tests---such as the Shapiro--Wilk test \citep{shapiro1965analysis, shapiro1972approximate}, the Jarque--Bera test \citep{jarque1987test}, and the Kolmogorov--Smirnov test (in particular, the Lilliefors version) \citep{lilliefors1967kolmogorov}---directly to ANOVA residuals; see, for example, \cite{bonett1990testing, bonett2002test, hwang2006novel}. While convenient in practice, such approaches suffer from two fundamental limitations. First, they ignore the estimation effect induced by fitting the ANOVA model, which may alter the distributional properties of the residuals. Second, they lack theoretical guarantees, especially in ANOVA models with a diverging number of groups, under which these procedures may become invalid. Consequently, they do not provide a rigorous assessment of normality within the ANOVA framework and fail to account for the inherent structural constraints of ANOVA.

In this article, following the spirit of \cite{neyman1937smooth}, we propose a unified Neyman's smooth test framework for assessing the normality of random errors in various types of ANOVA models where the number of groups is allowed to diverge. Neyman's smooth tests are widely used in diverse scientific fields due to their theoretical soundness and practical effectiveness; see, e.g., \cite{bera_neymans_2002}, Chapters 4 and 10 of \cite{thasComparingDistributions2010}, and Section 16.4 of  \cite{Lehmann_Romano_2022}. Specifically, we reformulate the normality testing problem as a uniformity problem via the Gaussian probability integral transform (PIT) and construct test statistics based on the Gaussian PIT of the ANOVA residuals. The use of residuals introduces parameter estimation effects whose structure depends on the underlying model and is therefore intrinsic to the ANOVA framework. Consequently, the resulting statistic takes the form of a quadratic expression involving the inverse of a covariance matrix, whose structure is determined by these estimation effects. Given mild conditions, the proposed test statistics are asymptotically Chi-square distributed under the null hypothesis of normality. We also analyze the power properties under both fixed and local alternatives. Furthermore, a modified Schwarz's selection rule is proposed to determine the test directions (i.e., the order of the test), yielding fully data-driven smooth tests that do not require additional tuning parameters. Numerical results demonstrate the good performance of our proposed methodology in finite samples.

The main contributions of this paper are twofold. First, we systematically investigate the effects of parameter estimation on ANOVA residuals-based normality testing. We show that heterogeneity in group means and variances alters the explicit form of the test statistics and the associated regularity conditions, while the convergence rate depends solely on the total sample size under both the null and alternative hypotheses. Second, we provide rigorous theoretical justification for data-driven smooth tests in the ANOVA setting. In particular, we establish a revised limiting null distribution for the corresponding statistics that remains valid in finite samples. As a result, the proposed tests are simple to implement, with critical values obtained directly from the limiting distributions without resorting to resampling. Overall, the proposed framework accommodates settings with a potentially diverging number of groups and thus overcomes key limitations of many classical methods.

The remainder of this paper is organized as follows. Section \ref{section_frame} introduces the general testing framework. Section \ref{section_prop} establishes theoretical results for three types of ANOVA models. Section \ref{section_datadriven} proposes a data-driven testing procedure. Section \ref{section_simu} reports simulation results. Section \ref{section_emp} provides an empirical application. Section \ref{section_end} concludes. Proofs and additional numerical results are provided in the \hyperlink{app}{Online Appendix}.

\section{The testing framework} \label{section_frame}

Our problem of interest is to assess whether the random error $\varepsilon = \sigma e$ in ANOVA models is normally distributed with mean zero and variance $\sigma^2$, that is, $\varepsilon \sim \mathcal{N} (0,\sigma^2)$ for some $\sigma>0$. Equivalently, this can be formulated as testing the null hypothesis about the standardized random error $H_0: e \sim \mathcal{N} (0,1)$ against alternatives of nonnormality. To motivate our methodology, consider the transformation  $Z=\Phi(e)$, where $\Phi(\cdot)$ denotes the cumulative distribution function (CDF) of the standard normal distribution. Under the null hypothesis of normality, the CDF of the transformed random variable $Z$ is given by
\begin{equation*}
	G(z)=\mathbb{P}(Z\le z)=\mathbb{P}\left(e \le \Phi^{-1}(z)\right)=\Phi(\Phi^{-1}(z))=z.
\end{equation*} 
Or, equivalently, under $H_0$, the probability density function (PDF) of $Z$ is $g(z)\equiv 1, z\in[0,1]$, that is, $Z \sim \mathcal{U}[0,1]$, where $\mathcal{U}[0,1]$ denotes the uniform distribution on the interval $[0,1]$. Under the alternative, the density $g(z)$ deviates from unity. This distinct behavior of $Z$ under $H_0$ and $H_1$ forms the basis of Neyman's smooth test. In particular, \cite{neyman1937smooth} introduced the following \emph{smooth} alternative to the uniform density:
\begin{equation}\label{pit_density1}
	h\left(z\right)  =c\left(  \bm\theta_K\right)  \exp\left[  \sum_{k=1}^{K}%
	\theta_k\pi_k\left(z\right)  \right], \quad 0\le z\le 1,
\end{equation}
where $c(\bm\theta_K)$ is a normalizing constant depending on
$\bm\theta_K = (\theta_1,\theta_2,...,\theta_K)^\top$, and $\{  \pi_k\left(z\right)\}  _{k=0}^{\infty}$ denotes an orthonormal system in $L_2[0,1]$ with $\pi_0(z)=1$ and
\begin{equation}
\int_0^1\pi_k\left(z\right)  \pi_{l}\left(z\right)  \mathrm{d}z=\delta
_{kl}, \text{ where }\delta_{kl}=\begin{cases}
1\text{, if }k=l,\\
0\text{, if }k\ne l.%
\end{cases}  \label{orth_cond1}%
\end{equation}
The null hypothesis $Z\sim \mathcal{U}[0,1]$ can thus be assessed by testing $\theta_1=\theta_2=  \ldots =\theta_K=0$ in \eqref{pit_density1}. 

If independent and identically distributed (i.i.d.) observations $\{\varepsilon_{i}\}_{i=1}^n$ are available and $\sigma$ is also known, the smooth test statistic for testing $H_0$ has the following quadratic form:
\begin{equation}\label{iFY}
	\Psi_K^2 = \sum_{k=1}^{K}\left(\frac{1}{n} \sum_{i=1}^{n} \pi_k(Z_{i})\right)^2,
\end{equation}
where $K$ is a fixed and given positive integer and $Z_{i}=\Phi(e_i)=\Phi(\varepsilon_i/\sigma)$. Under $H_0$, the statistic $n\Psi_K^2$ converges in distribution to a Chi-square law with $K$ degrees of freedom, denoted by $\chi_K^2$. Moreover, the test inherits the local optimal properties of Rao's score test.

However, in practice, the sequence $\{\varepsilon_{i}\}_{i=1}^n$ is unobserved and $\sigma$ is unknown. We instead rely on approximations $\{\widehat{\varepsilon}_i\}_{i=1}^n$ for $\{\varepsilon_{i}\}_{i=1}^n$ and an estimator $\widehat{\sigma}$ for $\sigma$ (e.g., residuals and the sample standard deviation, respectively). Accordingly, we consider the following approximation for $Z_i$:
\begin{equation*}
	\widehat{Z}_{i}=\Phi(\widehat{e}_i)=\Phi\left(\frac{\widehat\varepsilon_{i}}{\widehat\sigma}\right), \quad i=1,\ldots,n,
\end{equation*}
and construct a feasible test statistic of the following form:
\begin{equation}
	\sum_{k=1}^{K}\left(\frac{1}{n}\sum_{i=1}^{n}\pi_k\left(\widehat Z_{i}\right)\right)^2. \label{naive}
\end{equation}
The random variables $\{\widehat{Z}_{i}\}_{i=1}^n$ are not i.i.d. any longer due to estimation effects caused by $\widehat\varepsilon_i$ and $\widehat\sigma$. Therefore, it is crucial to study the asymptotic properties of ${n}^{-1}\sum_{i=1}^{n}\pi_k(\widehat Z_{i})$. Compared with $\Psi_K^2$ in \eqref{iFY}, the statistic based on $\widehat Z_{i}$ suffers from non-negligible estimation effects, as demonstrated in the theoretical results below. Consequently, the presence of estimation effects invalidates the form of the ``naive'' statistic \eqref{naive} and requires a different normalization matrix to restore the $\chi_K^2$ limiting null distribution. This motivates a detailed analysis of the effects of estimation that yields the correct quadratic-form test statistic.

Another important issue concerns sample sizes in ANOVA models. To highlight the essence of smooth tests, we first focus on the simplified setting of a single random sequence with sample size $n$ throughout this section. This simplified setting captures the core methodological ideas and facilitates the derivation of key asymptotic properties. In contrast, the complete ANOVA setting introduces additional complications due to multiple sample sizes and the associated indexing structure, necessitating a more refined analysis. From a practical standpoint, our method is designed to accommodate scenarios in which the number of groups increases with the total sample size. These issues will be systematically investigated in subsequent analysis of specific ANOVA models.

\section{Testing for normality in one-way fixed effects models} \label{section_prop}

In this section, we develop smooth tests for normality in three types of one-way fixed effects models, each imposing homogeneity on either the group means or the group variances. The testing procedures and their theoretical properties are studied separately for each case. Formally, let $\{Y_{ij}\}_{i=1,j=1}^{N_j,J}$ denote the observed responses from $J$ groups, where $N_j$ is the sample size of group $j$ for $1 \le j \le J$. The total sample size is $N = \sum_{j=1}^J N_j$. In our asymptotic framework, the number of groups $J$ is allowed to diverge as $N \to \infty$.

\subsection{Common group means and common group variances} \label{section_prop.1}

We begin with ANOVA models in which both the group means and group variances are equal. Specifically, consider the model 
\begin{equation}
	Y_{ij} = \mu + \varepsilon_{ij} = \mu + \sigma e_{ij}, \quad i = 1, \ldots, N_j, \quad j = 1, \ldots, J, \label{model_same}
\end{equation}
where $\{e_{ij}\}_{i=1,j=1}^{N_j,J}$ are i.i.d. standardized random errors with mean zero and unit variance, $\mu$ denotes the common group mean, and $\sigma$ is the standard deviation of random errors $\{\varepsilon_{ij}\}_{i=1,j=1}^{N_j,J}$. We also define $Z_{ij} = \Phi(e_{ij})$.

The parameters $\mu$ and $\sigma^2$ in model \eqref{model_same} are estimated by the sample mean $\widehat\mu=N^{-1}\sum_{j=1}^{J}\sum_{i=1}^{N_j}Y_{ij}$ and the sample variance $\widehat\sigma^2=N^{-1}\sum_{j=1}^{J}\sum_{i=1}^{N_j}\left(Y_{ij}-\widehat\mu\right)^2$, respectively. Define $\widehat{\varepsilon}_{ij}=Y_{ij}-\widehat{\mu}$, $\widehat{e}_{ij}=\widehat{\varepsilon}_{ij}/\widehat \sigma=(Y_{ij}-\widehat{\mu}) / \widehat{\sigma}$, and $\widehat Z_{ij} = \Phi (\widehat{e}_{ij})$. To derive the feasible test statistic and its properties, we first analyze the asymptotic behavior of $N^{-1}\sum_{j=1}^{J}\sum_{i=1}^{N_j}\pi_k(\widehat Z_{ij})$ for $k=1,\ldots,K$. For this purpose, we impose the following two assumptions.

\begin{assumption} \label{assumption1}
	$\{e_{ij}\}_{i=1,j=1}^{N_j,J}$ are i.i.d. with continuous CDF $F(x)$, PDF $f(x)$, mean zero, unit variance, and finite sixth moments.   
\end{assumption}
\begin{assumption}  \label{assumption2}
	For $k =1,\ldots, K$, $\pi_k(\cdot)$ are twice continuously differentiable with derivatives $\dot\pi_k(\cdot)$ and $\ddot\pi_k(\cdot)$, and they are both bounded.    
\end{assumption} 
The two mild assumptions are in line with similar ones adopted in the literature of Neyman’s smooth tests, including \cite{bera2013smooth} and \cite{song2022smooth}. For subsequent analysis, we introduce two constants:
\begin{equation*}
	c_{1k} = \int_0^1 \pi_k (z)  \Phi^{-1} (z)  \mathrm{d}z, \quad c_{2k} = \int_0^1 \pi_k (z)  \left( \Phi^{-1} (z) \right)^2  \mathrm{d}z, \quad k=1,\ldots,K.
\end{equation*}

\begin{theorem}\label{H0_same}
	Suppose Assumptions \ref{assumption1} and \ref{assumption2} hold. Then, under the null hypothesis $H_0: e \sim \Phi(x)$, for $k=1,\ldots,K$,
	\begin{equation*}
		\frac{1}{N}\sum_{j=1}^{J}\sum_{i=1}^{N_j}\pi_k\left(\widehat Z_{ij}\right)=\frac{1}{N}\sum_{j=1}^{J}\sum_{i=1}^{N_j}\left\{\pi_k(Z_{ij})-c_{1k}e_{ij}-\frac{c_{2k}}{2}\left(e_{ij}^2-1\right)\right\} + o_p\left(\frac{1}{\sqrt{N}}\right),
	\end{equation*}
	as $N\to\infty$.
\end{theorem}

Theorem \ref{H0_same} states that  $N^{-1}\sum_{j=1}^{J}\sum_{i=1}^{N_j}\pi_k(\widehat Z_{ij})$ is 
equivalent to the sum of three terms after neglecting the higher-order ones: $N^{-1}\sum_{j=1}^{J}\sum_{i=1}^{N_j}\pi_k(Z_{ij})$, $-c_{1k}N^{-1}\sum_{j=1}^{J}\sum_{i=1}^{N_j}e_{ij}$, and $-c_{2k}(2N)^{-1}\sum_{j=1}^{J}\sum_{i=1}^{N_j}(e_{ij}^2-1)$. The latter two represent the estimation effects due to estimating $\mu$ and $\sigma^2$, respectively, and they contribute to the limiting distribution of the feasible smooth test statistic based on $N^{-1}\sum_{j=1}^{J}\sum_{i=1}^{N_j}\pi_k(\widehat Z_{ij})$,  $k=1,\ldots,K$.

With the assistance of Theorem \ref{H0_same}, and by invoking the central limit theorem (CLT), we obtain the asymptotic normality of the $K$-dimensional vector
\begin{equation}\label{asym:thm1:K_dim_vec}
	\begin{aligned}
		\frac{1}{\sqrt{N}}\sum_{j=1}^{J}\sum_{i=1}^{N_j}\bm\pi_K\left(\widehat Z_{ij}\right) & =\sqrt{N}\left[\frac{1}{N}\sum_{j=1}^{J}\sum_{i=1}^{N_j}\pi_1\left(\widehat Z_{ij}\right),\ldots,\frac{1}{N}\sum_{j=1}^{J}\sum_{i=1}^{N_j}\pi_K\left(\widehat Z_{ij}\right)\right]^\top \\
		& \overset{d}{\to}  \mathcal{N}_K\left(\bm{0}, \bm{\Sigma}_K\right),
	\end{aligned}
\end{equation}
where the asymptotic covariance matrix $\bm{\Sigma}_K=\left(\sigma_{kl}\right)_{K\times K}$ is given by
\begin{align*}
	\sigma_{kl} &= \mathbb{E} \left[\left(\pi_k(Z)-c_{1k}e-\frac{c_{2k}}{2}\left(e^2-1\right)\right)\left(\pi_{l}(Z)-c_{1l}e-\frac{c_{2l}}{2}\left(e^2-1\right)\right)\right]\\
	& =\delta_{kl} - c_{1k} c_{1l} - \frac{1}{2} c_{2k} c_{2l}.
\end{align*}
Based on this result, we define the feasible smooth test statistic as 
\begin{equation}
	\widehat\Psi_K^2=\left(\frac{1}{N}\sum_{j=1}^{J}\sum_{i=1}^{N_j}\bm\pi_K\left(\widehat Z_{ij}\right)\right)^\top\bm{\Sigma}_K^{-1}\left(\frac{1}{N}\sum_{j=1}^{J}\sum_{i=1}^{N_j}\bm\pi_K\left(\widehat Z_{ij}\right)\right). \label{hat_Psi_k_1}
\end{equation}
The following corollary establishes the limiting null distribution of $\widehat\Psi_K^2$ defined in \eqref{hat_Psi_k_1}.

\begin{corollary}\label{corollary1}
	Suppose Assumptions \ref{assumption1} and \ref{assumption2} hold. Then, under the null hypothesis $H_0:e  \sim\Phi(x)$,
	\begin{equation*}
		N\widehat\Psi_K^2  \overset{d}{\to}  \chi_K^2,
	\end{equation*}
	as $N\to\infty$.
\end{corollary}

Corollary \ref{corollary1} establishes an asymptotic $\chi^2$ test for the normality of model \eqref{model_same} based on the statistic $N\widehat\Psi_K^2$. Given the asymptotic significance level of $\alpha$, we reject the null hypothesis if
\begin{equation*}
	N\widehat\Psi_K^2 >  \chi_{K, 1 - \alpha}^2,
\end{equation*}
where $\chi_{K, 1 - \alpha}^2$ denotes the $(1 - \alpha)$-th quantile of the $\chi_K^2$ distribution.

We now investigate the asymptotic behavior of $\widehat\Psi_K^2$ under the fixed alternatives as well as under a Pitman-type sequence of local alternatives. For the fixed alternatives, we consider the following form:
\begin{equation}
	H_1: F(x)\ne\Phi(x)\text{, and there exists at least one }1\le k \le K\text{ such that } \mathbb{E}[\pi_k (Z)] \ne 0. \label{H1_fixed}
\end{equation}
Under $H_1$ given by \eqref{H1_fixed}, the random error $e$ is no longer normally distributed. Consequently, the asymptotic behavior of $N^{-1}\sum_{j=1}^{J}\sum_{i=1}^{N_j}\pi_k(\widehat Z_{ij})$ and $\widehat\Psi_K^2$ differs from that under $H_0$. To formally state the corresponding theoretical results under $H_1$, we introduce the following notations. Let
\begin{equation*}
	d_{1k}=\mathbb{E} \left[\dot\pi_k(Z)\phi\left(e\right)\right]=\int_{- \infty}^{\infty}\dot\pi_k\left(\Phi\left(x\right)\right)\phi\left(x\right)f(x)\mathrm{d}x,
\end{equation*} 
\begin{equation*}
	d_{2k}=\mathbb{E} \left[\dot\pi_k(Z)\phi\left(e\right)e\right]=\int_{- \infty}^{\infty}\dot\pi_k\left(\Phi\left(x\right)\right)\phi\left(x\right)xf(x)\mathrm{d}x,
\end{equation*} 
\begin{equation*}
	d_{3k}=\mathbb{E} \left[\pi_k(Z)e\right]=\int_{- \infty}^{\infty}\pi_k\left(\Phi\left(x\right)\right)xf(x)\mathrm{d}x,
\end{equation*} 
and
\begin{equation*}
	d_{4k}=\mathbb{E} \left[\pi_k(Z)e^2\right]=\int_{- \infty}^{\infty}\pi_k\left(\Phi\left(x\right)\right)x^2f(x)\mathrm{d}x,
\end{equation*} 
for $k=1,\ldots,K$, where $\phi$ denotes the PDF of standard normal distribution. Note that under $H_0$, the constants $d_{1k}=d_{3k}=c_{1k}$ and $d_{2k}=d_{4k}=c_{2k}$ since $f(x) = \phi(x)$. The following theorem characterizes the asymptotic properties of $\widehat\Psi_K^2$ under the alternative hypothesis $H_1$.
\begin{theorem}\label{H1_same}
	Suppose Assumptions \ref{assumption1} and \ref{assumption2} hold. Then, under the alternative hypothesis $H_1$ in \eqref{H1_fixed}, for $k=1,\ldots,K$, 
	\begin{equation}\label{H1_same_decomp}
		\begin{aligned}
			\frac{1}{N}\sum_{j=1}^{J}\sum_{i=1}^{N_j}\pi_k\left(\widehat Z_{ij}\right)=&\frac{1}{N}\sum_{j=1}^{J}\sum_{i=1}^{N_j}\left\{\left(\pi_k(Z_{ij})-\mathbb{E} \left[\pi_k(Z)\right]\right)-d_{1k}e_{ij}-\frac{d_{2k}}{2}\left(e_{ij}^2-1\right)\right\}\\
			&+\mathbb{E} \left[\pi_k(Z)\right]+o_p\left(\frac{1}{\sqrt N}\right),
		\end{aligned}
	\end{equation}
	as $N\to\infty$. Furthermore, 
	\begin{equation*}
		\widehat{\Psi}_K^2   \overset{p}{\to}  \bm{a}_K^\top \bm{\Sigma}_K^{-1} \bm{a}_K,
	\end{equation*}
	and 
	\begin{equation}
		\sqrt{N} \left( \widehat{\Psi}_K^2 - \bm{a}_K^\top \bm{\Sigma}_K^{-1} \bm{a}_K \right)  \overset{d}{\to}  \mathcal{N}\left(0, 4 \bm{a}_K^\top \bm{\Sigma}_K^{-1} \bm{\Xi}_K \bm{\Sigma}_K^{-1} \bm{a}_K\right), \label{sandwitch}
	\end{equation}
	where $\bm{a}_K\equiv (a_1,\ldots,a_K)^\top = ( \mathbb{E}[\pi_1 (Z)], \ldots, \mathbb{E}[\pi_K (Z)])^\top$ and $\bm{\Xi}_K=\left(\xi_{kl}\right)_{K\times K}$ is given by
	\begin{align*}
		\xi_{kl} = & \mathbb{E} \left[ \pi_k (Z) \pi_l (Z) \right]-a_k a_l-\left[d_{1k}d_{3l} + d_{1l}d_{3k} \right] + d_{1k} d_{1l} + \frac{1}{2} \left[ a_k d_{2l} + a_l d_{2k} \right] \\
		&-\frac{1}{2}\left[ d_{2k}d_{4l} + d_{2l}d_{4k}\right] + \frac{1}{2}\left[ d_{1k}d_{2l} + d_{1l}d_{2k}\right] \mathbb{E} \left[e^3\right] +\frac{d_{2k} d_{2l} }{4} \left[\mathbb{E} \left[e^4\right]-1 \right].
	\end{align*}
\end{theorem}

The decomposition \eqref{H1_same_decomp} exhibits different properties from that in Theorem \ref{H0_same}. Specifically, the first term on the right-hand side $N^{-1}\sum_{j=1}^{J}\sum_{i=1}^{N_j} \pi_k(Z_{ij})$ now has a nonzero expectation under $H_1$ given by \eqref{H1_fixed}, in contrast to its zero mean under $H_0$, which constitutes the primary source of power for the proposed test. The latter two terms, $-d_{1k}N^{-1}\sum_{j=1}^{J}\sum_{i=1}^{N_j}e_{ij}$ and $-d_{2k}(2N)^{-1}\sum_{j=1}^{J}\sum_{i=1}^{N_j}(e_{ij}^2-1)$, represent the estimation effects of $\widehat \mu$ and $\widehat \sigma^2$ under $H_1$, respectively. Moreover, when $H_0$ is true, $d_{1k} = c_{1k}$ and $d_{2k} = c_{2k}$, then the decomposition \eqref{H1_same_decomp} coincides with that in Theorem \ref{H0_same}. 

From Theorem \ref{H1_same}, it follows that under $H_1$ given by \eqref{H1_fixed}, if there exists at least one $1\le k\le K$ such that 
\begin{equation*}
	a_k \equiv \mathbb{E} [\pi_k(Z)]=\int_{-\infty}^{\infty} \pi_k(\Phi(x))f(x)\mathrm{d}x
	=\int_0^1\pi_k(z)\frac{f(\Phi^{-1}(z))}{\phi(\Phi^{-1}(z))}\mathrm{d}z\ne 0, 
\end{equation*}
then the smooth test statistic satisfies $N \widehat{\Psi}_K^2 \to \infty$ in probability as $N \to \infty$, implying that the asymptotic power of the test is $1$. Specifically, the power function is given by
\begin{equation*}
	1 - \Phi \left( \left(4 N \bm{a}_K^\top \bm{\Sigma}_K^{-1} \bm{\Xi}_K \bm{\Sigma}_K^{-1} \bm{a}_K\right)^{-1/2} \chi_{K, 1 - \alpha}^2 - \left( N \bm{a}_K^\top \bm{\Sigma}_K^{-1} \bm{\Xi}_K \bm{\Sigma}_K^{-1} \bm{a}_K/4 \right)^{1/2}\right).
\end{equation*}

For further investigation of the test's power, we consider a Pitman-type sequence of local alternatives that converges to the null hypothesis at an appropriate rate, which is specified as follows:
\begin{equation}
	H_{1L}: F(x) = (1-\delta_N)\Phi(x)+\delta_N Q(x), \label{H1L}
\end{equation}
where $Q\left(x\right)$ (which admits PDF $q\left(x\right)$) represents some distribution function that is different from $\Phi(x)$, and $\delta_N \to 0$ as $N\to \infty$. Define
\begin{equation*}
	\Delta_k = \int_0^1\pi_k(z)\frac{q(\Phi^{-1}(z))}{\phi(\Phi^{-1}(z))} \mathrm{d}z, \quad k=1,\ldots,K.    
\end{equation*}
The following theorem presents the theoretical properties of $N\widehat{\Psi}_K^2$ under the local alternatives. Specifically, the rate of $\delta_N$ tending to $0$ as $N \to \infty$ is crucial to the nontrivial local power of the test.

\begin{theorem} \label{H1L_same}
	Suppose Assumptions \ref{assumption1} and \ref{assumption2} hold. Then, under the local alternative hypothesis $H_{1L}$ in \eqref{H1L}, with $\delta_N = N^{-1/2}$, for $k=1,\ldots,K$, 
	\begin{equation} \label{decomp_H1L}
		\begin{aligned}
			\frac{1}{N}\sum_{j=1}^{J}\sum_{i=1}^{N_j}\pi_k\left(\widehat Z_{ij}\right)=&\frac{1}{N}\sum_{j=1}^{J}\sum_{i=1}^{N_j}\left\{\left(\pi_k(Z_{ij})-\mathbb{E} \left[\pi_k(Z)\right]\right)-c_{1k}e_{ij}-\frac{c_{2k}}{2}\left(e_{ij}^2-1\right)\right\}\\
			&+\delta_N \Delta_k+o_p\left(\frac{1}{\sqrt N}\right),
		\end{aligned}
	\end{equation}
	as $N\to\infty$. Furthermore, 
	\begin{equation*}
		N\widehat{\Psi}_K^2   \overset{d}{\to}  \chi_K^2\left(\bm{\Delta}_K^\top\bm{\Sigma}_K^{-1}\bm{\Delta}_K\right),
	\end{equation*}
	where $\bm{\Delta}_K = ( \Delta_1, \ldots, \Delta_K)^\top$, and $\chi^2_K \left(\tau\right)$ denotes the noncentral $\chi^2$ distribution with $K$ degrees of freedom and nonnegative noncentrality parameter $\tau$.
\end{theorem}

The asymptotic decomposition \eqref{decomp_H1L} is closely related to those obtained under $H_0$ and $H_1$, while exhibiting a key difference. In particular, compared with the decomposition under $H_0$, the leading term $N^{-1}\sum_{j=1}^{J}\sum_{i=1}^{N_j}\pi_k(Z_{ij})$ becomes noncentered under $H_{1L}$, with mean shift $\mathbb{E}[\pi_k(Z)] = \delta_N \Delta_k$, which introduces an additional deterministic drift term of order $\delta_N$. The two estimation effect terms $-c_{1k}N^{-1}\sum_{j=1}^{J}\sum_{i=1}^{N_j}e_{ij}$ and $-c_{2k}(2N)^{-1}\sum_{j=1}^{J}\sum_{i=1}^{N_j}(e_{ij}^2-1)$ coincide with those under $H_0$. When $\delta_N = N^{-1/2}$, the drift term $\delta_N \Delta_k$ is of the same order as the stochastic fluctuations, leading to a nondegenerate limit distribution. Consequently, the statistic $N\widehat{\Psi}_K^2$ converges to a noncentral Chi-square distribution. This also shows that $N^{-1/2}$ is the critical rate for detecting Pitman-type local alternatives: if $\delta_N = o(N^{-1/2})$, the alternatives become asymptotically indistinguishable from the null. Under $H_{1L}$ in \eqref{H1L}, as long as there exists at least one $1 \le k \le K$ such that $\Delta_k \ne 0$, the proposed test statistic $N \widehat{\Psi}_K^2$ attains nontrivial asymptotic power against the local alternatives because the noncentrality parameter $\bm{\Delta}_K^\top\bm{\Sigma}_K^{-1}\bm{\Delta}_K>0$.

\begin{remark} \emph{
		We emphasize that the aforementioned smooth test is not ``\textit{consistent}'' in the strict sense; that is, the power of the test does not necessarily approach $1$ as $N \to \infty$ under all directions of fixed alternatives. Only under alternatives where $\mathbb{E} [\pi_k(Z)]\ne 0$ for at least one $1 \leq k \leq K$ does the asymptotic power converge to $1$. Otherwise, if $\mathbb{E} [\pi_k(Z)]=0$ for all $1 \leq k \leq K$ (but $\mathbb{E} [\pi_{K+1}(Z)]\ne 0$ maybe), $\widehat \Psi_K^2$ will fail to detect the discrepancy between $F$ and $\Phi$. For example, let $e\sim \mathcal{U}[-\sqrt{3},\sqrt{3}]$ so that $\mathbb{E}[e]=0$ and $\var[e]=1$, and consider the first-order orthonormal Legendre polynomial $\pi_1(z)=\sqrt{3}(2z-1)$. For $Z=\Phi(e)$, we have
		\begin{align*}
			\mathbb{E}[Z] &=\frac{1}{2\sqrt{3}} \int_{-\sqrt{3}}^{\sqrt{3}} \Phi(z) \mathrm{d}z \\
			&= \frac{1}{2\sqrt{3}} z\Phi(z)\Big|_{-\sqrt{3}}^{\sqrt{3}}-\frac{1}{2\sqrt{3}} \int_{-\sqrt{3}}^{\sqrt{3}}z \mathrm{d} \Phi(z) \\
			&=\frac{1}{2} \left(\Phi\left(\sqrt{3}\right)+\Phi\left(-\sqrt{3}\right)\right) \\
			& = \frac{1}{2},
		\end{align*}
		which implies $\mathbb{E}[\pi_1(Z)]=0$ and thus no power
under this choice. In fact, smooth tests are neither \textit{directional} nor \textit{omnibus}; that is, they maintain reasonable power across a broad---but not universal---range of alternatives, and generally exhibit good finite-sample power properties; see \cite{bera2013smooth} for further discussion.}
\end{remark}

\subsection{Heterogeneous group means with common variances} \label{section_prop.2}

Next, we consider ANOVA models with heterogeneous group means while maintaining common variances:
\begin{equation}
	Y_{ij} = \mu_j + \varepsilon_{ij}=\mu_j+ \sigma e_{ij},\quad i=1,\ldots,N_j,\quad j=1,\ldots,J, \label{model_diff_mean}
\end{equation}
where $\{\mu_j\}_{j=1}^J$ denote the potentially distinct means of each group. Let $\widehat\mu_j=N_j^{-1}\sum_{i=1}^{N_j}Y_{ij}$ and $\widehat\sigma^2=N^{-1}\sum_{j=1}^{J}\sum_{i=1}^{N_j}\left(Y_{ij}-\widehat\mu_j\right)^2$ be the estimators of $\mu_j$ ($j = 1,\ldots,J$) and $\sigma^2$, respectively. Correspondingly we define $\widehat \varepsilon
_{ij} = Y_{ij}-\widehat\mu_j$, $\widehat{e}_{ij}=\widehat \varepsilon
_{ij}/\widehat \sigma=(Y_{ij}-\widehat\mu_j) / \widehat{\sigma}$, and $\widehat Z_{ij}=\Phi(\widehat{e}_{ij})$. The following theorem characterizes the properties of $N^{-1}\sum_{j=1}^{J}\sum_{i=1}^{N_j}\pi_k(\widehat Z_{ij})$ under the model \eqref{model_diff_mean}.

\begin{theorem}\label{H0_different_mean}
	Suppose Assumptions \ref{assumption1} and \ref{assumption2} hold. Then, under the null hypothesis $H_0: e \sim\Phi(x)$, for $k=1,\ldots,K$, 
	\begin{align*}
		\frac{1}{N}\sum_{j=1}^{J}\sum_{i=1}^{N_j}\pi_k\left(\widehat Z_{ij}\right)=&\frac{1}{N}\sum_{j=1}^{J}\sum_{i=1}^{N_j}\left\{\pi_k(Z_{ij})-c_{1k}e_{ij}-\frac{c_{2k}}{2}\left(e_{ij}^2-1\right)\right\}+  o_p\left(\frac{1}{\sqrt{N}}\right),
	\end{align*}
	as $\min\{N_1,\ldots,N_J\}\to\infty$, $J = o( N^{1/2})$ and $\sum_{j=1}^J N_j^{-1}=o(1)$. Furthermore, for $\widehat{\Psi}_K^2$ defined in \eqref{hat_Psi_k_1}, we have
	\begin{equation*}
		N\widehat\Psi_K^2  \overset{d}{\to}  \chi_K^2.
	\end{equation*}
\end{theorem}

Theorem \ref{H0_different_mean} parallels Theorem \ref{H0_same} and Corollary \ref{corollary1}, except that it requires additional conditions on the group sample sizes and the number of groups. These conditions are imposed for technical reasons and to account for heterogeneity across group means in model \eqref{model_diff_mean}. The intuition behind the conditions is illustrated as follows. First, the sample size of each group, $N_j$, must tend to infinity. In model \eqref{model_diff_mean}, where $\mu_j$ can be distinct, only the observations within group $j$ can be used to estimate $\mu_j$, making $N_j \to \infty$ necessary for the consistency of $\widehat \mu_j$. Second, the number of groups $J$ is required to be finite or to diverge no faster than $N^{1/2}$, and $\sum_{j=1}^J N_j^{-1}=o(1)$ must hold jointly for $\{N_j\}_{j=1}^J$ and $J$. This rules out scenarios where the number of groups grows faster than the sample sizes within each group. In contrast, for model \eqref{model_same} in the previous section, it suffices that the total sample size $N \to \infty$ to derive asymptotic properties, regardless of the magnitudes of $N_j$ and $J$. This is because in model \eqref{model_same}, all groups share the common mean $\mu$ and common variance $\sigma^2$, so the observations $\{Y_{ij}\}_{i=1,j=1}^{N_j,J}$ are i.i.d., allowing them to be pooled for efficient estimation and inference.

The following two theorems establish formally the asymptotic properties of $N^{-1}\sum_{j=1}^{J}\sum_{i=1}^{N_j}\pi_k(\widehat Z_{ij})$ and $\widehat{\Psi}_K^2$ (defined in \eqref{hat_Psi_k_1}) under the alternatives. The conditions are identical to those in Theorem \ref{H0_different_mean}, and the results parallel those in Theorems \ref{H1_same} and \ref{H1L_same}.

\begin{theorem}\label{H1_different_mean}
	Suppose Assumptions \ref{assumption1} and \ref{assumption2} hold. Then, under the alternative hypothesis $H_1$ in \eqref{H1_fixed}, for $k=1,\ldots,K$, 
	\begin{align*}
		\frac{1}{N}\sum_{j=1}^{J}\sum_{i=1}^{N_j}\pi_k\left(\widehat Z_{ij}\right)=&\frac{1}{N}\sum_{j=1}^{J}\sum_{i=1}^{N_j}\left\{\left(\pi_k(Z_{ij})-\mathbb{E} \left[\pi_k(Z)\right]\right)-d_{1k}e_{ij}-\frac{d_{2k}}{2}\left(e_{ij}^2-1\right)\right\}\\
		&+\mathbb{E} \left[\pi_k(Z)\right]+o_p\left(\frac{1}{\sqrt N}\right),
	\end{align*}
	as $\min\{N_1,\ldots,N_J\}\to\infty$, $J = o( N^{1/2})$ and  $\sum_{j=1}^J N_j^{-1}=o(1)$. Furthermore, 
	\begin{equation*}
		\widehat{\Psi}_K^2   \overset{p}{\to}  \bm{a}_K^\top \bm{\Sigma}_K^{-1} \bm{a}_K,
	\end{equation*}
	and 
	\begin{equation*}
		\sqrt{N} \left( \widehat{\Psi}_K^2 - \bm{a}_K^\top \bm{\Sigma}_K^{-1} \bm{a}_K \right)  \overset{d}{\to}  \mathcal{N}\left(0, 4\bm{a}_K^\top \bm{\Sigma}_K^{-1} \bm{\Xi}_K \bm{\Sigma}_K^{-1} \bm{a}_K\right).
	\end{equation*}
\end{theorem}

\begin{theorem} \label{H1L_different_mean}
	Suppose Assumptions \ref{assumption1} and \ref{assumption2} hold. Then, under the local alternative hypothesis $H_{1L}$ in \eqref{H1L}, with $\delta_N = N^{-1/2}$, for $k=1,\ldots,K$, 
	\begin{align*}
		\frac{1}{N}\sum_{j=1}^{J}\sum_{i=1}^{N_j}\pi_k\left(\widehat Z_{ij}\right)=&\frac{1}{N}\sum_{j=1}^{J}\sum_{i=1}^{N_j}\left\{\left(\pi_k(Z_{ij})-\mathbb{E} \left[\pi_k(Z)\right]\right)-c_{1k}e_{ij}-\frac{c_{2k}}{2}\left(e_{ij}^2-1\right)\right\}\\
		&+\delta_N \Delta_k+o_p\left(\frac{1}{\sqrt N}\right),
	\end{align*}
	as $\min\{N_1,\ldots,N_J\}\to\infty$, $J = o( N^{1/2})$ and $\sum_{j=1}^J N_j^{-1}=o(1)$. Furthermore, 
	\begin{equation*}
		N\widehat{\Psi}_K^2   \overset{d}{\to}  \chi_K^2\left(\bm{\Delta}_K^\top\bm{\Sigma}_K^{-1}\bm{\Delta}_K\right).
	\end{equation*}
\end{theorem}

Theorems \ref{H0_different_mean}, \ref{H1_different_mean}, and \ref{H1L_different_mean} provide an asymptotic $\chi^2$ test for the normality of model \eqref{model_diff_mean} and demonstrate its power properties. Such a test is quite similar to that in Section \ref{section_prop.1}. When the group means in \eqref{model_diff_mean} are all equal in the sense that $\mu_1=\ldots=\mu_J=\mu$, the methodology in this section can be reduced to that in Section \ref{section_prop.1} for model \eqref{model_same}, which validates the unified inferential framework and reflects the effects of group heterogeneity in means.

\subsection{Common group means with heterogeneous variances} \label{section_prop.3}

We now extend model \eqref{model_same} to allow for heterogeneous group variances, that is,
\begin{equation}
	Y_{ij} = \mu + \varepsilon_{ij} = \mu + \sigma_j e_{ij}, \quad i = 1, \ldots, N_j, \quad j = 1, \ldots, J, \label{model_diff_var}
\end{equation}
where $\{\sigma_j\}_{j=1}^J$ represents the potentially distinct standard deviations of each group. For estimation, we adopt
$\widehat{\mu} = J^{-1} \sum_{j = 1}^J N_j^{-1} \sum_{i = 1}^{N_j} Y_{ij}$ and $\widehat{\sigma}_j^2 = N_j^{-1} \sum_{i = 1}^{N_j} ( Y_{ij} - \widehat{\mu})^2$ for $\mu$ and $\sigma_j^2$ ($j=1,\ldots, J$), respectively. Note that the estimator of the population mean here is different from the sample mean in Section \ref{section_prop.1} due to the heterogeneity of group variances. Define $\widehat \varepsilon_{ij} = Y_{ij}-\widehat\mu$, $\widehat{e}_{ij} = \widehat \varepsilon_{ij}/ \widehat{\sigma}_j= (Y_{ij} - \widehat{\mu}) / \widehat{\sigma}_j$ and $\widehat Z_{ij} = \Phi (\widehat{e}_{ij})$. Let $p_j=\lim N_j/N$ and $q_j=\lim J N_j / N$ be quantities characterizing the relative proportions of the group sample sizes. Here we further impose the following conditions for model \eqref{model_diff_var}: (\romannumeral1) there exist $0<\underline{\sigma}\le \overline{\sigma}<\infty$ such that $\underline{\sigma}< \inf_{1\le j\le J} \sigma_j\le \sup_{1\le j\le J} \sigma_j<\overline{\sigma}$; (\romannumeral2) there exist $0<\underline{q}\le \overline{q}<\infty$ such that $\underline{q}< \inf_{1\le j\le J} q_j\le \sup_{1\le j\le J} q_j<\overline{q}$. These conditions imply that the group-specific sample sizes and standard deviations are of the same order of magnitude.

Following the approach in Sections \ref{section_prop.1} and \ref{section_prop.2}, we first examine the asymptotic behavior of $N^{-1}\sum_{j=1}^{J}\sum_{i=1}^{N_j}\pi_k(\widehat Z_{ij})$, which is summarized in the theorem below. 

\begin{theorem}\label{H0_different_var}
	Suppose Assumptions \ref{assumption1} and \ref{assumption2} hold. Then, given the conditions of model \eqref{model_diff_var}, under $H_0:e  \sim\Phi(x)$, for $k=1,\ldots,K$,
	\begin{equation}\label{H0_different_var_decomp}
		\begin{aligned}
			\frac{1}{N}\sum_{j=1}^{J}\sum_{i=1}^{N_j}\pi_k\left(\widehat Z_{ij}\right) =& \frac{1}{N} \sum_{j = 1}^J \sum_{i = 1}^{N_j} \left\{  \pi_k (Z_{ij}) - c_{1k}\left( \sum_{\ell = 1}^J \frac{ p_{\ell}}{\sigma_{\ell}} \right) \frac{\sigma_j e_{ij}}{q_j} - \frac{c_{2k}}{2} \left( e_{ij}^2 - 1 \right)\right\} \\
			&+ o_p\left(\frac{1}{\sqrt N}\right),
		\end{aligned}
	\end{equation}
	as $\min\{N_1,\ldots,N_J\}\to\infty$ and $J = o( N^{1/2})$.
\end{theorem}

Note that the asymptotic decomposition of $N^{-1}\sum_{j=1}^{J}\sum_{i=1}^{N_j}\pi_k(\widehat Z_{ij})$ in Theorem \ref{H0_different_var} is no longer the same as that in Theorems \ref{H0_same} and \ref{H0_different_mean}. The difference lies in the parameter estimation effect of $\widehat \mu$, which is reflected by the term $-c_{1k}N^{-1}(\sum_{\ell =1}^J p_{\ell}\sigma^{-1}_{\ell})\sum_{j=1}^J\sum_{i=1}^{N_j}\sigma_j e_{ij}/q_j$ on the right-hand side of \eqref{H0_different_var_decomp}. Unlike model \eqref{model_same}, which uses the sample mean to estimate $\mu$, model \eqref{model_diff_var} employs a different estimator, $\widehat \mu$, resulting in this distinct estimation effect. In particular, when all groups share the common variance, i.e., $\sigma_1=\ldots=\sigma_J$, the term reduces to $-c_{1k}N^{-1} \sum_{j =1}^J q_j^{-1} \sum_{i=1}^{N_j} e_{ij} = -c_{1k} \sum_{j =1}^J N_j^{-1} \sum_{i=1}^{N_j} e_{ij}$; further if the group sample sizes are equal, i.e., $N_1=\ldots=N_J$, then the estimation effect simplifies to that in Theorem \ref{H0_same}, $-c_{1k}N^{-1} \sum_{j =1}^J\sum_{i=1}^{N_j} e_{ij}$. In contrast, the estimation effect of the variances always aggregates into the term $-c_{2k}(2N)^{-1}\sum_{j=1}^{J}\sum_{i=1}^{N_j}(e_{ij}^2-1)$ regardless of the equality of group variances or group sample sizes. 

The asymptotic normality of the vector $N^{-1} \sum_{j = 1}^J \sum_{i = 1}^{N_j} \bm\pi_K (\widehat Z_{ij})$ follows directly from \eqref{H0_different_var_decomp}. Under the conditions of Theorem \ref{H0_different_var}, an application of the CLT yields that
\begin{align*}
	\frac{1}{\sqrt{N}}\sum_{j=1}^{J}\sum_{i=1}^{N_j}\bm\pi_K\left(\widehat Z_{ij}\right) & =\sqrt{N}\left[\frac{1}{N}\sum_{j=1}^{J}\sum_{i=1}^{N_j}\pi_1\left(\widehat Z_{ij}\right),\ldots,\frac{1}{N}\sum_{j=1}^{J}\sum_{i=1}^{N_j}\pi_K\left(\widehat Z_{ij}\right)\right]^\top \\
	& \overset{d}{\to}  \sum_{j = 1}^J \sqrt{p_j} \bm{Z}_j,
\end{align*}
where $\bm{Z}_1, \ldots, \bm{Z}_J$ are independently distributed as
\begin{equation*}
	\bm{Z}_j  \sim  \mathcal{N}_K \left( \bm{0}, \bm{\Omega}_K^{(j)} \right),
\end{equation*}
with $\bm{\Omega}_K^{(j)}=\left(\omega_{kl}^{(j)}\right)_{K\times K}$ given by
\begin{align*}
	\omega_{kl}^{(j)} & = \mathbb{E} \left\{  \pi_k (Z) -c_{1k} \left( \sum_{\ell = 1}^J \frac{p_{\ell}}{\sigma_{\ell}} \right) \frac{\sigma_j e}{q_j} - \frac{c_{2k}}{2} \left( e^2 - 1 \right)\right\}  \\
	& \qquad \left\{  \pi_l (Z) - c_{1l}\left( \sum_{\ell = 1}^J \frac{ p_{\ell}}{\sigma_{\ell}} \right) \frac{\sigma_j e}{q_j} - \frac{c_{2l}}{2} \left( e^2 - 1 \right)\right\} \\
	& = \delta_{kl} - \frac{2 c_{1k} c_{1l} \sigma_j}{q_j} \sum_{\ell = 1}^J \frac{p_{\ell}}{\sigma_{\ell}} + \frac{c_{1k}c_{1 l} \sigma_j^2}{q_j^2} \left( \sum_{\ell = 1}^J \frac{p_{\ell}}{\sigma_{\ell}}\right)^2  - \frac{c_{2k} c_{2l}}{2}.
\end{align*}
Therefore, under $H_0$, we have
\begin{equation*}
	\frac{1}{\sqrt{N}}\sum_{j = 1}^J \sum_{i = 1}^{N_j} \bm\pi_K \left(\widehat Z_{ij}\right)  \overset{d}{\to}  \mathcal{N}_K \left( \bm{0}, \sum_{j = 1}^J p_j \bm{\Omega}_K^{(j)}\right).
\end{equation*}
By replacing $\sigma_j$, $p_j$ and $q_j$ with their sample analogues $\widehat\sigma_j$, $\widehat{p}_j = N_j / N$ and $\widehat{q}_j = J N_j / N$ respectively, we obtain a consistent estimator of $\bm{\Omega}_K^{(j)}$, denoted by $\widehat{\bm{\Omega}}_K^{(j)}$. Accordingly, the test statistic is defined as
\begin{equation} \label{hat_Psi_k_2}
	\widehat{\Psi}_K^2 = \left(  \frac{1}{N}\sum_{j = 1}^J \sum_{i = 1}^{N_j} \bm\pi_K \left(\widehat Z_{ij}\right) \right)^\top \left(\sum_{j = 1}^J \widehat{p}_j \widehat{\bm{\Omega}}_K^{(j)} \right)^{-1} \left(  \frac{1}{N}\sum_{j = 1}^J \sum_{i = 1}^{N_j} \bm\pi_K \left(\widehat Z_{ij}\right) \right).
\end{equation}
We also introduce the infeasible version,
\begin{equation} \label{tilde_Psi_k}
	\widetilde{\Psi}_K^2 = \left(  \frac{1}{N}\sum_{j = 1}^J \sum_{i = 1}^{N_j} \bm\pi_K \left(\widehat Z_{ij}\right) \right)^\top \left(\sum_{j = 1}^J p_j \bm{\Omega}_K^{(j)} \right)^{-1} \left(  \frac{1}{N}\sum_{j = 1}^J \sum_{i = 1}^{N_j} \bm\pi_K \left(\widehat Z_{ij}\right) \right).
\end{equation}

\begin{corollary} \label{corollary 2}
	Suppose Assumptions \ref{assumption1} and \ref{assumption2} hold. Then, given the conditions of model \eqref{model_diff_var}, under $H_0:e  \sim\Phi(x)$,
	\begin{equation*}
		N\widehat\Psi_K^2  \overset{d}{\to}  \chi_K^2 \text{ and } N\widetilde\Psi_K^2  \overset{d}{\to}  \chi_K^2,
	\end{equation*}
	as $\min\{N_1,\ldots,N_J\}\to\infty$ and $J = o( N^{1/2})$.
\end{corollary}

Corollary \ref{corollary 2} shows that under the null hypothesis, both the feasible test statistic $\widehat{\Psi}_K^2$ (defined in \eqref{hat_Psi_k_2}) and its infeasible counterpart $\widetilde{\Psi}_K^2$ (in \eqref{tilde_Psi_k}) are asymptotically $\chi_K^2$ distributed when multiplied by the total sample size $N$. Although these test statistics take a slightly different form from those in the previous Sections \ref{section_prop.1} and \ref{section_prop.2}, the limiting Chi-square distribution is retained, and the convergence rate remains $N$, which further supports the validity of our unified theoretical framework. The following two theorems characterize the asymptotic behavior of $\widetilde{\Psi}_K^2$ under both fixed and local alternatives, in a manner analogous to that of $\widehat{\Psi}_K^2$ discussed in Sections \ref{section_prop.1} and \ref{section_prop.2}.

\begin{theorem} \label{H1_different_var}
	Suppose Assumptions \ref{assumption1} and \ref{assumption2} hold. Then,  given the conditions of model \eqref{model_diff_var}, under the alternative hypothesis $H_1$ in \eqref{H1_fixed}, for $k=1,\ldots,K$, 
	\begin{align}
		\frac{1}{N}\sum_{j=1}^{J}\sum_{i=1}^{N_j}\pi_k\left(\widehat Z_{ij}\right) =& \frac{1}{N} \sum_{j = 1}^J \sum_{i = 1}^{N_j} \left\{  \left(\pi_k (Z_{ij})-\mathbb{E} \left[\pi_k(Z)\right]\right) - d_{1k}\left( \sum_{\ell = 1}^J \frac{ p_{\ell}}{\sigma_{\ell}} \right) \frac{\sigma_j e_{ij}}{q_j} \right. \nonumber\\
        &-\left.\frac{d_{2k}}{2} \left( e_{ij}^2 - 1 \right)\right\}+\mathbb{E} \left[\pi_k(Z)\right]+ o_p\left(\frac{1}{\sqrt N}\right), \label{H1_different_var_decomp}
	\end{align}
	as $\min\{N_1,\ldots,N_J\}\to\infty$ and $J = o( N^{1/2})$. Furthermore, 
	\begin{equation*}
		\widetilde{\Psi}_K^2   \overset{p}{\to}  \bm{a}_K^\top \left(\sum_{j = 1}^J p_j \bm{\Omega}_K^{(j)} \right)^{-1}\bm{a}_K,
	\end{equation*}
	and 
	\begin{equation}
		\sqrt{N} \left( \widetilde{\Psi}_K^2 - \bm{a}_K^\top \left(\sum_{j = 1}^J p_j \bm{\Omega}_K^{(j)} \right)^{-1} \bm{a}_K \right)  \overset{d}{\to}  \mathcal{N}\left(0, 4\bm{a}_K^\top\bm{\Upsilon}_K \bm{a}_K\right), \label{Psi_tilde_asym_normal}
	\end{equation}
	where 
	\begin{equation*}
		\bm{\Upsilon}_K= \left(\sum_{j = 1}^J p_j \bm{\Omega}_K^{(j)} \right)^{-1}\left(\sum_{j = 1}^J p_j \bm{\Lambda}_K^{(j)} \right) \left(\sum_{j = 1}^J p_j \bm{\Omega}_K^{(j)} \right)^{-1},    
	\end{equation*}
	and $\bm{\Lambda}_K^{(j)}=\left(\lambda_{kl}^{(j)}\right)_{K\times K}$ is given by
	\begin{align*}
		\lambda_{kl}^{(j)}
		= &\mathbb{E} \left[ \pi_k (Z) \pi_l (Z) \right]-a_k a_l- \frac{ (d_{1k} d_{3l}+d_{1l} d_{3k}) \sigma_j}{q_j} \sum_{\ell = 1}^J \frac{p_{\ell}}{\sigma_{\ell}} + \frac{d_{1k}d_{1 l} \sigma_j^2}{q_j^2} \left( \sum_{\ell = 1}^J \frac{p_{\ell}}{\sigma_{\ell}}\right)^2 \\
		&+ \frac{1}{2} \left[ a_k d_{2l} + a_l d_{2k} \right]-\frac{1}{2}\left[ d_{2k}d_{4l} + d_{2l}d_{4k}\right]+ \frac{ (d_{1k} d_{2l}+d_{1l} d_{2k}) \sigma_j}{2q_j} \sum_{\ell = 1}^J \frac{p_{\ell}}{\sigma_{\ell}}\mathbb{E} \left[e^3\right]\\
		&+\frac{d_{2k} d_{2l} }{4} \left[\mathbb{E} \left[e^4\right]-1 \right].
	\end{align*}
\end{theorem}

\begin{theorem} \label{H1L_different_var}
	Suppose Assumptions \ref{assumption1} and \ref{assumption2} hold. Then,  given the conditions of model \eqref{model_diff_var}, under the local alternative hypothesis $H_{1L}$ in \eqref{H1L}, with $\delta_N = N^{-1/2}$, for $k=1,\ldots,K$, 
	\begin{align*}
		\frac{1}{N}\sum_{j=1}^{J}\sum_{i=1}^{N_j}\pi_k\left(\widehat Z_{ij}\right) =& \frac{1}{N} \sum_{j = 1}^J \sum_{i = 1}^{N_j} \left\{  \left(\pi_k (Z_{ij})-\mathbb{E} \left[\pi_k(Z)\right]\right) - c_{1k}\left( \sum_{\ell = 1}^J \frac{ p_{\ell}}{\sigma_{\ell}} \right) \frac{\sigma_j e_{ij}}{q_j} \right.\\
        &-\left. \frac{c_{2k}}{2} \left( e_{ij}^2 - 1 \right)\right\}+\delta_N\Delta_k+ o_p\left(\frac{1}{\sqrt N}\right),
	\end{align*}
	as $\min\{N_1,\ldots,N_J\}\to\infty$ and $J = o( N^{1/2})$. Furthermore,
	\begin{equation*}
		N\widetilde{\Psi}_K^2   \overset{d}{\to}  \chi_K^2\left(\bm{\Delta}_K^\top\left(\sum_{j = 1}^J p_j \bm{\Omega}_K^{(j)} \right)^{-1}\bm{\Delta}_K\right).
	\end{equation*}
\end{theorem}

\begin{remark} \emph{
	To conclude this section, we summarize the regularity conditions on sample size structure and cross-group heterogeneity for the three models \eqref{model_same}, \eqref{model_diff_mean}, and \eqref{model_diff_var} in Table \ref{tb_summary_conditions}. On the one hand, more complex model structures naturally impose additional restrictions on within-group variability and group sample sizes, as heterogeneity across groups introduces additional challenges for theoretical analysis. On the other hand, under these mild conditions, an asymptotic Chi-square test with convergence rate $N$ can be established for each model. This highlights that our proposed methodology provides a unified theoretical framework while remaining flexible enough to accommodate the distinctive features of various ANOVA settings.}
\end{remark}

\begin{table}[!ht] 
	\caption{Regularity conditions on sample sizes and group heterogeneity for the three one-way models}
	\centering
	\begin{tabular}{c|c}
		\toprule
		Model & Conditions \\
		\midrule
		Model \eqref{model_same} &  $N \to \infty$  \\ \hline
		Model \eqref{model_diff_mean} &  \makecell{$\min\{N_1,\ldots,N_J\}\to\infty$,  \\$J = o( N^{1/2})$,  $\sum_{j=1}^J N_j^{-1}=o(1)$} \\ \hline
		Model \eqref{model_diff_var} & \makecell{$\min\{N_1,\ldots,N_J\}\to\infty$,\\ $J = o( N^{1/2})$ ,  $\sum_{j=1}^J N_j^{-1}=o(1)$,\\ $\underline{\sigma}< \inf_{1\le j\le J} \sigma_j\le \sup_{1\le j\le J} \sigma_j<\overline{\sigma}$, \\ $\underline{q}< \inf_{1\le j\le J} q_j\le \sup_{1\le j\le J} q_j<\overline{q}$} \\
		\bottomrule
	\end{tabular}
	
	\label{tb_summary_conditions}
\end{table}

\section{Data-driven choice of \texorpdfstring{$K$}{K}} \label{section_datadriven}

The methodology in Section \ref{section_prop} relies on the fact that the order of smooth alternatives \eqref{pit_density1}, $K$, is fixed before testing. The choice of $K$ critically affects the performance of smooth tests \citep{ledwina1994data, kallenberg1995consistency,  kallenberg1995data, inglot1996asymptotic, inglot1997data, kallenberg1997dataA, kallenberg1997dataB, kallenberg1999data, janic2000data, ducharme2004smooth, ducharme2004goodness, kraus2007data, duchesne2016estimating}. A large $K$ introduces redundant components into the alternatives that contribute little to the test statistic but inflate the degrees of freedom, thereby diluting power. Conversely, a small $K$ increases the risk that $\mathbb{E}[\pi_k (Z)] = 0$ for all $1\le k \le K$, rendering the test powerless. Hence, an appropriate choice of $K$ is crucial. In this section, we focus on the data-driven selection of $K$ for the three tests developed in Section \ref{section_prop}.

In Neyman's smooth test literature, the data-driven selection of $K$ was first proposed by \cite{ledwina1994data} for testing uniformity. The data-driven testing procedure consists of two steps. First, Schwarz's selection rule (also known as the Bayesian Information Criterion, BIC) is applied to determine the dimension of the smooth model that best fits the data. Second, Neyman's smooth test is performed within the selected model space, yielding a data-driven test statistic. The data-driven smooth test thus combines preliminary model selection with a more precise inferential procedure, serving as Neyman's test in the ``right'' direction. Desirable theoretical properties and extensive numerical results of the selection rule and the induced test statistic were established in \cite{kallenberg1995consistency} and \cite{kallenberg1995data}, \cite{inglot1996asymptotic}. Later, modifications of Schwarz's rule addressed more complex problems, including testing composite hypotheses \citep{kallenberg1997dataA, kallenberg1997dataB, inglot1997data}, testing independence \citep{kallenberg1999data}, and two-sample testing \citep{janic2000data}, among others. These modified rules leverage smooth test statistics directly, avoiding the computation of the maximized log-likelihood, and are easier to implement.

Motivated by these strategies, we adopt a modified Schwarz's rule to determine the order $K$ for smooth tests in ANOVA models. The selection rule is specified as follows:
\begin{equation}
	\widehat K = \min\left\{ \mathop{\arg\max}\limits_{1 \le k \le D} \left(N \widehat{\Psi}_k^2 - k \log N \right)\right\}, \label{bic}
\end{equation}
where $D$ is a fixed positive integer, and $\widehat{\Psi}_k^2$ takes the form of \eqref{hat_Psi_k_1} or \eqref{hat_Psi_k_2}  depending on the specific scenario. The resulting data-driven test statistic is $N\widehat \Psi_{\widehat K}^2$, with $\widehat K$ given by \eqref{bic}. The form of \eqref{bic} is mainly inspired by \cite{ducharme2004smooth}, \cite{ducharme2004goodness}, \cite{kraus2007data} and \cite{duchesne2016estimating} where the upper bound of selection $D$ is fixed. An alternative is to let $D = D(N) \to \infty$ as $N \to \infty$ \citep[see, for example,][]{kallenberg1995consistency, kallenberg1995data, inglot1996asymptotic, inglot1997data, kallenberg1997dataA, kallenberg1997dataB, kallenberg1999data, janic2000data}. Although a diverging upper bound can, in principle, improve the consistency of smooth tests, the required rate of divergence is slow, and simulation studies show that empirical power levels off rapidly as $D$ increases. Hence, our proposed selection rule \eqref{bic} with fixed $D$ is both reasonable and easy to implement in practice.

To establish the theoretical properties of $\widehat K$ and $N\widehat{\Psi}_{\widehat K}^2$, we introduce a revised version of the alternative hypothesis $H_1$ \eqref{H1_fixed} and Assumption \ref{assumption2}. Consider
\begin{equation}
	H_1^\prime: \mathbb{E}[\pi_1(Z)]=\ldots=\mathbb{E}[\pi_{K_0-1}(Z)]=0, \mathbb{E}[\pi_{K_0}(Z)]\ne 0, \quad K_0 \le D,  \label{H1_data_driven}
\end{equation}
which encompasses a broader range of alternatives than $H_1$ by replacing the fixed order $K$ with the larger $D$. Correspondingly, the condition on the orthonormal system is extended to include all functions up to order $D$, as follows.
\begin{assumption}  \label{assumption2'}
	For $k =1,\ldots, D$, $\pi_k(\cdot)$ are two times differentiable with derivatives $\dot\pi_k(\cdot)$ and $\ddot\pi_k(\cdot)$, and they are both bounded.    
\end{assumption} 
The following theorem establishes the properties of data-driven tests under the null and the revised alternatives.
\begin{theorem} \label{thm_data driven}
	Suppose Assumptions \ref{assumption1} and \ref{assumption2'} hold.
	\begin{itemize}
		\item[(1).] For model \eqref{model_same} in Section \ref{section_prop.1}, as $N \to \infty$, under $H_0: e \sim \Phi(x)$, $\mathbb{P}(\widehat K = 1) \to 1$ and $N \widehat{\Psi}_{\widehat K}^2 \overset{d}{\to} \chi_1^2$; under $H_1^\prime$ in \eqref{H1_data_driven}, $\mathbb{P}( \widehat K  \ge K_0) \to 1$ and $\mathbb{P}(N\widehat\Psi^2_{\widehat K}\le x) \to 0$ for any $x\in \mathbb{R}$.
		\item[(2).] For model \eqref{model_diff_mean} in Section \ref{section_prop.2}, as $\min\{N_1,\ldots,N_J\}\to\infty$, $J = o( N^{1/2})$, and $\sum_{j=1}^J N_j^{-1}=o(1)$, under $H_0: e \sim \Phi(x)$, $\mathbb{P}(\widehat K = 1) \to 1$ and $N \widehat{\Psi}_{\widehat K}^2 \overset{d}{\to} \chi_1^2$; under $H_1^\prime$ in \eqref{H1_data_driven}, $\mathbb{P}( \widehat K  \ge K_0 ) \to 1$ and $\mathbb{P}(N\widehat\Psi^2_{\widehat K}\le x) \to 0$ for any $x\in \mathbb{R}$.
		\item[(3).] For model \eqref{model_diff_var} in Section \ref{section_prop.3}, with additional conditions such that (\romannumeral1) there exist $0<\underline{\sigma}\le \overline{\sigma}<\infty$ such that $\underline{\sigma}< \inf_{1\le j\le J} \sigma_j\le \sup_{1\le j\le J} \sigma_j<\overline{\sigma}$; (\romannumeral2) there exist $0<\underline{q}\le \overline{q}<\infty$ such that $\underline{q}< \inf_{1\le j\le J} q_j\le \sup_{1\le j\le J} q_j<\overline{q}$,  as $\min\{N_1,\ldots,N_J\}\to\infty$ and $J = o( N^{1/2})$, under $H_0: e \sim \Phi(x)$, $\mathbb{P}(\widehat K = 1) \to 1$, $N \widehat{\Psi}_{\widehat K}^2 \overset{d}{\to} \chi_1^2$ and $N \widetilde{\Psi}_{\widehat K}^2 \overset{d}{\to} \chi_1^2$; under $H_1^\prime$ in \eqref{H1_data_driven}, $\mathbb{P}( \widehat K  \ge K_0 ) \to 1$ and $\mathbb{P}(N\widetilde\Psi^2_{\widehat K}\le x) \to 0$ for any $x\in \mathbb{R}$.
	\end{itemize}
\end{theorem}

Theorem \ref{thm_data driven} presents the unified results of data-driven smooth tests for the three cases in Section \ref{section_prop}. Under the null hypothesis, the probability of $\{\widehat K = 1\}$ tends to $1$ asymptotically, implying that the first-order smooth model provides the best fit to $\{\widehat Z_{ij}\}_{i=1,j=1}^{N_j,J}$ among the $D$ candidate models. From the perspective of hypothesis testing, this means that $\widehat \Psi_1^2$ is informative and sufficient for assessing uniformity or normality. Meanwhile, the data-driven selection procedure allows identification of the mechanism underlying nonuniformity (equivalently, nonnormality) and enhances the test's power against a broader class of alternatives, $H_1^\prime$. These findings are consistent with previous results on data-driven smooth tests.

In practice, however, the $\chi_1^2$ limiting null distribution of data-driven smooth test statistics often performs poorly in finite samples \citep{kallenberg1995data, inglot1997data, kallenberg1997dataA, kallenberg1997dataB, kallenberg1999data, janic2000data, ducharme2004smooth, ducharme2004goodness, kraus2007data, duchesne2016estimating}. To address this issue in our problem, and provided the nested orthonormal system (i.e. $\{\pi_k\}_{k=1}^K \subseteq \{\pi_k\}_{k=1}^{K+1}$ for $1\le K\le D-1$), we follow the approach of \cite{kallenberg1995data}, \cite{inglot1997data}, \cite{kallenberg1997dataA}, \cite{kallenberg1999data}, \cite{janic2000data}, \citet{kraus2007data} and adopt the following finite-sample approximation for the null distribution of $N \widehat{\Psi}_{\widehat K}^2$:
\begin{equation}
	\begin{aligned}
		&H(x) = 
		\begin{cases}
			(2 \Phi(\sqrt{x})-1)(2 \Phi(\sqrt{\log N})-1), & x \le \log N, \\
			H(\log N)+(x-\log N)(H(2 \log N)-H(\log N))/(\log N), & \log N < x < 2 \log N, \\ 
			(2 \Phi(\sqrt{x})-1)(2 \Phi(\sqrt{\log N})-1)+2(1-\Phi(\sqrt{\log N})), & x \ge 2 \log N.
		\end{cases}    
	\end{aligned} \label{null_approx}    
\end{equation}
Such an approximation primarily accounts for the selection uncertainty of $\widehat K$. In finite samples, under the null, the empirical probability of the event $\{\widehat{K} = 1\}$ may not be exactly $1$, as $\{\widehat{K} = 2\}$ can occur with a small but non-negligible probability. Consequently, \eqref{null_approx} is derived from the approximation $\mathbb{P} (N \widehat{\Psi}_{\widehat K}^2 \le x) \approx \mathbb{P} (N \widehat{\Psi}_1^2 \le x, \widehat K = 1) + \mathbb{P} (N \widehat{\Psi}_2^2 \le x, \widehat K = 2)$. The reliability and accuracy of $H(x)$ are further illustrated through simulation studies in the next section and in the \hyperlink{app}{Online Appendix}.

\section{Simulations} \label{section_simu}

In this section, we conduct Monte Carlo experiments to examine the finite-sample performance of the smooth tests developed in Sections \ref{section_prop} and \ref{section_datadriven}. Following \cite{neyman1937smooth}, \cite{bera2013smooth}, \cite{song2022smooth}, and \cite{beutner2025two}, we adopt the orthonormal Legendre polynomials on the interval $[0,1]$ for $\{\pi_k\}_{k = 1}^{\infty}$. Under this choice, the analytical properties of the constants $c_{1k}$ and $c_{2k}$ are provided in Proposition 1 of \cite{duchesne2016estimating}, and numerical values for $1 \le k \le 10$ are available in Table 1 of their Supplementary Material. From a computational perspective, the test statistic $N\widehat{\Psi}_K^2$, introduced in Sections \ref{section_prop.1} and \ref{section_prop.2}, can be computed using a simplified expression analogous to equation~(14) in \cite{duchesne2016estimating}, which leverages the closed form of $\bm{\Sigma}_K^{-1}$ (see equations (12) and (13) therein). For comparison, we include three classical normalty tests in the literature---Shapiro--Wilk (SW), Jarque--Bera (JB), and Kolmogorov--Smirnov (KS) tests---applied to the standardized residuals $\{\widehat e_{ij}\}_{i=1, j=1}^{N_j, J}$ under the corresponding model, without accounting for the parameter estimation effects.

We first study the tests for model \eqref{model_same} through two experiments. The observations are generated as follows.
\begin{itemize}
    \item \textbf{Experiment \uppercase\expandafter{\romannumeral1}}:
\end{itemize}
\begin{equation*}
	Y_{ij}^{(0,1)}  \sim  \mathcal{N}\left( 5, 4\right),  \quad Y_{ij}^{(1,1)}  \sim  \chi_2^2 + 3, \quad i = 1, \ldots, jm, \quad j = 1, \ldots, 5.
\end{equation*}
\begin{itemize}
    \item \textbf{Experiment \uppercase\expandafter{\romannumeral2}}:
\end{itemize}
\begin{equation*}
	Y_{ij}^{(0,2)}  \sim  \mathcal{N}\left( 8, 1\right), \ Y_{ij}^{(1,2)}  \sim \mathcal{U} \left[ 8 - \sqrt{3}, 8 +  \sqrt{3}\right], \ i = 1, \ldots, jm, \ j = 1, \ldots, 5.
\end{equation*}
Here $\{Y_{ij}^{(0,1)}\}_{i=1,j=1}^{jm,5}$ and $\{Y_{ij}^{(0,2)}\}_{i=1,j=1}^{jm,5}$ are generated under $H_0$, while $\{Y_{ij}^{(1,1)}\}_{i=1,j=1}^{jm,5}$ and $\{Y_{ij}^{(1,2)}\}_{i=1,j=1}^{jm,5}$ are generated under $H_1$. For the proposed smooth tests, we evaluate the test statistic $N\widehat \Psi_K^2$ from Section \ref{section_prop.1} with $1 \leq K \leq 5$, as well as the data-driven statistic $N\widehat \Psi_{\widehat K}^2$ from Section \ref{section_datadriven}, where $\widehat K$ is determined by \eqref{bic} with $D=5$. The sample size parameter $m$ ranges from 10 to 150 in increments of 10, and the significance level is set at $\alpha = 5\%$ with $500$ replications performed.

\begin{figure}[!ht]
	\centering
	\includegraphics[scale=0.65]{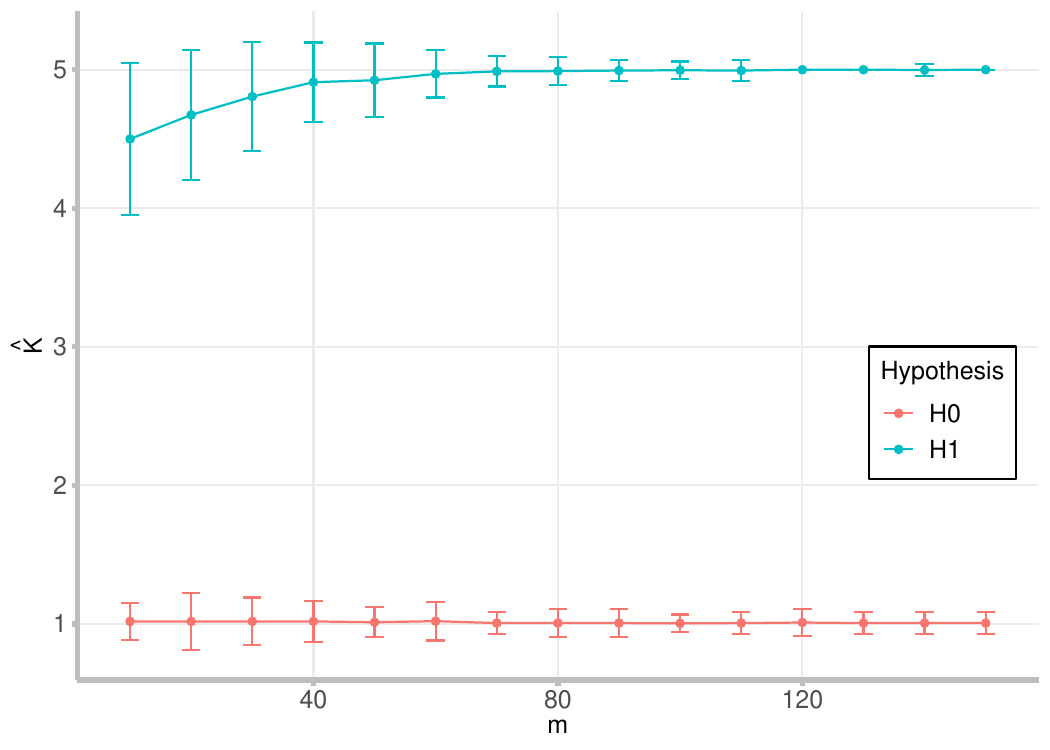}
	\caption{The sample means and error bars of $\widehat K$ in Experiment \uppercase\expandafter{\romannumeral1}}
	\label{Fig3.1-1}    
\end{figure}

Results of Experiment \uppercase\expandafter{\romannumeral1} are presented in Tables \ref{rejection_rate_3.1-1}, \ref{k_freq_3.1-1}, and Figure \ref{Fig3.1-1}. Table \ref{rejection_rate_3.1-1} reports the empirical rejection rates under $H_0$ for the proposed smooth tests and the classical normality tests. The smooth tests comprise both fixed-$K$ versions ($K=1,\ldots,5$) and data-driven procedures, implemented using the limiting $\chi^2_1$ null distribution ($\widehat K\&\chi^2_1$) or the approximated null distribution $H(x)$ ($\widehat K\&H(x)$). For fixed $K$, under $H_0$, the statistic $N\widehat \Psi_K^2$ maintains the nominal level well, even for small sample sizes such as $m=10$ (corresponding to a total sample size of $N=150$). For the data-driven test statistic $N\widehat \Psi_{\widehat K}^2$, the limiting $\chi^2_1$ distribution fails to control the rejection rate under $H_0$, whereas the approximated distribution $H(x)$ achieves better size control, yielding rejection rates close to the nominal $5\%$ level. Under $H_1$, all versions of the proposed smooth tests exhibit high power across the considered settings. As for the classical methods, although the Shapiro--Wilk and Jarque--Bera tests are applied directly to the standardized residuals without accounting for the estimation effects, their empirical rejection rates under $H_0$ remain close to the nominal $5\%$ level. Under $H_1$, both tests achieve high power, indicating that they still provide reasonably reliable numerical performance despite the lack of formal adjustment for the estimation effect. In contrast, under $H_0$, the Kolmogorov--Smirnov test exhibits almost zero rejection across all settings, indicating the practical failure of the test. Table \ref{k_freq_3.1-1} presents the empirical frequency of $\widehat K$ under $H_0$ and $H_1$, respectively. Under $H_0$, the frequency of $\{\widehat K=1\}$ approaches $1$ as $m$ increases, whereas under $H_1$, $\widehat K$ is consistently greater than $1$, in line with theoretical expectations. Figure \ref{Fig3.1-1} displays the sample means of $\widehat K$ across different values of $m$, with error bars representing one sample standard deviation around the sample means. Under $H_0$, the sample mean of $\widehat K$ stays close to $1$ with diminishing variability as $m$ grows, indicating consistent model selection. Under $H_1$, $\widehat K$ increases steadily and approaches the maximum $D=5$, with reduced variability, demonstrating the adaptiveness of the data-driven procedure.

\begin{figure}[!ht]
	\centering
	\includegraphics[scale=0.65]{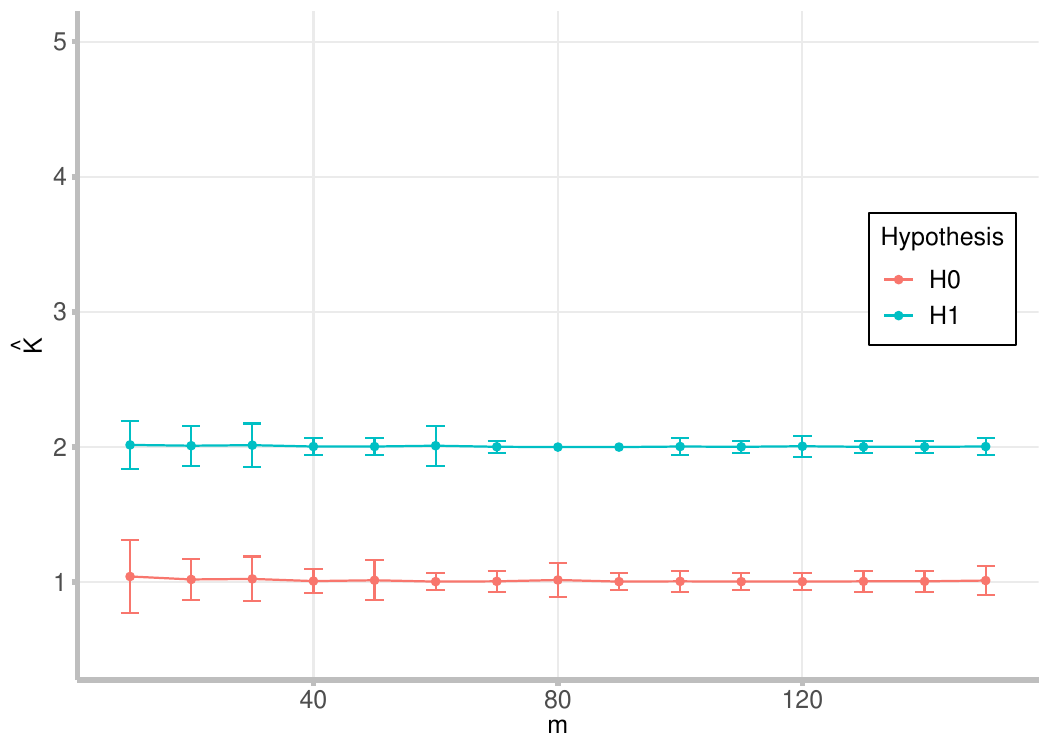}
	\caption{The sample means and error bars of $\widehat K$ in Experiment \uppercase\expandafter{\romannumeral2}}
	\label{Fig3.1-2}    
\end{figure}

The results of Experiment \uppercase\expandafter{\romannumeral2} are quite similar; see Tables \ref{rejection_rate_3.1-2_H0}, \ref{rejection_rate_3.1-2_H1}, \ref{k_freq_3.1-2}, and Figure \ref{Fig3.1-2}. However, from Table \ref{k_freq_3.1-2}, under $H_1$, the test statistic $N\widehat \Psi_1^2$ fails to exhibit power as a result of $\mathbb{E}[\pi_1(Z)] = \mathbb{E}[\pi_1(\Phi(Y_{ij}^{(1,2)} - 8))] = 0$ in this case. This illustrates that $N\widehat \Psi_K^2$ may be inconsistent against certain alternatives, especially when $K$ is small. Correspondingly, both Table \ref{k_freq_3.1-2} and Figure \ref{Fig3.1-2} show that under $H_1$, $\{\widehat K = 1\}$ does not occur, consistent with Theorem \ref{thm_data driven} (1), which ensures that $\mathbb{P}(\widehat K \ge K_0) \to 1$ under $H_1^\prime$ with $K_0 = 2$. 

\begin{table}[!ht]
\small
\centering
\caption{
Empirical rejection rates under $H_0$ in Experiment \uppercase\expandafter{\romannumeral1}. The empirical power under $H_1$ equals 1 for all sample sizes and all considered tests.
}
\begin{tabular}{c|ccccccc|ccc}
\toprule
& \multicolumn{7}{c|}{Smooth tests} & \multirow{2}{*}{SW} & \multirow{2}{*}{JB} & \multirow{2}{*}{KS} \\
\cmidrule(lr){2-8}
$m$  & $K = 1$ & $K = 2$ & $K = 3$ & $K = 4$ & $K = 5$ 
     & $\widehat K\&\chi^2_1$ & $\widehat K\&H(x)$ 
     \\
\midrule
10  & 0.050 & 0.048 & 0.048 & 0.044 & 0.040 & 0.056 & 0.034 & 0.054 & 0.044 & 0     \\
20  & 0.046 & 0.044 & 0.034 & 0.040 & 0.048 & 0.066 & 0.046 & 0.054 & 0.042 & 0     \\
30  & 0.042 & 0.036 & 0.044 & 0.068 & 0.068 & 0.060 & 0.046 & 0.040 & 0.032 & 0.002 \\
40  & 0.058 & 0.040 & 0.036 & 0.058 & 0.052 & 0.058 & 0.050 & 0.044 & 0.038 & 0     \\
50  & 0.064 & 0.044 & 0.054 & 0.054 & 0.062 & 0.068 & 0.060 & 0.036 & 0.030 & 0     \\
60  & 0.054 & 0.034 & 0.038 & 0.044 & 0.050 & 0.070 & 0.062 & 0.056 & 0.058 & 0     \\
70  & 0.056 & 0.038 & 0.048 & 0.046 & 0.052 & 0.056 & 0.048 & 0.072 & 0.062 & 0     \\
80  & 0.038 & 0.050 & 0.072 & 0.042 & 0.048 & 0.058 & 0.048 & 0.046 & 0.038 & 0     \\
90  & 0.062 & 0.066 & 0.052 & 0.064 & 0.058 & 0.048 & 0.040 & 0.028 & 0.038 & 0     \\
100 & 0.034 & 0.050 & 0.050 & 0.060 & 0.054 & 0.052 & 0.038 & 0.038 & 0.040 & 0     \\
110 & 0.054 & 0.032 & 0.054 & 0.060 & 0.050 & 0.038 & 0.036 & 0.042 & 0.044 & 0     \\
120 & 0.052 & 0.052 & 0.038 & 0.044 & 0.052 & 0.052 & 0.050 & 0.050 & 0.046 & 0     \\
130 & 0.052 & 0.050 & 0.064 & 0.060 & 0.058 & 0.068 & 0.064 & 0.050 & 0.060 & 0     \\
140 & 0.062 & 0.040 & 0.026 & 0.046 & 0.030 & 0.038 & 0.030 & 0.068 & 0.050 & 0     \\
150 & 0.054 & 0.044 & 0.062 & 0.034 & 0.058 & 0.056 & 0.048 & 0.064 & 0.066 & 0     \\
\bottomrule
\end{tabular}
\label{rejection_rate_3.1-1}
\end{table}

\begin{table}[!ht]
\small
\centering
\caption{Empirical frequency of $\widehat K$ in Experiment \uppercase\expandafter{\romannumeral1}}
\begin{tabular}{c|ccccc|ccccc}
\toprule
 & \multicolumn{5}{c|}{Under $H_0$} & \multicolumn{5}{c}{Under $H_1$} \\
\midrule
$m$ & $\widehat K=1$ & $\widehat K=2$ & $\widehat K=3$ & $\widehat K=4$ & $\widehat K=5$
    & $\widehat K=1$ & $\widehat K=2$ & $\widehat K=3$ & $\widehat K=4$ & $\widehat K=5$ \\
\midrule
10  & 0.982 & 0.018 & 0     & 0     & 0     & 0.004 & 0     & 0     & 0.484 & 0.512 \\
20  & 0.988 & 0.010 & 0     & 0     & 0.002 & 0     & 0     & 0     & 0.326 & 0.674 \\
30  & 0.986 & 0.012 & 0     & 0.002 & 0     & 0     & 0     & 0     & 0.194 & 0.806 \\
40  & 0.984 & 0.014 & 0.002 & 0     & 0     & 0     & 0     & 0     & 0.090 & 0.910 \\
50  & 0.988 & 0.012 & 0     & 0     & 0     & 0     & 0     & 0     & 0.076 & 0.924 \\
60  & 0.980 & 0.020 & 0     & 0     & 0     & 0     & 0     & 0     & 0.030 & 0.970 \\
70  & 0.994 & 0.006 & 0     & 0     & 0     & 0     & 0     & 0     & 0.012 & 0.988 \\
80  & 0.996 & 0.002 & 0.002 & 0     & 0     & 0     & 0     & 0     & 0.010 & 0.990 \\
90  & 0.996 & 0.002 & 0.002 & 0     & 0     & 0     & 0     & 0     & 0.006 & 0.994 \\
100 & 0.996 & 0.004 & 0     & 0     & 0     & 0     & 0     & 0     & 0.004 & 0.996 \\
110 & 0.994 & 0.006 & 0     & 0     & 0     & 0     & 0     & 0     & 0.006 & 0.994 \\
120 & 0.990 & 0.010 & 0     & 0     & 0     & 0     & 0     & 0     & 0     & 1 \\
130 & 0.994 & 0.006 & 0     & 0     & 0     & 0     & 0     & 0     & 0     & 1 \\
140 & 0.994 & 0.006 & 0     & 0     & 0     & 0     & 0     & 0     & 0.002 & 0.998 \\
150 & 0.994 & 0.006 & 0     & 0     & 0     & 0     & 0     & 0     & 0     & 1 \\
\bottomrule
\end{tabular}
\label{k_freq_3.1-1}
\end{table}

\begin{table}[!ht]
\small
\centering
\caption{
Empirical rejection rates under $H_0$ in Experiment \uppercase\expandafter{\romannumeral2}
}
\begin{tabular}{c|ccccccc|ccc}
\toprule
 & \multicolumn{7}{c|}{Smooth tests} & \multirow{2}{*}{SW} & \multirow{2}{*}{JB} & \multirow{2}{*}{KS} \\
\cmidrule(lr){2-8}
$m$  & $K = 1$ & $K = 2$ & $K = 3$ & $K = 4$ & $K = 5$ 
     & $\widehat K\&\chi^2_1$ & $\widehat K\&H(x)$ 
     \\
\midrule
10  & 0.042 & 0.050 & 0.046 & 0.042 & 0.052 & 0.076 & 0.048 & 0.050 & 0.042 & 0     \\
20  & 0.034 & 0.042 & 0.026 & 0.050 & 0.060 & 0.070 & 0.054 & 0.032 & 0.042 & 0     \\
30  & 0.028 & 0.048 & 0.036 & 0.046 & 0.046 & 0.070 & 0.062 & 0.064 & 0.052 & 0.002 \\
40  & 0.054 & 0.054 & 0.046 & 0.050 & 0.060 & 0.062 & 0.052 & 0.044 & 0.048 & 0     \\
50  & 0.064 & 0.062 & 0.042 & 0.060 & 0.050 & 0.068 & 0.058 & 0.042 & 0.044 & 0     \\
60  & 0.032 & 0.054 & 0.060 & 0.040 & 0.040 & 0.078 & 0.064 & 0.056 & 0.062 & 0     \\
70  & 0.076 & 0.038 & 0.052 & 0.052 & 0.036 & 0.052 & 0.042 & 0.054 & 0.054 & 0     \\
80  & 0.040 & 0.046 & 0.034 & 0.048 & 0.058 & 0.074 & 0.068 & 0.044 & 0.044 & 0     \\
90  & 0.044 & 0.052 & 0.050 & 0.054 & 0.042 & 0.060 & 0.050 & 0.052 & 0.052 & 0     \\
100 & 0.038 & 0.068 & 0.050 & 0.046 & 0.046 & 0.064 & 0.056 & 0.044 & 0.038 & 0     \\
110 & 0.052 & 0.052 & 0.046 & 0.052 & 0.048 & 0.050 & 0.046 & 0.046 & 0.060 & 0     \\
120 & 0.052 & 0.042 & 0.038 & 0.034 & 0.052 & 0.040 & 0.036 & 0.044 & 0.052 & 0     \\
130 & 0.048 & 0.052 & 0.044 & 0.036 & 0.048 & 0.062 & 0.054 & 0.054 & 0.038 & 0     \\
140 & 0.062 & 0.058 & 0.056 & 0.038 & 0.038 & 0.046 & 0.042 & 0.032 & 0.044 & 0     \\
150 & 0.038 & 0.056 & 0.046 & 0.052 & 0.060 & 0.060 & 0.054 & 0.050 & 0.050 & 0     \\
\bottomrule
\end{tabular}
\label{rejection_rate_3.1-2_H0}
\end{table}

\begin{table}[!ht]
\small
\centering
\caption{
Empirical rejection rates under $H_1$ in Experiment \uppercase\expandafter{\romannumeral2}
}
\begin{tabular}{c|ccccccc|ccc}
\toprule
& \multicolumn{7}{c|}{Smooth tests} & \multirow{2}{*}{SW} & \multirow{2}{*}{JB} & \multirow{2}{*}{KS} \\
\cmidrule(lr){2-8}
$m$  & $K = 1$ & $K = 2$ & $K = 3$ & $K = 4$ & $K = 5$ 
     & $\widehat K\&\chi^2_1$ & $\widehat K\&H(x)$ 
     \\
\midrule
10  & 0.046 & 0.998 & 0.996 & 0.998 & 0.996 & 0.998 & 0.998 & 1 & 0.990 & 0.824 \\
20  & 0.018 & 1 & 1 & 1 & 1 & 1 & 1 & 1 & 1 & 0.994 \\
30  & 0.022 & 1 & 1 & 1 & 1 & 1 & 1 & 1 & 1 & 1 \\
40  & 0.042 & 1 & 1 & 1 & 1 & 1 & 1 & 1 & 1 & 1 \\
50  & 0.032 & 1 & 1 & 1 & 1 & 1 & 1 & 1 & 1 & 1 \\
60  & 0.026 & 1 & 1 & 1 & 1 & 1 & 1 & 1 & 1 & 1 \\
70  & 0.022 & 1 & 1 & 1 & 1 & 1 & 1 & 1 & 1 & 1 \\
80  & 0.028 & 1 & 1 & 1 & 1 & 1 & 1 & 1 & 1 & 1 \\
90  & 0.032 & 1 & 1 & 1 & 1 & 1 & 1 & 1 & 1 & 1 \\
100 & 0.028 & 1 & 1 & 1 & 1 & 1 & 1 & 1 & 1 & 1 \\
110 & 0.032 & 1 & 1 & 1 & 1 & 1 & 1 & 1 & 1 & 1 \\
120 & 0.030 & 1 & 1 & 1 & 1 & 1 & 1 & 1 & 1 & 1 \\
130 & 0.040 & 1 & 1 & 1 & 1 & 1 & 1 & 1 & 1 & 1 \\
140 & 0.044 & 1 & 1 & 1 & 1 & 1 & 1 & 1 & 1 & 1 \\
150 & 0.034 & 1 & 1 & 1 & 1 & 1 & 1 & 1 & 1 & 1 \\
\bottomrule
\end{tabular}
\label{rejection_rate_3.1-2_H1}
\end{table}

\begin{table}[!ht]
\small
\centering
\caption{Empirical frequency of $\widehat K$ in Experiment \uppercase\expandafter{\romannumeral2}}
\begin{tabular}{c|ccccc|ccccc}
\toprule
 & \multicolumn{5}{c|}{Under $H_0$} & \multicolumn{5}{c}{Under $H_1$} \\
\midrule
$m$ & $\widehat K=1$ & $\widehat K=2$ & $\widehat K=3$ & $\widehat K=4$ & $\widehat K=5$
    & $\widehat K=1$ & $\widehat K=2$ & $\widehat K=3$ & $\widehat K=4$ & $\widehat K=5$ \\
\midrule
10  & 0.968 & 0.026 & 0.004 & 0     & 0.002 & 0.002 & 0.984 & 0.012 & 0     & 0.002 \\
20  & 0.982 & 0.016 & 0.002 & 0     & 0     & 0     & 0.994 & 0.004 & 0     & 0.002 \\
30  & 0.978 & 0.020 & 0.002 & 0     & 0     & 0     & 0.990 & 0.008 & 0     & 0.002 \\
40  & 0.992 & 0.008 & 0     & 0     & 0     & 0     & 0.996 & 0.004 & 0     & 0 \\
50  & 0.990 & 0.006 & 0.004 & 0     & 0     & 0     & 0.996 & 0.004 & 0     & 0 \\
60  & 0.996 & 0.004 & 0     & 0     & 0     & 0     & 0.994 & 0.004 & 0     & 0.002 \\
70  & 0.994 & 0.006 & 0     & 0     & 0     & 0     & 0.998 & 0.002 & 0     & 0 \\
80  & 0.984 & 0.016 & 0     & 0     & 0     & 0     & 1     & 0     & 0     & 0 \\
90  & 0.996 & 0.004 & 0     & 0     & 0     & 0     & 1     & 0     & 0     & 0 \\
100 & 0.994 & 0.006 & 0     & 0     & 0     & 0     & 0.996 & 0.004 & 0     & 0 \\
110 & 0.996 & 0.004 & 0     & 0     & 0     & 0     & 0.998 & 0.002 & 0     & 0 \\
120 & 0.996 & 0.004 & 0     & 0     & 0     & 0     & 0.994 & 0.006 & 0     & 0 \\
130 & 0.994 & 0.006 & 0     & 0     & 0     & 0     & 0.998 & 0.002 & 0     & 0 \\
140 & 0.994 & 0.006 & 0     & 0     & 0     & 0     & 0.998 & 0.002 & 0     & 0 \\
150 & 0.988 & 0.012 & 0     & 0     & 0     & 0     & 0.996 & 0.004 & 0     & 0 \\
\bottomrule
\end{tabular}
\label{k_freq_3.1-2}
\end{table}

\begin{figure}[!ht]
	\centering
	\includegraphics[scale=0.65]{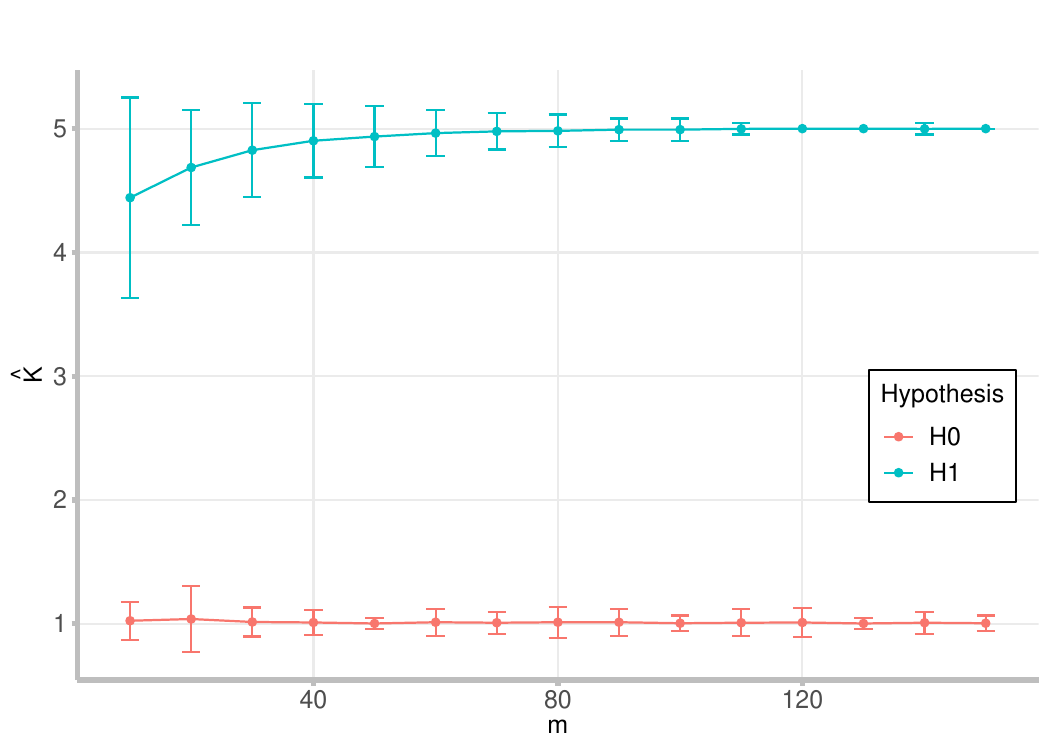}
	\caption{The sample means and error bars of $\widehat K$ in Experiment \uppercase\expandafter{\romannumeral3}}
	\label{Fig3.2}
\end{figure}

\begin{figure}[!ht]
	\centering
	\includegraphics[scale=0.65]{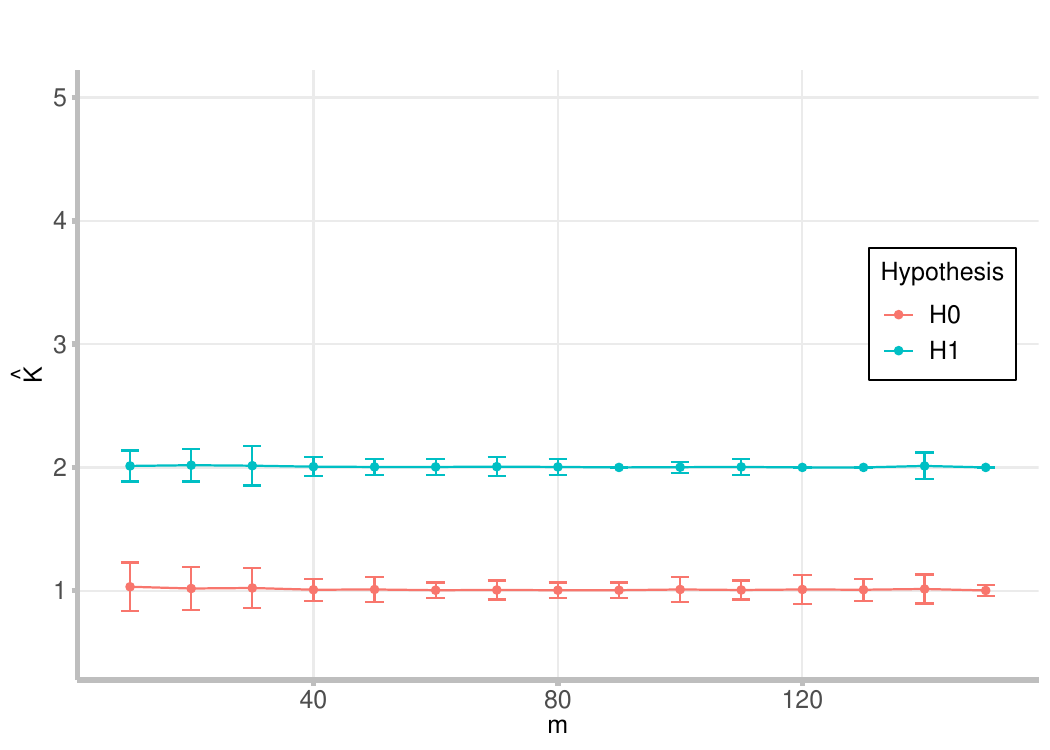}
	\caption{The sample means and error bars of $\widehat K$ in Experiment \uppercase\expandafter{\romannumeral4}}
	\label{Fig3.3}    
\end{figure}

The next two experiments, focusing on models \eqref{model_diff_mean} and \eqref{model_diff_var}, follow simulation settings similar to the previous ones. The data-generating process for each model is specified as follows.
\begin{itemize}
    \item \textbf{Experiment \uppercase\expandafter{\romannumeral3}}:
\end{itemize}
\begin{equation*}
	Y_{ij}^{(0,3)}  \sim  \mathcal{N}\left( 5j, 4\right),  \quad Y_{ij}^{(1,3)}  \sim  \chi_2^2 + (5j - 2), \quad i = 1, \ldots, jm, \quad j = 1, \ldots, 5.
\end{equation*}
\begin{itemize}
    \item \textbf{Experiment \uppercase\expandafter{\romannumeral4}}:
\end{itemize}
\begin{equation*}
	Y_{ij}^{(0,4)}  \sim  \mathcal{N}\left( 8, j^2\right), \ Y_{ij}^{(1,4)}  \sim \mathcal{U} \left[ 8 - \sqrt{3} j, 8 +  \sqrt{3} j\right], \ i = 1, \ldots, jm, \ j = 1, \ldots, 5.
\end{equation*}
Here $\{Y_{ij}^{(0,3)}\}_{i=1,j=1}^{jm,5}$ and $\{Y_{ij}^{(0,4)}\}_{i=1,j=1}^{jm,5}$ are generated under $H_0$, while $\{Y_{ij}^{(1,3)}\}_{i=1,j=1}^{jm,5}$ and $\{Y_{ij}^{(1,4)}\}_{i=1,j=1}^{jm,5}$ are generated under $H_1$. Note that in Experiment \uppercase\expandafter{\romannumeral3}, $\mathbb{E} [Y_{ij}^{(0,3)}] = \mathbb{E} [Y_{ij}^{(1,3)}] = 5j$ and $\var [Y_{ij}^{(0,3)}] = \var[Y_{ij}^{(1,3)}] = 4$, which corresponds to model \eqref{model_diff_mean}; in Experiment \uppercase\expandafter{\romannumeral4}, $\mathbb{E} [Y_{ij}^{(0,4)}] = \mathbb{E} [Y_{ij}^{(1,4)}] = 8$ and $\var[Y_{ij}^{(0,4)}] = \var [Y_{ij}^{(1,4)}] = j^2$, which corresponds to model \eqref{model_diff_var}. For the smooth tests, the results of Experiments \uppercase\expandafter{\romannumeral3} and \uppercase\expandafter{\romannumeral4} are broadly consistent with those of Experiments \uppercase\expandafter{\romannumeral1} and \uppercase\expandafter{\romannumeral2} (see Tables \ref{rejection_rate_3.2}--\ref{k_freq_3.3} and Figures \ref{Fig3.2}--\ref{Fig3.3}), further supporting the validity of our unified testing framework. For the other three classical methods, a notable phenomenon arises in Experiment \uppercase\expandafter{\romannumeral4}: under $H_0$, the Jarque--Bera test exhibits undersizing, suggesting that it is not well suited for models with heterogeneous group variances. In contrast, our proposed test remains well-calibrated and maintains stable size control.

To further assess robustness with respect to the number of groups $J$, we also consider a variant of Experiment \uppercase\expandafter{\romannumeral3} with $J=10$ (denoted as Experiment \uppercase\expandafter{\romannumeral3}$^\prime$). The results corroborate our main conclusions; see Section \ref{section_add_simu} of the \hyperlink{app}{Online Appendix} for details. In summary, the experiments demonstrate that the proposed smooth tests in Section \ref{section_prop} maintain the nominal significance level well under the null, and exhibit high power under the alternatives except when $\mathbb{E}[\pi_k(Z)] = 0$ for all $1 \le k \le K$. The data-driven tests with the approximated null distribution $H(x)$ reliably control the Type \uppercase\expandafter{\romannumeral1} error and achieve excellent power performance. In contrast with the three classical methods, our approach is supported by solid theoretical guarantees and demonstrates uniformly stable and competitive performance across all settings considered.

\begin{table}[!ht]
\small
\centering
\caption{
Empirical rejection rates under $H_0$ in Experiment \uppercase\expandafter{\romannumeral3}. The empirical power under $H_1$ equals 1 for all sample sizes and all considered tests.
}
\begin{tabular}{c|ccccccc|ccc}
\toprule
& \multicolumn{7}{c|}{Smooth tests} & \multirow{2}{*}{SW} & \multirow{2}{*}{JB} & \multirow{2}{*}{KS} \\
\cmidrule(lr){2-8}
$m$  & $K = 1$ & $K = 2$ & $K = 3$ & $K = 4$ & $K = 5$ 
     & $\widehat K\&\chi^2_1$ & $\widehat K\&H(x)$ 
     \\
\midrule
10  & 0.050 & 0.046 & 0.048 & 0.048 & 0.052 & 0.076 & 0.050 & 0.046 & 0.038 & 0 \\
20  & 0.072 & 0.060 & 0.050 & 0.046 & 0.050 & 0.054 & 0.048 & 0.064 & 0.048 & 0 \\
30  & 0.054 & 0.046 & 0.054 & 0.038 & 0.056 & 0.074 & 0.062 & 0.064 & 0.062 & 0 \\
40  & 0.046 & 0.048 & 0.060 & 0.054 & 0.048 & 0.054 & 0.050 & 0.050 & 0.046 & 0 \\
50  & 0.044 & 0.060 & 0.052 & 0.052 & 0.048 & 0.040 & 0.038 & 0.042 & 0.050 & 0 \\
60  & 0.066 & 0.044 & 0.048 & 0.052 & 0.058 & 0.048 & 0.042 & 0.048 & 0.052 & 0 \\
70  & 0.044 & 0.048 & 0.038 & 0.046 & 0.048 & 0.072 & 0.064 & 0.052 & 0.044 & 0 \\
80  & 0.050 & 0.044 & 0.048 & 0.038 & 0.042 & 0.074 & 0.068 & 0.042 & 0.044 & 0 \\
90  & 0.040 & 0.060 & 0.046 & 0.054 & 0.046 & 0.048 & 0.042 & 0.058 & 0.046 & 0 \\
100 & 0.062 & 0.056 & 0.042 & 0.050 & 0.048 & 0.058 & 0.052 & 0.048 & 0.044 & 0 \\
110 & 0.060 & 0.046 & 0.066 & 0.052 & 0.050 & 0.068 & 0.054 & 0.058 & 0.060 & 0 \\
120 & 0.034 & 0.044 & 0.038 & 0.038 & 0.036 & 0.058 & 0.056 & 0.046 & 0.052 & 0 \\
130 & 0.038 & 0.050 & 0.042 & 0.052 & 0.050 & 0.058 & 0.048 & 0.052 & 0.056 & 0 \\
140 & 0.038 & 0.050 & 0.048 & 0.042 & 0.044 & 0.044 & 0.040 & 0.058 & 0.068 & 0 \\
150 & 0.064 & 0.056 & 0.048 & 0.058 & 0.058 & 0.046 & 0.042 & 0.046 & 0.048 & 0 \\
\bottomrule
\end{tabular}
\label{rejection_rate_3.2}
\end{table}

\begin{table}[!ht]
\small
\centering
\caption{Empirical frequency of $\widehat K$ in Experiment \uppercase\expandafter{\romannumeral3}}
\begin{tabular}{c|ccccc|ccccc}
\toprule
 & \multicolumn{5}{c|}{Under $H_0$} & \multicolumn{5}{c}{Under $H_1$} \\
\midrule
$m$ & $\widehat K=1$ & $\widehat K=2$ & $\widehat K=3$ & $\widehat K=4$ & $\widehat K=5$
    & $\widehat K=1$ & $\widehat K=2$ & $\widehat K=3$ & $\widehat K=4$ & $\widehat K=5$ \\
\midrule
10  & 0.976 & 0.024 & 0     & 0     & 0     & 0.032 & 0.004 & 0     & 0.418 & 0.546 \\
20  & 0.974 & 0.018 & 0.006 & 0     & 0.002 & 0     & 0     & 0     & 0.314 & 0.686 \\
30  & 0.986 & 0.014 & 0     & 0     & 0     & 0     & 0     & 0     & 0.174 & 0.826 \\
40  & 0.990 & 0.010 & 0     & 0     & 0     & 0     & 0     & 0     & 0.098 & 0.902 \\
50  & 0.998 & 0.002 & 0     & 0     & 0     & 0     & 0     & 0     & 0.064 & 0.936 \\
60  & 0.988 & 0.012 & 0     & 0     & 0     & 0     & 0     & 0     & 0.036 & 0.964 \\
70  & 0.992 & 0.008 & 0     & 0     & 0     & 0     & 0     & 0     & 0.022 & 0.978 \\
80  & 0.990 & 0.008 & 0.002 & 0     & 0     & 0     & 0     & 0     & 0.018 & 0.982 \\
90  & 0.988 & 0.012 & 0     & 0     & 0     & 0     & 0     & 0     & 0.008 & 0.992 \\
100 & 0.996 & 0.004 & 0     & 0     & 0     & 0     & 0     & 0     & 0.008 & 0.992 \\
110 & 0.994 & 0.004 & 0.002 & 0     & 0     & 0     & 0     & 0     & 0.002 & 0.998 \\
120 & 0.992 & 0.006 & 0.002 & 0     & 0     & 0     & 0     & 0     & 0     & 1     \\
130 & 0.998 & 0.002 & 0     & 0     & 0     & 0     & 0     & 0     & 0     & 1     \\
140 & 0.992 & 0.008 & 0     & 0     & 0     & 0     & 0     & 0     & 0.002 & 0.998 \\
150 & 0.996 & 0.004 & 0     & 0     & 0     & 0     & 0     & 0     & 0     & 1     \\
\bottomrule
\end{tabular}
\label{k_freq_3.2}
\end{table}


\begin{table}[!ht]
\small
\centering
\caption{
Empirical rejection rates under $H_0$ in Experiment \uppercase\expandafter{\romannumeral4}
}
\begin{tabular}{c|ccccccc|ccc}
\toprule
& \multicolumn{7}{c|}{Smooth tests} & \multirow{2}{*}{SW} & \multirow{2}{*}{JB} & \multirow{2}{*}{KS} \\
\cmidrule(lr){2-8}
$m$  & $K = 1$ & $K = 2$ & $K = 3$ & $K = 4$ & $K = 5$ 
     & $\widehat K\&\chi^2_1$ & $\widehat K\&H(x)$ 
     \\
\midrule
10  & 0.034 & 0.044 & 0.040 & 0.030 & 0.028 & 0.062 & 0.042 & 0.046 & 0.008 & 0 \\
20  & 0.062 & 0.064 & 0.060 & 0.056 & 0.058 & 0.060 & 0.050 & 0.038 & 0.026 & 0.002 \\
30  & 0.040 & 0.040 & 0.038 & 0.042 & 0.048 & 0.060 & 0.044 & 0.042 & 0.032 & 0 \\
40  & 0.038 & 0.042 & 0.044 & 0.028 & 0.032 & 0.052 & 0.040 & 0.042 & 0.034 & 0 \\
50  & 0.044 & 0.042 & 0.042 & 0.042 & 0.044 & 0.060 & 0.048 & 0.052 & 0.048 & 0 \\
60  & 0.062 & 0.080 & 0.052 & 0.064 & 0.074 & 0.082 & 0.066 & 0.056 & 0.044 & 0 \\
70  & 0.062 & 0.066 & 0.060 & 0.050 & 0.062 & 0.052 & 0.042 & 0.040 & 0.046 & 0 \\
80  & 0.046 & 0.048 & 0.042 & 0.058 & 0.036 & 0.040 & 0.038 & 0.044 & 0.046 & 0 \\
90  & 0.060 & 0.058 & 0.050 & 0.052 & 0.046 & 0.052 & 0.042 & 0.048 & 0.046 & 0 \\
100 & 0.052 & 0.038 & 0.046 & 0.058 & 0.042 & 0.054 & 0.048 & 0.062 & 0.046 & 0 \\
110 & 0.040 & 0.034 & 0.038 & 0.032 & 0.040 & 0.052 & 0.046 & 0.042 & 0.034 & 0 \\
120 & 0.056 & 0.044 & 0.066 & 0.046 & 0.046 & 0.066 & 0.058 & 0.052 & 0.038 & 0 \\
130 & 0.052 & 0.058 & 0.054 & 0.056 & 0.050 & 0.058 & 0.052 & 0.068 & 0.050 & 0 \\
140 & 0.050 & 0.060 & 0.054 & 0.042 & 0.040 & 0.074 & 0.068 & 0.062 & 0.058 & 0 \\
150 & 0.056 & 0.064 & 0.054 & 0.058 & 0.060 & 0.054 & 0.052 & 0.042 & 0.046 & 0 \\
\bottomrule
\end{tabular}
\label{rejection_rate_3.3_H0}
\end{table}

\begin{table}[!ht]
\small
\centering
\caption{
Empirical rejection rates under $H_1$ in Experiment \uppercase\expandafter{\romannumeral4}
}
\begin{tabular}{c|ccccccc|ccc}
\toprule
& \multicolumn{7}{c|}{Smooth tests} & \multirow{2}{*}{SW} & \multirow{2}{*}{JB} & \multirow{2}{*}{KS} \\
\cmidrule(lr){2-8}
$m$  & $K = 1$ & $K = 2$ & $K = 3$ & $K = 4$ & $K = 5$ 
     & $\widehat K\&\chi^2_1$ & $\widehat K\&H(x)$ 
     \\
\midrule
10  & 0.040 & 0.998 & 0.998 & 0.996 & 0.994 & 0.998 & 0.998 & 0.994 & 0.896 & 0.036 \\
20  & 0.034 & 1 & 1 & 1 & 1 & 1 & 1 & 1 & 1 & 0.394 \\
30  & 0.034 & 1 & 1 & 1 & 1 & 1 & 1 & 1 & 1 & 0.834 \\
40  & 0.024 & 1 & 1 & 1 & 1 & 1 & 1 & 1 & 1 & 0.978 \\
50  & 0.036 & 1 & 1 & 1 & 1 & 1 & 1 & 1 & 1 & 1 \\
60  & 0.016 & 1 & 1 & 1 & 1 & 1 & 1 & 1 & 1 & 1 \\
70  & 0.050 & 1 & 1 & 1 & 1 & 1 & 1 & 1 & 1 & 1 \\
80  & 0.042 & 1 & 1 & 1 & 1 & 1 & 1 & 1 & 1 & 1 \\
90  & 0.032 & 1 & 1 & 1 & 1 & 1 & 1 & 1 & 1 & 1 \\
100 & 0.018 & 1 & 1 & 1 & 1 & 1 & 1 & 1 & 1 & 1 \\
110 & 0.032 & 1 & 1 & 1 & 1 & 1 & 1 & 1 & 1 & 1 \\
120 & 0.040 & 1 & 1 & 1 & 1 & 1 & 1 & 1 & 1 & 1 \\
130 & 0.028 & 1 & 1 & 1 & 1 & 1 & 1 & 1 & 1 & 1 \\
140 & 0.028 & 1 & 1 & 1 & 1 & 1 & 1 & 1 & 1 & 1 \\
150 & 0.028 & 1 & 1 & 1 & 1 & 1 & 1 & 1 & 1 & 1 \\
\bottomrule
\end{tabular}
\label{rejection_rate_3.3_H1}
\end{table}

\begin{table}[!ht]
\small
\centering
\caption{Empirical frequency of $\widehat K$ in Experiment \uppercase\expandafter{\romannumeral4}}
\begin{tabular}{c|ccccc|ccccc}
\toprule
 & \multicolumn{5}{c|}{Under $H_0$} & \multicolumn{5}{c}{Under $H_1$} \\
\cmidrule(lr){2-6}\cmidrule(lr){7-11}
$m$ & $\widehat K=1$ & $\widehat K=2$ & $\widehat K=3$ & $\widehat K=4$ & $\widehat K=5$
    & $\widehat K=1$ & $\widehat K=2$ & $\widehat K=3$ & $\widehat K=4$ & $\widehat K=5$ \\
\midrule
10  & 0.972 & 0.024 & 0.004 & 0     & 0     & 0     & 0.990 & 0.008 & 0.002 & 0     \\
20  & 0.986 & 0.012 & 0     & 0.002 & 0     & 0     & 0.982 & 0.018 & 0     & 0     \\
30  & 0.980 & 0.018 & 0.002 & 0  & 0   & 0     & 0.990 & 0.008 & 0      & 0.002 \\
40  & 0.992 & 0.008 & 0     & 0     & 0     & 0     & 0.994 & 0.006 & 0     & 0     \\
50  & 0.990 & 0.010 & 0     & 0     & 0     & 0     & 0.996 & 0.004 & 0     & 0     \\
60  & 0.996 & 0.004 & 0     & 0     & 0     & 0     & 0.996 & 0.004 & 0     & 0     \\
70  & 0.994 & 0.006 & 0     & 0     & 0     & 0     & 0.994 & 0.006 & 0     & 0     \\
80  & 0.996 & 0.004 & 0     & 0     & 0     & 0     & 0.996 & 0.004 & 0     & 0     \\
90  & 0.996 & 0.004 & 0     & 0     & 0     & 0     & 1     & 0     & 0     & 0     \\
100 & 0.990 & 0.010 & 0     & 0     & 0     & 0     & 0.998 & 0.002 & 0     & 0     \\
110 & 0.994 & 0.006 & 0     & 0     & 0     & 0     & 0.996 & 0.004 & 0     & 0     \\
120 & 0.992 & 0.006 & 0.002 & 0     & 0     & 0     & 1     & 0     & 0     & 0     \\
130 & 0.992 & 0.008 & 0     & 0     & 0     & 0     & 1     & 0     & 0     & 0     \\
140 & 0.986 & 0.014 & 0     & 0     & 0     & 0     & 0.988 & 0.012 & 0     & 0     \\
150 & 0.998 & 0.002 & 0     & 0     & 0     & 0     & 1     & 0     & 0     & 0     \\
\bottomrule
\end{tabular}
\label{k_freq_3.3}
\end{table}

\section{An empirical application} \label{section_emp}
In this section, we illustrate the proposed smooth tests using data from the OECD Programme for International Student Assessment (PISA) 2018. PISA is a large-scale international assessment coordinated by the Organisation for Economic Co-operation and Development (OECD) that evaluates educational systems by measuring students' learning outcomes and school characteristics across countries \citep{schleicher2019pisa}. The dataset used in this study is publicly available at \url{https://www.kaggle.com/datasets/dilaraahan/pisa-2018-school-questionnaire}. Since PISA data are widely used to compare educational resources across countries, statistical inference for group means is commonly employed, the validity of which depends crucially on distributional assumptions such as normality. However, educational indicators, including school size and student–teacher ratios, often exhibit substantial cross-country heterogeneity and dispersion, and prior studies have documented skewed distributions in school-level resource measures \citep{hanushek2011economics,chandir2022student}. Deviations from normality may therefore affect variance estimation and the reliability of ANOVA-based inference, making it important to formally assess this assumption.

We consider two response variables, \texttt{STRATIO} (student--teacher ratio) and \texttt{SCHSIZE} (school size), measured at the school level, and use \texttt{CNTRYID} (country identifier) to define the grouping structure. Table \ref{tab:pisa_used} summarizes key sample characteristics after removing missing values, including the total number of observations $N$, the number of groups $J$, and the minimum group size $\min_j N_j$. The resulting group structure satisfies the conditions required by the asymptotic framework in Section \ref{section_prop} (see Table \ref{tb_summary_conditions}), thereby supporting the application of the proposed models and normality tests. For each response variable (\texttt{STRATIO} and \texttt{SCHSIZE}) under the grouping structure defined by \texttt{CNTRYID}, we fit the ANOVA models \eqref{model_same}, \eqref{model_diff_mean}, and \eqref{model_diff_var}, and apply the corresponding smooth tests for normality. For each model, we consider two implementations of the smooth test: one with fixed $K=4$, and one based on the data-driven procedure with $D=5$.


\begin{table}[ht]
\centering
\caption{Summary of the response and grouping variables and sample characteristics in the PISA 2018 dataset}
\label{tab:pisa_used}
\begin{tabular}{cccccc}
\toprule
Response variable & Grouping variable & $N$ & $J$ & $\min_j N_j$ \\
\hline
\texttt{STRATIO} & \texttt{CNTRYID} & 18042 &  76  & 39  \\ 
\texttt{SCHSIZE} & \texttt{CNTRYID} & 18321 &  76  & 42 \\ 
\bottomrule
\end{tabular}
\end{table}

The results show that, for both response variables, \texttt{STRATIO} and \texttt{SCHSIZE}, the null hypothesis of normality is rejected at the $0.1\%$ significance level across all three model specifications and both implementations of the proposed tests. This uniform rejection pattern indicates strong robustness of the findings. Overall, there is compelling evidence that the distributions of \texttt{STRATIO} and \texttt{SCHSIZE} deviate substantially from normality. These deviations have important implications for PISA data, where ANOVA models are commonly used to compare educational resources across countries. For variables such as student--teacher ratios and school size, pronounced skewness or heavy-tailed behavior likely reflects substantial heterogeneity in educational systems, with a small number of countries or schools exhibiting extreme values. In such settings, classical $F$-tests may be sensitive to these distributional features, potentially leading to distorted inference when comparing group means. Consequently, ANOVA-based comparisons of educational resources should be interpreted with caution, and more robust approaches, such as transformations, rank-based methods, or other distribution-free procedures, may be preferable in practice. More broadly, these findings underscore the importance of formally assessing distributional assumptions prior to conducting inference in cross-country educational studies.

\section{Concluding remarks} \label{section_end}
In this paper, we propose and examine data-driven Neyman's smooth tests for assessing the normality assumption in ANOVA models with a potentially diverging number of groups. For three types of one-way fixed effects models, we derive the asymptotic properties of the proposed tests and validate their finite-sample performance through extensive numerical studies. Our results provide a rigorous and practical tool for evaluating normality in a broad range of ANOVA settings.

Several directions remain open for future research. First, while the proposed tests are developed for specific ANOVA models, the true data-generating mechanism may not be known in practice. A natural extension is to combine our procedures with preliminary structural tests to identify the most appropriate ANOVA specification before applying the proposed methodology. Second, the assumption of independent random errors can be relaxed. For example, repeated clinical trials often involve correlated errors \citep{ma2012beyond, schober2018repeated, langenberg2022repeated}. In such cases, our smooth test framework may be adapted by incorporating estimated correlation structures. Finally, for the selection rule \eqref{bic}, it is of theoretical interest to explore the scenario where the upper bound $D = D(N)$ diverges slowly with $N$, that is, $D \to \infty$ as $N \to \infty$.

\bibliographystyle{apalike_revised_nonumber}
\bibliography{ref}

@article{shapiro1972approximate,
  title={An Approximate Analysis of Variance Test for Normality},
  author={Shapiro, Samuel S. and Francia, R. S.},
  journal={Journal of the American Statistical Association},
  volume={67},
  number={337},
  pages={215--216},
  year={1972},
  publisher={Taylor \& Francis}
}

@article{shapiro1965analysis,
  title={An Analysis of Variance Test for Normality (Complete Samples)},
  author={Shapiro, Samuel S. and Wilk, Martin B.},
  journal={Biometrika},
  volume={52},
  number={3/4},
  pages={591--611},
  year={1965},
  publisher={JSTOR}
}

@article{jarque1987test,
  title={A Test for Normality of Observations and Regression Residuals},
  author={Jarque, Carlos M. and Bera, Anil K.},
  journal={International Statistical Review/Revue Internationale de Statistique},
  volume={55},
  number={2},
  pages={163--172},
  year={1987},
  publisher={JSTOR}
}

@book{dagostino1986goodness,
  title={Goodness-of-fit Techniques},
  editor={D'Agostino, Ralph B. and Stephens, Michael A.},
  volume={68},
  year={1986},
  publisher={Marcel Dekker},
  address={New York},
  isbn={0824774874},
  series={Statistics: Textbooks and Monographs}
}

@article{lilliefors1967kolmogorov,
  title={On the Kolmogorov-Smirnov Test for Normality with Mean and Variance Unknown},
  author={Lilliefors, Hubert W.},
  journal={Journal of the American Statistical Association},
  volume={62},
  number={318},
  pages={399--402},
  year={1967},
  publisher={Taylor \& Francis}
}

@article{beutner2025two,
  title={Two-sample smooth test for the equality of distributions for dependent data and its bootstrap consistency},
  author={Beutner, Eric},
  journal={Electronic Journal of Statistics},
  volume={19},
  number={1},
  pages={982--1033},
  year={2025},
  publisher={The Institute of Mathematical Statistics and the Bernoulli Society}
}

@article{akritas2004heteroscedastic,
  title={Heteroscedastic one-way {ANOVA} and lack-of-fit tests},
  author={Akritas, Michael G and Papadatos, Nikolaos},
  journal={Journal of the American Statistical Association},
  volume={99},
  number={466},
  pages={368--382},
  year={2004},
  publisher={Taylor \& Francis}
}

@article{wang2006two,
  title={Two-way heteroscedastic {ANOVA} when the number of levels is large},
  author={Wang, Lan and Akritas, Michael G},
  journal={Statistica Sinica},
  pages={1387--1408},
  year={2006},
  publisher={JSTOR}
}

@book{van2000asymptotic,
  title={Asymptotic statistics},
  author={Van der Vaart, Aad W},
  year={2000},
  publisher={Cambridge university press}
}

@book{wickens2004design,
  title={Design and analysis: A researcher's handbook},
  author={Wickens, Thomas D and Keppel, Geoffrey},
  year={2004},
  publisher={Pearson Prentice-Hall Upper Saddle River, NJ}
}

@book{dean2017design,
  title={Design and Analysis of Experiments},
  author={Dean, Angela and Voss, Daniel and Dragulji{\'c}, Danel},
  year={2017},
  publisher={Springer}
}

@book{hirotsu2017advanced,
  title={Advanced analysis of variance},
  author={Hirotsu, Chihiro},
  year={2017},
  publisher={John Wiley \& Sons}
}

@book{montgomery2017design,
  title={Design and analysis of experiments},
  author={Montgomery, Douglas C},
  year={2017},
  publisher={John wiley \& sons}
}

@article{bonett2002test,
  title={A test of normality with high uniform power},
  author={Bonett, Douglas G and Seier, Edith},
  journal={Computational statistics \& data analysis},
  volume={40},
  number={3},
  pages={435--445},
  year={2002},
  publisher={Elsevier}
}

@article{hwang2006novel,
  title={A novel method for testing normality in a mixed model of a nested classification},
  author={Hwang, Yi-Ting and Wei, Peir Feng},
  journal={Computational statistics \& data analysis},
  volume={51},
  number={2},
  pages={1163--1183},
  year={2006},
  publisher={Elsevier}
}

@article{tiku1971power,
  title={Power function of the {F}-test under non-normal situations},
  author={Tiku, Moti Lal},
  journal={Journal of the American Statistical Association},
  volume={66},
  number={336},
  pages={913--916},
  year={1971},
  publisher={Taylor \& Francis}
}

@article{tiku1964approximating,
  title={Approximating the general non-normal variance-ratio sampling distributions},
  author={Tiku, ML},
  journal={Biometrika},
  volume={51},
  number={1-2},
  pages={83--95},
  year={1964},
  publisher={Oxford University Press}
}

@article{donaldson1968robustness,
  title={Robustness of the {F}-test to errors of both kinds and the correlation between the numerator and denominator of the {F}-ratio},
  author={Donaldson, Theodore S},
  journal={Journal of the American Statistical Association},
  volume={63},
  number={322},
  pages={660--676},
  year={1968},
  publisher={Taylor \& Francis}
}

@article{srivastava1959effect,
  title={Effect of non-normality on the power of the analysis of variance test},
  author={Srivastava, ABL},
  journal={Biometrika},
  volume={46},
  number={1/2},
  pages={114--122},
  year={1959},
  publisher={JSTOR}
}

@article{atiqullah1962estimation,
  title={The estimation of residual variance in quadratically balanced least-squares problems and the robustness of the {F}-test},
  author={Atiqullah, M},
  journal={Biometrika},
  volume={49},
  number={1-2},
  pages={83--91},
  year={1962},
  publisher={Oxford University Press}
}

@article{david1951effect,
  title={The effect of non-normality on the power function of the {F}-test in the analysis of variance},
  author={David, Florence N and Johnson, NL},
  journal={Biometrika},
  volume={38},
  number={1/2},
  pages={43--57},
  year={1951},
  publisher={JSTOR}
}

@article{david1951method,
  title={A method of investigating the effect of nonnormality and heterogeneity of variance on tests of the general linear hypothesis},
  author={David, FN and Johnson, NL},
  journal={The Annals of Mathematical Statistics},
  pages={382--392},
  year={1951},
  publisher={JSTOR}
}

@article{gayen1950distribution,
  title={The distribution of the variance ratio in random samples of any size drawn from non-normal universes},
  author={Gayen, AK},
  journal={Biometrika},
  volume={37},
  number={3/4},
  pages={236--255},
  year={1950},
  publisher={JSTOR}
}

@article{pearson1931analysis,
  title={The analysis of variance in cases of non-normal variation},
  author={Pearson, Egon S},
  journal={Biometrika},
  pages={114--133},
  year={1931},
  publisher={JSTOR}
}

@article{ali1996robustness,
  title={Robustness to nonnormality of regression {F}-tests},
  author={Ali, Mukhtar M and Sharma, Subhash C},
  journal={Journal of Econometrics},
  volume={71},
  number={1-2},
  pages={175--205},
  year={1996},
  publisher={Elsevier}
}

@article{bera_neymans_2002,
	title = {Neyman's smooth test and its applications in econometrics},
	volume = {165},
	url = {https://ink.library.smu.edu.sg/soe_research/587},
	journal = {Handbook of Applied Econometrics and Statistical Inference},
	author = {Bera, Anil K. and Ghosh, Aurobindo},
	month = {jan},
	year = {2002},
	pages = {170--230},
}

@book{thasComparingDistributions2010,
  title = {Comparing Distributions},
  author = {Thas, Olivier},
  year = {2010},
  publisher = {Springer},
  address = {New York},
  isbn = {978-0-387-92709-1 978-0-387-92710-7}
}

@book{Lehmann_Romano_2022,
  title = {Testing Statistical Hypotheses},
  author = {E.L. Lehmann and Joseph P. Romano},
  year = {2022},
  publisher = {Springer},
  address = {New York},
  isbn = {978-3-030-70577-0 978-3-030-70580-0 978-3-030-70578-7}
}

@article{kallenberg1997dataB,
  title={Data driven smooth tests for composite hypotheses comparison of powers},
  author={Kallenberg, Wilbert CM and Teresa, Ledwina},
  journal={Journal of Statistical Computation and Simulation},
  volume={59},
  number={2},
  pages={101--121},
  year={1997},
  publisher={Taylor \& Francis}
}

@article{kallenberg1997dataA,
  title={Data-driven smooth tests when the hypothesis is composite},
  author={Kallenberg, Wilbert CM and Ledwina, Teresa},
  journal={Journal of the American Statistical Association},
  volume={92},
  number={439},
  pages={1094--1104},
  year={1997},
  publisher={Taylor \& Francis}
}

@article{kallenberg1995consistency,
  title={Consistency and Monte Carlo simulation of a data driven version of smooth goodness-of-fit tests},
  author={Kallenberg, Wilbert CM and Ledwina, Teresa},
  journal={The Annals of Statistics},
  pages={1594--1608},
  year={1995},
  publisher={JSTOR}
}

@article{inglot1996asymptotic,
  title={Asymptotic optimality of data-driven Neyman's tests for uniformity},
  author={Inglot, Tadeusz and Ledwina, Teresa},
  journal={The Annals of Statistics},
  volume={24},
  number={5},
  pages={1982--2019},
  year={1996},
  publisher={Institute of Mathematical Statistics}
}

@book{miller1997beyond,
  title={Beyond {ANOVA}: basics of applied statistics},
  author={Miller Jr, Rupert G},
  year={1997},
  publisher={CRC press}
}

@article{bera2013smooth,
  title={A smooth test for the equality of distributions},
  author={Bera, Anil K and Ghosh, Aurobindo and Xiao, Zhijie},
  journal={Econometric Theory},
  volume={29},
  number={2},
  pages={419--446},
  year={2013},
  publisher={Cambridge University Press}
}

@article{janic2000data,
  title={Data driven rank test for two-sample problem},
  author={Janic-Wr{\'o}blewska, Alicja and Ledwina, Teresa},
  journal={Scandinavian Journal of Statistics},
  volume={27},
  number={2},
  pages={281--297},
  year={2000},
  publisher={Wiley Online Library}
}

@article{neyman1937smooth,
  title={Smooth Test for Goodness of Fit},
  author={Neyman, Jerzy},
  journal={Scandinavian Actuarial Journal},
  volume={1937},
  number={3-4},
  pages={149--199},
  year={1937},
  publisher={Taylor \& Francis}
}

@article{song2022smooth,
  title={On smooth tests for the equality of distributions},
  author={Song, Xiaojun and Xiao, Zhijie},
  journal={Econometric Theory},
  volume={38},
  number={1},
  pages={194--208},
  year={2022},
  publisher={Cambridge University Press}
}

@article{akritas2000asymptotics,
  title={Asymptotics for analysis of variance when the number of levels is large},
  author={Akritas, Michael and Arnold, Steven},
  journal={Journal of the American Statistical association},
  volume={95},
  number={449},
  pages={212--226},
  year={2000},
  publisher={Taylor \& Francis Group}
}

@article{bonett1990testing,
  title={Testing residual normality in the {ANOVA} model},
  author={Bonett, Douglas G and Woodward, J Arthur},
  journal={Journal of Applied Statistics},
  volume={17},
  number={3},
  pages={383--387},
  year={1990},
  publisher={Taylor \& Francis}
}

@article{ledwina1994data,
  title={Data-driven version of {N}eyman's smooth test of fit},
  author={Ledwina, Teresa},
  journal={Journal of the American Statistical Association},
  volume={89},
  number={427},
  pages={1000--1005},
  year={1994},
  publisher={Taylor \& Francis Group}
}

@article{kallenberg1999data,
  title={Data-driven rank tests for independence},
  author={Kallenberg, Wilbert CM and Ledwina, Teresa},
  journal={Journal of the American Statistical Association},
  volume={94},
  number={445},
  pages={285--301},
  year={1999},
  publisher={Taylor \& Francis Group}
}

@article{kallenberg1995data,
  title={On data driven Neyman's tests},
  author={Kallenberg, Willibrordes Cornelis Maria and Ledwina, Teresa},
  journal={PROBABILITY AND MATHEMATICAL STATISTICS-WROCLAW UNIVERSITY},
  volume={15},
  pages={409--426},
  year={1995}
}

@article{inglot1997data,
  title={Data driven smooth tests for composite hypotheses},
  author={Inglot, Tadeusz and Kallenberg, Wilbert CM and Ledwina, Teresa},
  journal={The Annals of Statistics},
  volume={25},
  number={3},
  pages={1222--1250},
  year={1997},
  publisher={Institute of Mathematical Statistics}
}

@article{kraus2007data,
  title={Data-driven smooth tests of the proportional hazards assumption},
  author={Kraus, David},
  journal={Lifetime data analysis},
  volume={13},
  number={1},
  pages={1--16},
  year={2007},
  publisher={Springer}
}

@article{duchesne2016estimating,
  title={Estimating the mean and its effects on Neyman smooth tests of normality for ARMA models},
  author={Duchesne, Pierre and Lafaye De Micheaux, Pierre and Tagne Tatsinkou, Joseph},
  journal={Canadian Journal of Statistics},
  volume={44},
  number={3},
  pages={241--270},
  year={2016},
  publisher={Wiley Online Library}
}

@article{ducharme2004goodness,
  title={Goodness-of-fit tests of normality for the innovations in ARMA models},
  author={Ducharme, Gilles R and Lafaye de Micheaux, Pierre},
  journal={Journal of Time Series Analysis},
  volume={25},
  number={3},
  pages={373--395},
  year={2004},
  publisher={Wiley Online Library}
}

@article{ducharme2004smooth,
  title={A smooth test of goodness-of-fit for growth curves and monotonic nonlinear regression models},
  author={Ducharme, Gilles R and Fontez, B{\'e}n{\'e}dicte},
  journal={Biometrics},
  volume={60},
  number={4},
  pages={977--986},
  year={2004},
  publisher={Oxford University Press}
}

@book{scheffe1959anova,
  author    = {Henry Scheffé},
  title     = {The Analysis of Variance},
  year      = {1959},
  publisher = {John Wiley \& Sons},
  address   = {New York}
}

@article{gelman2005analysis,
  title={Analysis of Variance: Why It Is More Important than Ever},
  author={Gelman, Andrew},
  journal={The Annals of Statistics},
  volume={33},
  number={1},
  pages={1--53},
  year={2005}
}

@article{ma2012beyond,
  title={Beyond repeated-measures analysis of variance: advanced statistical methods for the analysis of longitudinal data in anesthesia research},
  author={Ma, Yan and Mazumdar, Madhu and Memtsoudis, Stavros G},
  journal={Regional Anesthesia \& Pain Medicine},
  volume={37},
  number={1},
  pages={99--105},
  year={2012},
  publisher={BMJ Publishing Group Ltd}
}

@article{schober2018repeated,
  title={Repeated measures designs and analysis of longitudinal data: if at first you do not succeed—try, try again},
  author={Schober, Patrick and Vetter, Thomas R},
  journal={Anesthesia \& Analgesia},
  volume={127},
  number={2},
  pages={569--575},
  year={2018},
  publisher={LWW}
}

@article{langenberg2022repeated,
  title={Repeated measures {ANOVA} with latent variables to analyze interindividual differences in contrasts},
  author={Langenberg, Benedikt and Helm, Jonathan L and Mayer, Axel},
  journal={Multivariate Behavioral Research},
  volume={57},
  number={1},
  pages={2--19},
  year={2022},
  publisher={Taylor \& Francis}
}

@article{hanushek2011economics,
  title={The economics of international differences in educational achievement},
  author={Hanushek, Eric A and Woessmann, Ludger},
  journal={Handbook of the Economics of Education},
  volume={3},
  pages={89--200},
  year={2011},
  publisher={Elsevier}
}

@article{chandir2022student,
  title={Student responses on the survey of global competence in PISA 2018},
  author={Chandir, Harsha},
  journal={Discourse: Studies in the Cultural Politics of Education},
  volume={43},
  number={4},
  pages={526--542},
  year={2022},
  publisher={Taylor \& Francis}
}

@article{schleicher2019pisa,
  title={PISA 2018: Insights and interpretations.},
  author={Schleicher, Andreas},
  journal={oecd Publishing},
  year={2019},
  publisher={ERIC}
}

\clearpage

\hypertarget{app}{}
\begin{appendix}
\renewcommand{\thesection}{A}
\section*{Online Appendix} \label{appendix} 
We provide proofs of our main theoretical results and additional numerical results in this online appendix. The symbol ``$\lesssim$'' means that the left side is bounded by a positive constant times the right side. The symbol ``$\asymp$'' means that both sides are asymptotically equivalent.

\subsection{Proof of Theorem \ref{H0_same}}

We first study the estimation effects of $\widehat \mu-\mu$ and $\widehat \sigma^2-\sigma^2$. Note that 
\begin{equation}
\widehat\mu-\mu=\frac{1}{N}\sum_{j=1}^{J}\sum_{i=1}^{N_j}\left(Y_{ij}-\mu\right)=\frac{1}{N}\sum_{j=1}^{J}\sum_{i=1}^{N_j}\varepsilon_{ij}=\frac{\sigma}{N}\sum_{j=1}^{J}\sum_{i=1}^{N_j}e_{ij}. \label{mu_hat-mu_same}
\end{equation}
Under $H_0$, $e_{ij} \sim \mathcal{N}(0,1) i.i.d$., then $\sqrt{N}(\widehat\mu-\mu)\sim \mathcal{N}(0,\sigma^2)$, and $\widehat \mu-\mu=O_p(N^{-1/2})$ as $N\to\infty$. Besides, 
\begin{align}
\widehat\sigma^2-\sigma^2=&\frac{1}{N}\sum_{j=1}^{J}\sum_{i=1}^{N_j}\left(\left(Y_{ij}-\widehat\mu\right)^2-\sigma^2\right) \nonumber\\
=&\frac{1}{N}\sum_{j=1}^{J}\sum_{i=1}^{N_j}\left(\varepsilon_{ij}^2-\sigma^2\right)-\left(\widehat\mu-\mu\right)^2\nonumber\\
=&\frac{\sigma^2}{N}\sum_{j=1}^{J}\sum_{i=1}^{N_j}\left(e_{ij}^2-1\right)+O_p\left(\frac{1}{N}\right). \label{sigma^2_hat-sigma^2_same}
\end{align}
Under $H_0$, by CLT, $\sqrt{N}(\widehat\sigma^2-\sigma^2)\overset{d}{\to} \mathcal{N}(0,2\sigma^4)$ as $N\to \infty$, then $\widehat\sigma^2-\sigma^2=O_p(N^{-1/2})$.

Now we deal with $N^{-1}\sum_{j=1}^{J}\sum_{i=1}^{N_j}\pi_k(\widehat Z_{ij})$. By second order Taylor expansion of $\pi_k(\widehat Z_{ij})$ with respect to $\widehat Z_{ij}$ at $Z_{ij}$, we obtain
\begin{align}
\frac{1}{N}\sum_{j=1}^{J}\sum_{i=1}^{N_j}\pi_k\left(\widehat Z_{ij}\right)=&\frac{1}{N}\sum_{j=1}^{J}\sum_{i=1}^{N_j}\pi_k(Z_{ij})+\frac{1}{N}\sum_{j=1}^{J}\sum_{i=1}^{N_j}\dot\pi_k(Z_{ij})\left(  \widehat Z_{ij}-Z_{ij}\right) \nonumber\\
&+\frac{1}{2N}\sum_{j=1}^{J}\sum_{i=1}^{N_j}\ddot\pi_k\left(\widetilde Z_{ij}\right)\left(  \widehat Z_{ij}-Z_{ij}\right)^2, \label{decomp}
\end{align}
where $\widetilde Z_{ij}$ lies between $\widehat Z_{ij}$ and $Z_{ij}$. Recall that $\widehat Z_{ij}=\Phi({\widehat{\varepsilon}_{ij}} / {\widehat{\sigma}})$ and $Z_{ij}=\Phi(\varepsilon_{ij}/\sigma)=\Phi(e_{ij})$. Then, by Taylor expansion again, the second term on the right-hand side of \eqref{decomp} can be further decomposed as
\begin{align}
&\frac{1}{N}\sum_{j=1}^{J}\sum_{i=1}^{N_j}\dot\pi_k(Z_{ij})\phi\left(e_{ij}\right)\left(\frac{\widehat\varepsilon_{ij}}{\widehat\sigma}- \frac{\varepsilon_{ij}}{\sigma}\right)+\frac{1}{2N}\sum_{j=1}^{J}\sum_{i=1}^{N_j}\dot\pi_k(Z_{ij})\dot\phi\left(\frac{\widetilde\varepsilon_{ij}}{\widetilde\sigma}\right)\left(\frac{\widehat\varepsilon_{ij}}{\widehat\sigma}-\frac{\varepsilon_{ij}}{\sigma}\right)^2, \label{main_term}
\end{align}
where $\widetilde\varepsilon_{ij}/\widetilde\sigma$ an intermediate point between ${\widehat{\varepsilon}_{ij}} / {\widehat{\sigma}}$ and $\varepsilon_{ij}/\sigma$, whose exact value may vary across different scenarios.

We focus on the first term in \eqref{main_term}. Straightforward calculation leads to
\begin{align}
&\frac{1}{N}\sum_{j=1}^{J}\sum_{i=1}^{N_j}\dot\pi_k(Z_{ij})\phi\left(e_{ij}\right)\left(\frac{\widehat\varepsilon_{ij}}{\widehat\sigma}-\frac{\varepsilon_{ij}}{\sigma}\right)\nonumber\\
=&\frac{1}{N}\sum_{j=1}^{J}\sum_{i=1}^{N_j}\dot\pi_k(Z_{ij})\phi\left(e_{ij}\right)\frac{\widehat\varepsilon_{ij}-\varepsilon_{ij}}{\widehat\sigma}-\frac{1}{N}\sum_{j=1}^{J}\sum_{i=1}^{N_j}\dot\pi_k(Z_{ij})\phi\left(e_{ij}\right)\frac{\varepsilon_{ij}(\widehat\sigma^2-\sigma^2)}{\widehat\sigma\sigma(\widehat\sigma+\sigma)}\nonumber\\
=&-\frac{\widehat\mu-\mu}{\widehat\sigma}\frac{1}{N}\sum_{j=1}^{J}\sum_{i=1}^{N_j}\dot\pi_k(Z_{ij})\phi\left(e_{ij}\right)-\frac{\widehat\sigma^2-\sigma^2}{\widehat\sigma(\widehat\sigma+\sigma)}\frac{1}{N}\sum_{j=1}^{J}\sum_{i=1}^{N_j}\dot\pi_k(Z_{ij})\phi\left(e_{ij}\right)e_{ij}\nonumber\\
=&-\frac{\widehat\mu-\mu}{\sigma}\mathbb{E} \left[\dot\pi_k(Z)\phi\left(e\right)\right]-\frac{\widehat\sigma^2-\sigma^2}{2\sigma^2}\mathbb{E} \left[\dot\pi_k(Z)\phi\left(e\right)e\right]\nonumber\\
&+o_p\left(\widehat\mu-\mu\right)+o_p\left(\widehat\sigma^2-\sigma^2\right), \label{a1} 
\end{align}
where the last step follows due to the law of large numbers (LLN) for $N^{-1}\sum_{j=1}^J\sum_{i=1}^{N_j}\dot\pi_k(Z_{ij})\phi(e_{ij})=\mathbb{E} [\dot\pi_k(Z)\phi(e)]+o_p(1)$ 
and $N^{-1}\sum_{j=1}^{J}\sum_{i=1}^{N_j}\dot\pi_k(Z_{ij})\phi(e_{ij})e_{ij}=\mathbb{E} [\dot\pi_k(Z)\phi(e)e]+o_p(1)$
as $N\to\infty$, as well as the consistency of $\widehat\sigma$ to $\sigma$ from the arguments above. Under $H_0$, through integration by parts, we have
\begin{equation*}
\mathbb{E} \left[\dot\pi_k(Z)\phi\left(e\right)\right]=\int_0^1\pi_k(z)\Phi^{-1}(z)\mathrm{d}z=c_{1k},
\end{equation*}
and
\begin{equation*}
\mathbb{E} \left[\dot\pi_k(Z)\phi\left(e\right)e\right]=\int_0^1\pi_k(z) \left[ \left(\Phi^{-1}(z)\right)^2  - 1\right]\mathrm{d}z = \int_0^1\pi_k(z) \left(\Phi^{-1}(z)\right)^2  \mathrm{d}z=c_{2k},
\end{equation*}
where we use the fact that $\mathrm{d} \Phi^{-1} (z) = \mathrm{d}z/\phi( \Phi^{-1} (z))$
and $\int_0^1 \pi_k (z) \mathrm{d}z = 0$. Thus, along with \eqref{mu_hat-mu_same} and \eqref{sigma^2_hat-sigma^2_same}, \eqref{a1} can be re-expressed as
\begin{equation}
\frac{1}{N}\sum_{j=1}^{J}\sum_{i=1}^{N_j}\pi_k(Z_{ij})-c_{1k}\frac{1}{N}\sum_{j=1}^{J}\sum_{i=1}^{N_j}e_{ij}-\frac{c_{2k}}{2}\frac{1}{N}\sum_{j=1}^{J}\sum_{i=1}^{N_j}\left(e_{ij}^2-1\right)+o_p\left(\frac{1}{\sqrt N}\right).  \label{main_same_H0}
\end{equation}

It suffices to show that
\begin{equation}
R_{N1}\equiv \frac{1}{2N}\sum_{j=1}^{J}\sum_{i=1}^{N_j}\ddot\pi_k\left(\widetilde Z_{ij}\right)\left(  \widehat Z_{ij}-Z_{ij}\right)^2=o_p\left(\frac{1}{\sqrt N}\right), \label{RN1}
\end{equation}  
and
\begin{equation}
R_{N2}\equiv\frac{1}{2N}\sum_{j=1}^{J}\sum_{i=1}^{N_j}\dot\pi_k(Z_{ij})\dot\phi\left(\frac{\widetilde\varepsilon_{ij}}{\widetilde\sigma}\right)\left(\frac{\widehat\varepsilon_{ij}}{\widehat\sigma}-\frac{\varepsilon_{ij}}{\sigma}\right)^2=o_p\left(\frac{1}{\sqrt N}\right).\label{RN2}
\end{equation}
For $R_{N1}$, by the boundedness of $\ddot\pi_k$ (in Assumption \ref{assumption2}) and $\phi$,
\begin{align*}
\vert R_{N1}\vert & \lesssim \frac{1}{N}\sum_{j=1}^{J}\sum_{i=1}^{N_j} \left[ 
\Phi \left( \frac{\widehat{\varepsilon}_{ij}}{\widehat{\sigma}} \right) - \Phi \left( \frac{{\varepsilon}_{ij}}{\sigma}\right)\right]^2 \\
& =\frac{1}{N}\sum_{j=1}^{J}\sum_{i=1}^{N_j} \phi^2 \left( \frac{\widetilde{\varepsilon}_{ij}}{\widetilde{\sigma}} \right) \left( 
\frac{\widehat{\varepsilon}_{ij}}{\widehat{\sigma}}  -  \frac{{\varepsilon}_{ij}}{\sigma}\right)^2 \\
& \lesssim \frac{1}{N} \sum_{j=1}^{J}\sum_{i=1}^{N_j} \left( 
\frac{\widehat{\varepsilon}_{ij}}{\widehat{\sigma}}  -  \frac{{\varepsilon}_{ij}}{\sigma}\right)^2 \\
& = \frac{1}{N} \sum_{j=1}^{J}\sum_{i=1}^{N_j}\left( \frac{\widehat{\varepsilon}_{i j}-\varepsilon_{i j}}{\widehat{\sigma}} - \frac{\varepsilon_{i j}(\widehat{\sigma}^2-\sigma^2)}{\widehat{\sigma} \sigma(\widehat{\sigma}+\sigma)}  \right)^2.
\end{align*}
By the inequality $(a-b)^2 \le 2(a^2+b^2)$ for any $a,b \in \mathbb{R}$,
\begin{align*}
& \frac{1}{N} \sum_{j=1}^{J}\sum_{i=1}^{N_j}\left( \frac{\widehat{\varepsilon}_{i j}-\varepsilon_{i j}}{\widehat{\sigma}} - \frac{\varepsilon_{i j}(\widehat{\sigma}^2-\sigma^2)}{\widehat{\sigma} \sigma(\widehat{\sigma}+\sigma)}  \right)^2 \\
\lesssim & \frac{1}{N} \sum_{j=1}^{J}\sum_{i=1}^{N_j} \left(\frac{\widehat{\varepsilon}_{i j}-\varepsilon_{i j}}{\widehat{\sigma}} \right)^2 + \frac{1}{N} \sum_{j=1}^{J}\sum_{i=1}^{N_j} \left( \frac{\varepsilon_{i j}(\widehat{\sigma}^2-\sigma^2)}{\widehat{\sigma} \sigma(\widehat{\sigma}+\sigma)}  \right)^2 \\
\asymp &  \left(\widehat{\mu} - \mu \right)^2 + (\widehat{\sigma}^2-\sigma^2)^2 \frac{1}{N} \sum_{j=1}^{J}\sum_{i=1}^{N_j} \varepsilon_{i j}^2 \\
= &  O_p \left( \frac{1}{N} \right) + O_p \left(  \frac{1}{N} \right) \left( \sigma^2 + o_p(1) \right) = o_p \left( \frac{1}{\sqrt{N}}\right),
\end{align*}
where the last step is the result of \eqref{mu_hat-mu_same}, \eqref{sigma^2_hat-sigma^2_same} and LLN for $N^{-1}\sum_{j=1}^{J}\sum_{i=1}^{N_j} \varepsilon_{i j}^2=\sigma^2+o_p(1)$. For $R_{N2}$, we have  
\begin{align*}
\vert R_{N2}\vert & = \left\vert \frac{1}{2N} \sum_{j=1}^{J}\sum_{i=1}^{N_j} \dot\pi_k (Z_{ij}) \left[ -\frac{\widetilde{\varepsilon}_{ij}}{\widetilde{\sigma}} \phi \left( \frac{\widetilde{\varepsilon}_{ij}}{\widetilde{\sigma}} \right)\right]  \left( 
\frac{\widehat{\varepsilon}_{ij}}{\widehat{\sigma}}  -  \frac{{\varepsilon}_{ij}}{\sigma}\right)^2  \right\vert  \\
& \lesssim \frac{1}{N} \sum_{j=1}^{J}\sum_{i=1}^{N_j}  \left\vert \frac{\widetilde{\varepsilon}_{ij}}{\widetilde{\sigma}} \right\vert \left( 
\frac{\widehat{\varepsilon}_{ij}}{\widehat{\sigma}}  -  \frac{{\varepsilon}_{ij}}{\sigma}\right)^2 \\
& \le \left( \frac{1}{N} \sum_{j=1}^{J}\sum_{i=1}^{N_j} \frac{\widetilde{\varepsilon}_{ij}^2}{\widetilde{\sigma}^2} \right)^{1/2} \left( \frac{1}{N} \sum_{j=1}^{J}\sum_{i=1}^{N_j} \left( 
\frac{\widehat{\varepsilon}_{ij}}{\widehat{\sigma}}  -  \frac{{\varepsilon}_{ij}}{\sigma}\right)^4 \right)^{1/2},
\end{align*}
where the first ``$=$" is due to $\dot\phi(x) = -x\phi (x)$, and the last ``$\le$" is due to Cauchy--Schwarz inequality. Similar arguments as above yield that
\begin{align*}
&\frac{1}{N} \sum_{j=1}^{J}\sum_{i=1}^{N_j} \left( 
\frac{\widehat{\varepsilon}_{ij}}{\widehat{\sigma}}  -  \frac{{\varepsilon}_{ij}}{\sigma}\right)^4 \\
= & \frac{1}{N} \sum_{j=1}^{J}\sum_{i=1}^{N_j} \left( \frac{\widehat{\varepsilon}_{i j}-\varepsilon_{i j}}{\widehat{\sigma}} - \frac{\varepsilon_{i j}(\widehat{\sigma}^2-\sigma^2)}{\widehat{\sigma} \sigma(\widehat{\sigma}+\sigma)}  \right)^4 \\
\lesssim &\left[ \frac{1}{N} \sum_{j=1}^{J}\sum_{i=1}^{N_j} \left( 
\frac{\widehat{\varepsilon}_{ij}-\varepsilon_{ij}}{\widehat{\sigma}} \right)^4 + \frac{1}{N} \sum_{j=1}^{J}\sum_{i=1}^{N_j} \left( 
\frac{\varepsilon_{i j}(\widehat{\sigma}^2-\sigma^2)}{\widehat{\sigma} \sigma(\widehat{\sigma}+\sigma)} \right)^4 \right] \\
\asymp &\left( \widehat{\mu} - \mu \right)^4 + \left( \widehat{\sigma}^2-\sigma^2\right)^4 \left( \mathbb{E} \left[e^4\right] + o_p(1)\right) = O_p \left( \frac{1}{N^2}\right),
\end{align*}
and
\begin{align}
\frac{1}{N} \sum_{j=1}^{J}\sum_{i=1}^{N_j} \frac{\widetilde{\varepsilon}_{ij}^2}{\widetilde{\sigma}^2} & \le \frac{1}{N} \sum_{j=1}^{J}\sum_{i=1}^{N_j} \left( \frac{\widehat{\varepsilon}_{ij}}{\widehat{\sigma}}\right)^2 + \frac{1}{N} \sum_{j=1}^{J}\sum_{i=1}^{N_j} \left( \frac{{\varepsilon}_{ij}}{{\sigma}}\right)^2 \nonumber\\
& = 1 + \left( 1 + o_p(1)\right) = O_p(1). \label{RN2_Op1}
\end{align}
Thus, $\vert R_{N2}\vert = O_p(N^{-1})=o_p(N^{-1/2})$. Hence \eqref{RN1} and \eqref{RN2} are verified. Combining \eqref{decomp}, \eqref{main_term}, \eqref{a1}, \eqref{main_same_H0}, \eqref{RN1} and
\eqref{RN2}, we complete the proof of Theorem \ref{H0_same}. $\square$

\subsection{Proof of Theorem \ref{H1_same}} 

The proof of Theorem \ref{H1_same} is similar to that of Theorem \ref{H0_same}. Under $H_1$, \eqref{mu_hat-mu_same} still holds, and $\sqrt{N}(\widehat\mu-\mu) \overset{d}{\to} \mathcal{N}(0,\sigma^2)$ since $e_{ij}\sim F \ne \Phi$. Analogously, \eqref{sigma^2_hat-sigma^2_same} still holds, and $\sqrt{N}(\widehat\sigma^2-\sigma^2) \overset{d}{\to} \sigma^2\mathcal{N}(0,\mathbb{E}[e^4]-1)$. 

We start from \eqref{decomp}. Under $H_1$, the arguments and results in \eqref{decomp}, \eqref{main_term} and \eqref{a1} still hold. Unlike \eqref{main_same_H0}, here in the context of $H_1$ the last line of \eqref{a1} should be expressed as
\begin{equation}
\frac{1}{N}\sum_{j=1}^{J}\sum_{i=1}^{N_j}\pi_k(Z_{ij})-d_{1k}\frac{1}{N}\sum_{j=1}^{J}\sum_{i=1}^{N_j}e_{ij}-\frac{d_{2k}}{2}\frac{1}{N}\sum_{j=1}^{J}\sum_{i=1}^{N_j}\left(e_{ij}^2-1\right)+o_p\left(\frac{1}{\sqrt N}\right),  \label{main_same_H1}
\end{equation}
where we replace $c_{1k}$ with $d_{1k}$ and $c_{2k}$ with $d_{2k}$. Besides, the remainder terms $R_{N1}$ in \eqref{RN1} and $R_{N2}$ in \eqref{RN2} are still $O_p(N^{-1/2})$ through similar derivation above. Combining \eqref{decomp}, \eqref{main_term}, \eqref{a1}, \eqref{RN1}, \eqref{RN2} and \eqref{main_same_H1}, we obtain \eqref{H1_same_decomp} in Theorem \ref{H1_same}. From \eqref{H1_same_decomp}, we know that
\begin{equation*}
\frac{1}{N}\sum_{j=1}^{J}\sum_{i=1}^{N_j}\pi_k\left(\widehat Z_{ij}\right) = \mathbb{E} \left[\pi_k (Z)\right] + o_p(1), 
\end{equation*}
as $N\to\infty$, which implies $\widehat{\Psi}_K^2  \overset{p}{\to}  \bm{a}_K^\top \bm{\Sigma}_K^{-1} \bm{a}_K$. Furthermore,
\begin{align}
&\sqrt{N} \left( \widehat{\Psi}_K^2 - \bm{a}_K^\top \bm{\Sigma}_K^{-1} \bm{a}_K \right)  \nonumber\\
=& \sqrt{N} \left[ \bm{a}_K +  \frac{1}{N} \sum_{j = 1}^J \sum_{i = 1}^{N_j} \bm\pi_K\left(\widehat Z_{ij}\right)-\bm{a}_K\right]^\top \bm{\Sigma}_K^{-1} \left[ \bm{a}_K +  \frac{1}{N} \sum_{j = 1}^J \sum_{i = 1}^{N_j} \bm\pi_K\left(\widehat Z_{ij}\right)-\bm{a}_K\right] \nonumber\\
&- \sqrt{N}\bm{a}_K^\top \bm{\Sigma}_K^{-1} \bm{a}_K  \nonumber\\
=& 2a^\top \bm{\Sigma}_K^{-1} \frac{1}{\sqrt{N}} \sum_{j = 1}^J \sum_{i = 1}^{N_j} \left( \bm\pi_K \left(\widehat Z_{ij}\right) - \bm{a}_K \right) +\left[\frac{1}{\sqrt{N}} \sum_{j = 1}^J \sum_{i = 1}^{N_j} \left( \bm\pi_K \left(\widehat Z_{ij}\right) - \bm{a}_K \right)\right]^\top\bm{\Sigma}_K^{-1}\nonumber\\
&\left[\frac{1}{\sqrt{N}} \sum_{j = 1}^J \sum_{i = 1}^{N_j} \left( \bm\pi_K \left(\widehat Z_{ij}\right) - \bm{a}_K \right)\right]. \label{quadratic}
\end{align}
The following part demonstrates the asymptotic normality of \begin{equation*}
\frac{1}{\sqrt N}\sum_{j = 1}^J \sum_{i = 1}^{N_j}\left(\bm\pi_K (\widehat Z_{ij}) - \bm{a}_K \right).    
\end{equation*}
Indeed, the $k$-th component in it is
\begin{equation*}
\frac{1}{\sqrt{N}} \sum_{j = 1}^J \sum_{i = 1}^{N_j} \left( \pi_k (\widehat Z_{ij}) - a_k \right) = \frac{1}{\sqrt{N}} \sum_{j = 1}^J \sum_{i = 1}^{N_j} \left( \zeta_k (e_{ij}) - \mathbb{E} \left[\zeta_k (e_{ij})\right] \right) + o_p(1),
\end{equation*}
where
\begin{equation*}
\zeta_k(e) \equiv \pi_k(\Phi(e)) - d_{1k} e - \frac{d_{2k}}{2} (e^2 - 1).
\end{equation*}
Then, by CLT,
\begin{equation*}
    \frac{1}{\sqrt{N}} \sum_{j = 1}^J \sum_{i = 1}^{N_j} \left( \bm\pi_K (\widehat Z_{ij}) - \bm{a}_K \right)  \overset{d}{\to}  \mathcal{N}_K\left( \bm{0}, \bm{\Xi}_K\right),
\end{equation*}
with $\bm{\Xi}_K=\left(\xi_{kl}\right)_{K \times K}$ given by
\begin{align*}
    \xi_{kl} =& \mathbb{E} \left[\zeta_k(e)  \zeta_l (e) \right] \\
    =&\mathbb{E} \left[ \pi_k (Z) \pi_l (Z) \right]-a_k a_l-\left[d_{1k}d_{3l} + d_{1l}d_{3k} \right] + d_{1k} d_{1l} + \frac{1}{2} \left[ a_k d_{2l} + a_l d_{2k} \right] \\
    &-\frac{1}{2}\left[ d_{2k}d_{4l} + d_{2l}d_{4k}\right] + \frac{1}{2}\left[ d_{1k}d_{2l} + d_{1l}d_{2k}\right] \mathbb{E} \left[e^3\right] +\frac{d_{2k} d_{2l} }{4} \left[\mathbb{E} \left[e^4\right]-1 \right].
\end{align*}
Along with \eqref{quadratic}, \eqref{sandwitch} in Theorem \ref{H1_same} is validated. This completes the proof of Theorem \ref{H1_same}. $\square$

\subsection{Proof of Theorem \ref{H1L_same}}

The proof of Theorem \ref{H1L_same} is similar to that of Theorems \ref{H0_same} and \ref{H1_same}. We just follow the arguments in the previous proof and carefully specify the behavior of the estimation effects. Under $H_{1L}$ in \eqref{H1L}, the PDF of $e$ is given by
\begin{equation*}
f(x) = (1-\delta_N) \phi(x)+\delta_N q(x).
\end{equation*}

We first examine the convergence of $\widehat \mu-\mu$ and $\widehat\sigma^2-\sigma^2$ as before. For $\widehat \mu-\mu$, \eqref{mu_hat-mu_same} still holds under $H_{1L}$, and $\sqrt{N}(\widehat\mu-\mu) \overset{d}{\to} \mathcal{N}(0,\sigma^2)$ as $N\to \infty$. For $\widehat\sigma^2-\sigma^2$, under $H_{1L}$, \eqref{sigma^2_hat-sigma^2_same} still holds, and $\sqrt{N}(\widehat\sigma^2-\sigma^2) \overset{d}{\to} \mathcal{N}(0,2\sigma^4)$ \footnote{Here under $H_{1L}$, the asymptotic normality of $\sqrt{N}(\widehat\mu-\mu)$ and $\sqrt{N}(\widehat\sigma^2-\sigma^2)$ are both derived from CLT for triangular arrays since the distribution of $e_{ij}$ changes over $N$. It can be verified that the conditions of Lindeberg--Feller CLT hold for both random sequences, see, for example, Proposition 2.27 of \cite{van2000asymptotic}
.} since
\begin{align*}
\mathbb{E}\left[e^4\right]=&\int_{-\infty}^{\infty} x^4f(x)\mathrm{d}x \\
=&(1-\delta_N)\int_{-\infty}^{\infty}x^4\phi(x)\mathrm{d}x+\delta_N\int_{-\infty}^{\infty}x^4q(x)\mathrm{d}x\\
=&3+o(1).
\end{align*}

Next, we clarify the relationship of the denoted constants. Recall the definitions of the constants $d_{1k}=\mathbb{E} [\dot\pi_k(Z)\phi\left(e\right)]=\int_{- \infty}^{\infty}\dot\pi_k(\Phi\left(x\right))\phi\left(x\right)f(x)\mathrm{d}x$ and $d_{2k}=\mathbb{E} [\dot\pi_k(Z)\phi\left(e\right)e]=\int_{- \infty}^{\infty}\dot\pi_k(\Phi\left(x\right))\phi\left(x\right)xf(x)\mathrm{d}x$.
Then under $H_{1L}$, as $N\to \infty$, we have
\begin{align}
d_{1k}=&\int_{-\infty}^{\infty}\dot\pi_k(\Phi\left(x\right))\phi\left(x\right)f(x)\mathrm{d}x \nonumber \\
=&(1-\delta_N)\int_{-\infty}^{\infty}\dot\pi_k(\Phi\left(x\right))\phi^2\left(x\right)\mathrm{d}x+\delta_N\int_{-\infty}^{\infty}\dot\pi_k(\Phi\left(x\right))\phi\left(x\right)q(x)\mathrm{d}x \nonumber\\
=&c_{1k}+o(1), \label{c1_d1_H1L_same}
\end{align}
\begin{align}
d_{2k}=&\int_{- \infty}^{\infty}\dot\pi_k(\Phi\left(x\right))\phi\left(x\right)xf(x)\mathrm{d}x \nonumber\\
=&(1-\delta_N)\int_{-\infty}^{\infty}\dot\pi_k(\Phi\left(x\right))x\phi^2\left(x\right)\mathrm{d}x+\delta_N\int_{-\infty}^{\infty}\dot\pi_k(\Phi\left(x\right))x\phi\left(x\right)q(x)\mathrm{d}x \nonumber\\
=&c_{2k}+o(1), \label{c2_d2_H1L_same}
\end{align}
and
\begin{align}
\mathbb{E}[\pi_k (Z)] =& \int_{-\infty}^{\infty} \pi_k \left(\Phi(x)\right)f(x) \mathrm{d}x \nonumber\\
=&\left(1-\delta_N\right)\int_{-\infty}^{\infty} \pi_k \left(\Phi(x)\right)\phi(x) \mathrm{d}x+\delta_N\int_{-\infty}^{\infty} \pi_k \left(\Phi(x)\right)q(x) \mathrm{d}x \nonumber\\
=& \delta_N \Delta_k. \label{drift_H1L_same}
\end{align}

Now we start from \eqref{decomp} again. Under $H_{1L}$, we still have \eqref{decomp}, \eqref{main_term}, and \eqref{a1}. Also, \eqref{main_same_H0} holds as a result of \eqref{c1_d1_H1L_same} and \eqref{c2_d2_H1L_same}. The remainder terms $R_{N1}$ in \eqref{RN1} and $R_{N2}$ in \eqref{RN2} are still $O_p(N^{-1/2})$. Combining \eqref{decomp}, \eqref{main_term}, \eqref{a1}, \eqref{main_same_H0}, \eqref{RN1}, \eqref{RN2}, and \eqref{drift_H1L_same}, we obtain \eqref{decomp_H1L} in Theorem \ref{H1L_same}. 

It suffices to derive the asymptotic normality of 
$N^{-1/2}\sum_{j = 1}^J \sum_{i = 1}^{N_j}\bm\pi_K (\widehat Z_{ij})$. Under $H_{1L}$, straightforward calculation yields that
\begin{align*}
\mathbb{E}\left[\pi_k(Z)\pi_l(Z)\right]=&\int_{-\infty}^{\infty} \pi_k(\Phi(x))\pi_l(\Phi(x))f(x)\mathrm{d}x \\
=&(1-\delta_N)\int_{-\infty}^{\infty}\pi_k(\Phi(x))\pi_k(\Phi(x))\phi(x)\mathrm{d}x\\
&+\delta_N\int_{-\infty}^{\infty}\pi_k(\Phi(x))\pi_k(\Phi(x))q(x)\mathrm{d}x\\
=&\delta_{kl}+o(1),
\end{align*}
\begin{align*}
d_{3k}=&\mathbb{E}\left[\pi_k(Z)e\right]=\int_{-\infty}^{\infty} \pi_k(\Phi(x))xf(x)\mathrm{d}x \\
=&(1-\delta_N)\int_{-\infty}^{\infty}\pi_k(\Phi(x))x\phi(x)\mathrm{d}x+\delta_N\int_{-\infty}^{\infty}\pi_k(\Phi(x))xq(x)\mathrm{d}x\\
=&c_{1k}+o(1),
\end{align*}
\begin{align*}
d_{4k}=&\mathbb{E}\left[\pi_k(Z)e^2\right]=\int_{-\infty}^{\infty} \pi_k(\Phi(x))x^2f(x)\mathrm{d}x \\
=&(1-\delta_N)\int_{-\infty}^{\infty}\pi_k(\Phi(x))x^2\phi(x)\mathrm{d}x+\delta_N\int_{-\infty}^{\infty}\pi_k(\Phi(x))x^2q(x)\mathrm{d}x\\
=&c_{2k}+o(1),
\end{align*}
and
\begin{align*}
\mathbb{E}\left[e^3\right]=&\int_{-\infty}^{\infty} x^3f(x)\mathrm{d}x \\
=&(1-\delta_N)\int_{-\infty}^{\infty}x^3\phi(x)\mathrm{d}x+\delta_N\int_{-\infty}^{\infty}x^3q(x)\mathrm{d}x\\
=&o(1).
\end{align*}
Thus, by \eqref{decomp_H1L} and CLT, 
\begin{align*}
    \frac{1}{\sqrt{N}}\sum_{j=1}^{J}\sum_{i=1}^{N_j}\bm\pi_K\left(\widehat Z_{ij}\right) & =\sqrt{N}\left[\frac{1}{N}\sum_{j=1}^{J}\sum_{i=1}^{N_j}\pi_1\left(\widehat Z_{ij}\right),\ldots,\frac{1}{N}\sum_{j=1}^{J}\sum_{i=1}^{N_j}\pi_K\left(\widehat Z_{ij}\right)\right]^\top \\
    & \overset{d}{\to}  \mathcal{N}_K\left(\bm{\Delta}_K, \bm{\Sigma}_K\right),
\end{align*}
since the asymptotic covariance matrix is still $\bm{\Sigma}_K$ from the calculation above. Besides,
\begin{equation*}
        N\widehat{\Psi}_K^2   \overset{d}{\to}  \chi_K^2\left(\bm{\Delta}_K^\top\bm{\Sigma}_K^{-1}\bm{\Delta}_K\right),
\end{equation*}
by definition of the non-central $\chi^2$ distribution. This completes the proof of Theorem \ref{H1L_same}. $\square$

\subsection{Proof of Theorem \ref{H0_different_mean}, \ref{H1_different_mean} and \ref{H1L_different_mean}}

The proof is similar to that of Theorems \ref{H0_same}, \ref{H1_same}, and \ref{H1L_same}. Note that in the case of different means and same variance, we assume $\min\{N_1,\ldots,N_J\}\to\infty$, $J = o( N^{1/2})$ and $\sum_{j=1}^J N_j^{-1}=o(1)$. To simplify the procedure and notations, here we just clarify the asymptotic properties of $N^{-1}\sum_{j=1}^{J}\sum_{i=1}^{N_j}\pi_k(\widehat Z_{ij})$ without distinction of the null and alternatives. Our goal is to prove that
\begin{align}
&\frac{1}{N}\sum_{j=1}^{J}\sum_{i=1}^{N_j}\pi_k\left(\widehat Z_{ij}\right)\nonumber\\
=&\frac{1}{N}\sum_{j=1}^{J}\sum_{i=1}^{N_j}\left\{\pi_k(Z_{ij})-\mathbb{E} \left[\dot\pi_k(Z)\phi\left(e\right)\right]e_{ij}-\frac{\mathbb{E} \left[\dot\pi_k(Z)\phi\left(e\right)e\right]}{2}\left(e_{ij}^2-1\right)\right\}\nonumber\\
&+o_p\left(\frac{1}{\sqrt N}\right). \label{decomp_diff_mean}
\end{align}

We first focus on the estimation effects of $\widehat \mu_j-\mu_j$ and $\widehat\sigma^2-\sigma^2$. For $\widehat \mu_j-\mu_j$, by CLT, we have $\sqrt{N_j}(\widehat \mu_j-\mu_j)\overset{d}{\to} \mathcal{N}(0,\sigma^2)$ for each $1\le j\le J$ as $\min\{N_1,\ldots,N_J\}\to\infty$. Besides, since
\begin{equation}
\mathbb{E}\left[\frac{1}{N}\sum_{j=1}^{J}N_j\left(\widehat\mu_j-\mu_j\right)^2\right]=\frac{\sigma^2}{N}\sum_{j=1}^{J}\mathbb{E}\left[N_j\left(\frac{1}{N_j}\sum_{i=1}^{N_j}e_{ij}\right)^2\right]=\frac{J}{N}\sigma^2, \label{mu_j_hat-mu_j_sum_2}
\end{equation}
we have $N^{-1}\sum_{j=1}^{J}N_j\left(\widehat\mu_j-\mu_j\right)^2=O_p(J/N)=o_p(N^{-1/2})$ from $J = o( N^{1/2})$. Analogously, by
\begin{equation*}
\mathbb{E}\left[\left(\frac{1}{N_j}\sum_{i=1}^{N_j}e_{ij}\right)^4\right] =\frac{1}{N_j^4} \left[3N_j(N_j-1)+N_j\mathbb{E}\left[e^4\right]\right]=\frac{3}{N_j^2}+\frac{\mathbb{E}\left[e^4\right]-3}{N_j^3},      
\end{equation*} 
we obtain
\begin{align}
&\mathbb{E}\left[\frac{1}{N}\sum_{j=1}^{J}N_j\left(\widehat\mu_j-\mu_j\right)^4\right]\nonumber\\
=&\frac{\sigma^4}{N}\sum_{j=1}^{J}\mathbb{E}\left[N_j\left(\frac{1}{N_j}\sum_{i=1}^{N_j}e_{ij}\right)^4\right]=\frac{\sigma^4}{N}\sum_{j=1}^{J}\left(\frac{3}{N_j}+\frac{\mathbb{E}\left[e^4\right]-3}{N_j^2}\right), \label{mu_j_hat-mu_j_sum_4}
\end{align}
which implies $N^{-1}\sum_{j=1}^{J}N_j\left(\widehat\mu_j-\mu_j\right)^4=N^{-1}O_p(\sum_{j=1}^J N_j^{-1})=o_p(N^{-1})$ since $\sum_{j=1}^J N_j^{-1}=o_p(1)$. For $\widehat\sigma^2-\sigma^2$, we have
\begin{align*}
\widehat\sigma^2-\sigma^2=&\frac{1}{N}\sum_{j=1}^{J}\sum_{i=1}^{N_j}\left(\left(Y_{ij}-\widehat\mu_j\right)^2-\sigma^2\right)\\
=&\frac{1}{N}\sum_{j=1}^{J}\sum_{i=1}^{N_j}\left(\varepsilon_{ij}^2-\sigma^2\right)-\frac{1}{N}\sum_{j=1}^{J}N_j\left(\widehat\mu_j-\mu_j\right)^2\\
=&\frac{\sigma^2}{N}\sum_{j=1}^{J}\sum_{i=1}^{N_j}\left(e_{ij}^2-1\right)+o_p\left(\frac{1}{\sqrt{N}}\right),
\end{align*}
with the last step following \eqref{mu_j_hat-mu_j_sum_2}. Note that the main term in the last line is the same as \eqref{sigma^2_hat-sigma^2_same}.

Next, we will show that 
\begin{align}
&\frac{1}{N\sigma}\sum_{j=1}^{J}N_j\left(\widehat\mu_j-\mu_j\right)\left\{\frac{1}{N_j}\sum_{i=1}^{N_j}\dot\pi_k(Z_{ij})\phi\left(e_{ij}\right)\right\}\nonumber\\
=&\mathbb{E} \left[\dot\pi_k(Z)\phi\left(e\right)\right]\frac{1}{N}\sum_{j=1}^{J}\sum_{i=1}^{N_j}e_{ij}+o_p\left(\frac{1}{\sqrt{N}}\right).\label{c1k_different_mean}
\end{align} 
Indeed, since
\begin{align*}
&\frac{1}{N\sigma}\sum_{j=1}^{J}N_j\left(\widehat\mu_j-\mu_j\right)\left\{\frac{1}{N_j}\sum_{i=1}^{N_j}\dot\pi_k(Z_{ij})\phi\left(e_{ij}\right)\right\} \\
= & \frac{1}{N}\sum_{j=1}^{J}\frac{1}{N_j}\sum_{i_1=1}^{N_j}\sum_{i_2=1}^{N_j}\dot\pi_k(Z_{i_1j})\phi\left(e_{i_1j}\right)e_{i_2j},
\end{align*} 
and
\begin{equation*}
\frac{1}{N}\sum_{j=1}^{J}\sum_{i=1}^{N_j}e_{ij} =  \frac{1}{N}\sum_{j=1}^{J}\frac{1}{N_j}\sum_{i_1=1}^{N_j}\sum_{i_2=1}^{N_j}e_{i_2j},
\end{equation*}
then we have
\begin{align*}
&\frac{1}{N\sigma}\sum_{j=1}^{J}N_j\left(\widehat\mu_j-\mu_j\right)\left\{\frac{1}{N_j}\sum_{i=1}^{N_j}\dot\pi_k(Z_{ij})\phi\left(e_{ij}\right)\right\}- \mathbb{E} \left[\dot\pi_k(Z)\phi\left(e\right)\right]\frac{1}{N}\sum_{j=1}^{J}\sum_{i=1}^{N_j}e_{ij}\\
=&\frac{1}{N}\sum_{j=1}^{J}\frac{1}{N_j}\sum_{i_1=1}^{N_j}\sum_{i_2=1}^{N_j}\left(\dot\pi_k(Z_{i_1j})\phi\left(e_{i_1j}\right)-\mathbb{E} \left[\dot\pi_k(Z)\phi\left(e\right)\right]\right)e_{i_2j}\\
=&\frac{1}{N}\sum_{j=1}^{J}\frac{1}{N_j}\sum_{i=1}^{N_j}\left(\dot\pi_k(Z_{ij})\phi\left(e_{ij}\right)-\mathbb{E} \left[\dot\pi_k(Z)\phi\left(e\right)\right]\right)e_{ij}\\
&+\frac{1}{N}\sum_{j=1}^{J}\frac{1}{N_j}\sum_{i_1=1}^{N_j}\sum_{i_2\ne i_1}^{N_j}\left(\dot\pi_k(Z_{i_1j})\phi\left(e_{i_1j}\right)-\mathbb{E} \left[\dot\pi_k(Z)\phi\left(e\right)\right]\right)e_{i_2j}\\
\equiv& A_{N1}+A_{N2}.
\end{align*} 
For $A_{N1}$, we have 
\begin{equation*}
\mathbb{E}[A_{N1}] = \frac{J}{N}    \mathbb{E} \left[\dot\pi_k(Z)\phi\left(e\right)e\right],
\end{equation*}
and
\begin{equation*}
\var[A_{N1}] = \var \left[\left(\dot\pi_k(Z)\phi\left(e\right)-\mathbb{E} \left[\dot\pi_k(Z)\phi\left(e\right)\right]\right)e\right]\frac{1}{N^2}\sum_{j=1}^{J}\frac{1}{N_j},
\end{equation*}
then $A_{N1} = o_p(N^{-1/2})$ since $J=o(N^{1/2})$ as $N\to\infty$. For $A_{N2}$, we have $\mathbb{E}[A_{N2}]=0$, and
\begin{align*}
&\var[A_{N2}] \\
=& \frac{1}{N^2} \sum_{j=1}^{J} \frac{1}{N_j^2} \var\left[\sum_{i_1=1}^{N_j}\sum_{i_2\ne i_1}^{N_j}\left(\dot\pi_k(Z_{i_1j})\phi\left(e_{i_1j}\right)-\mathbb{E} \left[\dot\pi_k(Z)\phi\left(e\right)\right]\right)e_{i_2j}\right] \\
=&\frac{1}{N^2} \sum_{j=1}^{J} \frac{1}{N_j^2} \mathbb{E}\left[\left(\sum_{i_1=1}^{N_j}\sum_{i_2\ne i_1}^{N_j}\left(\dot\pi_k(Z_{i_1j})\phi\left(e_{i_1j}\right)-\mathbb{E} \left[\dot\pi_k(Z)\phi\left(e\right)\right]\right)e_{i_2j}\right)^2\right]\\
=&\frac{1}{N^2} \sum_{j=1}^{J} \frac{N_j (N_j-1)}{N_j^2}\left(\var \left[\dot\pi_k(Z)\phi\left(e\right)\right]+\mathbb{E}^2 \left[\dot\pi_k(Z)\phi\left(e\right)e\right]\right)\\
=&\left(\var \left[\dot\pi_k(Z)\phi\left(e\right)\right]+\mathbb{E}^2 \left[\dot\pi_k(Z)\phi\left(e\right)e\right]\right)\frac{1}{N^2}\left(J-\sum_{j=1}^J\frac{1}{N_j}\right),
\end{align*}
then $A_{N2} = o_p(N^{-1/2})$. Hence, \eqref{c1k_different_mean} is verified. 

Now we start from \eqref{decomp}. For model \eqref{model_diff_mean}, \eqref{decomp} and \eqref{main_term} still hold. The difference lies in that the first term of \eqref{main_term} should be 
further expressed as
\begin{align}
&\frac{1}{N}\sum_{j=1}^{J}\sum_{i=1}^{N_j}\dot\pi_k(Z_{ij})\phi\left(\frac{{\varepsilon}_{ij}}{\sigma}\right)\left(\frac{\widehat\varepsilon_{ij}}{\widehat\sigma}-e_{ij}\right)\nonumber\\
=&\frac{1}{N}\sum_{j=1}^{J}\sum_{i=1}^{N_j}\dot\pi_k(Z_{ij})\phi\left(e_{ij}\right)\frac{\widehat\varepsilon_{ij}-\varepsilon_{ij}}{\widehat\sigma}-\frac{1}{N}\sum_{j=1}^{J}\sum_{i=1}^{N_j}\dot\pi_k(Z_{ij})\phi\left(e_{ij}\right)\frac{\varepsilon_{ij}(\widehat\sigma^2-\sigma^2)}{\widehat\sigma\sigma(\widehat\sigma+\sigma)}\nonumber\\
=&-\frac{1}{\widehat\sigma}\frac{1}{N}\sum_{j=1}^{J}N_j\left(\widehat\mu_j-\mu_j\right)\left\{\frac{1}{N_j}\sum_{i=1}^{N_j}\dot\pi_k(Z_{ij})\phi\left(e_{ij}\right)\right\}\nonumber\\
&-\frac{\widehat\sigma^2-\sigma^2}{\widehat\sigma(\widehat\sigma+\sigma)}\frac{1}{N}\sum_{j=1}^{J}\sum_{i=1}^{N_j}\dot\pi_k(Z_{ij})\phi\left(e_{ij}\right)e_{ij}\nonumber\\
=&-\mathbb{E} \left[\dot\pi_k(Z)\phi\left(e\right)\right]\frac{1}{N}\sum_{j=1}^{J}\sum_{i=1}^{N_j}e_{ij}-\frac{1}{2\sigma^2}\mathbb{E} \left[\dot\pi_k(Z)\phi\left(e\right)e\right]\left(\widehat\sigma^2-\sigma^2\right)\nonumber\\
&+o_p\left(\frac{1}{N}\sum_{j=1}^{J}\sum_{i=1}^{N_j}e_{ij}\right)+o_p\left(\widehat\sigma^2-\sigma^2\right) \nonumber\\
=&-\mathbb{E} \left[\dot\pi_k(Z)\phi\left(e\right)\right]\frac{1}{N}\sum_{j=1}^{J}\sum_{i=1}^{N_j}e_{ij}\nonumber\\
&-\frac{1}{2N}\mathbb{E} \left[\dot\pi_k(Z)\phi\left(e\right)e\right]\sum_{j=1}^{J}\sum_{i=1}^{N_j}\left(e_{ij}^2-1\right)+o_p\left(\frac{1}{\sqrt{N}}\right), \label{a2}
\end{align}
where we apply \eqref{sigma^2_hat-sigma^2_same}, \eqref{c1k_different_mean}, LLN for $N^{-1}\sum_{j=1}^{J}\sum_{i=1}^{N_j}\dot\pi_k(Z_{ij})\phi\left(e_{ij}\right)e_{ij}$, and CLT for $N^{-1}\sum_{j=1}^{J}\sum_{i=1}^{N_j}e_{ij}$ as $N \to \infty$. 

It suffices to show $R_{N1}$ (in \eqref{RN1}) and $R_{N2}$ (in \eqref{RN2}) are $o_p( N^{-1/2})$ in order to obtain \eqref{decomp_diff_mean}. The technique is similar to that in the proof of Theorem \ref{H0_same}. For $R_{N1}$, we have 
\begin{align*}
\vert R_{N1}\vert & \lesssim \frac{1}{N} \sum_{j = 1}^J \sum_{i = 1}^{N_j} \left( \frac{\widehat{\varepsilon}_{ij} - \varepsilon_{ij}}{\widehat{\sigma}}\right)^2 + \frac{1}{N} \sum_{j=1}^{J} \sum_{i=1}^{N_{j}}\left(\frac{\varepsilon_{i j}(\widehat{\sigma}^2-\sigma^2)}{\widehat{\sigma} \sigma(\widehat{\sigma}+\sigma)}\right)^2 \\
& \lesssim \frac{1}{N \sigma^2} \sum_{j = 1}^J N_j \left( \widehat{\mu}_j - \mu_j\right)^2 + \left( \widehat{\sigma}^2  - \sigma^2 \right)^2 \frac{1}{N} \sum_{j = 1}^J \sum_{i = 1}^{N_j} \varepsilon_{ij}^2 \\
& \asymp O_{p}\left(\frac{J}{N}\right)  + O_p \left( \frac{1}{N}\right) \left( \sigma^2 + o_p (1)\right)=O_p \left( \frac{J}{N}\right) = o_p \left( \frac{1}{\sqrt{N}}\right),
\end{align*}
by \eqref{mu_j_hat-mu_j_sum_2} and the assumption $J = o(N^{1/2})$. For $R_{N2}$, we first have
\begin{align*}
&\frac{1}{N} \sum_{j=1}^{J} \sum_{i=1}^{N_{j}}\left(\frac{\widehat{\varepsilon}_{i j}}{\widehat{\sigma}}-\frac{\varepsilon_{i j}}{\sigma}\right)^{4}\\
\lesssim &\left[\frac{1}{N} \sum_{j=1}^{J} \sum_{i=1}^{N_{j}}\left(\frac{\widehat{\varepsilon}_{i j}-\varepsilon_{ij}}{\widehat{\sigma}}\right)^{4}+\frac{1}{N} \sum_{j=1}^{J} \sum_{i=1}^{N_{j}}\left(\frac{\varepsilon_{i j}(\widehat{\sigma}^2-\sigma^2)}{\widehat{\sigma} \sigma(\widehat{\sigma}+\sigma)}\right)^{4}\right] \\
\lesssim &\frac{1}{N} \sum_{j = 1}^J N_j \left( \widehat{\mu}_j - \mu_j \right)^4 + (\widehat{\sigma}^2-\sigma^2)^{4}\left(\mathbb{E} \left[e^{4}\right]+o_{p}(1)\right) \\
\asymp & o_p \left( \frac{1}{N}\right) + O_p \left( \frac{1}{N^2}\right) = o_p \left( \frac{1}{N}\right),
\end{align*}
by \eqref{mu_j_hat-mu_j_sum_4}. Along with $N^{-1} \sum_{j=1}^{J} \sum_{i=1}^{N_{j}} \widetilde{\varepsilon}_{i j}^2/\widetilde{\sigma}^2 = O_p(1)$ from \eqref{RN2_Op1}, we obtain
\begin{equation*}
\left\vert R_{N 2}\right\vert \lesssim\left[\frac{1}{N} \sum_{j=1}^{J} \sum_{i=1}^{N_{j}} \frac{\widetilde{\varepsilon}_{i j}^2}{\widetilde{\sigma}^2}\right]^{1 / 2}\left[\frac{1}{N} \sum_{j=1}^{J} \sum_{i=1}^{N_{j}}\left(\frac{\widehat{\varepsilon}_{i j}}{\widehat{\sigma}}-\frac{\varepsilon_{i j}}{\sigma}\right)^{4}\right]^{1 / 2}  = o_p \left( \frac{1}{\sqrt{N}}\right).
\end{equation*}
Thus, \eqref{decomp_diff_mean} is verified. The form of \eqref{decomp_diff_mean} can be further refined under $H_0$, $H_1$ and $H_{1L}$ respectively. For example, under $H_0$, we can write  $\mathbb{E}\left[\dot\pi_k(Z)\phi\left(e\right)\right]=c_{1k}$ and
$\mathbb{E}\left[\dot\pi_k(Z)\phi\left(e\right)e\right]=c_{2k}$ in \eqref{decomp_diff_mean}, and then obtain the results of Theorem \ref{H0_different_mean}. $\square$

\subsection{Proof of Theorem \ref{H0_different_var}}

The technical arguments remain the same as in the previous cases. Recall that $p_j=N_j/N$ and $q_j=JN_j/N$ for $1\le j \le J$. And for model \eqref{model_diff_var}, we further assume that as $\min\{N_1,\ldots,N_J\}\to\infty$ and $J = o( N^{1/2})$ (\romannumeral1) there exist $0<\underline{\sigma}\le \overline{\sigma}<\infty$ such that $\underline{\sigma}< \inf_{1\le j\le J} \sigma_j\le \sup_{1\le j\le J} \sigma_j<\overline{\sigma}$; (\romannumeral2) there exist $0<\underline{q}\le \overline{q}<\infty$ such that $\underline{q}< \inf_{1\le j\le J} q_j\le \sup_{1\le j\le J} q_j<\overline{q}$. 

We first tackle the estimation effects of $\widehat\mu-\mu$ and $\widehat\sigma^2_j-\sigma^2_j$. For $\widehat\mu-\mu$, note that $\widehat{\mu} = J^{-1}\sum_{j = 1}^J N_j^{-1}\sum_{i = 1}^{N_j} Y_{ij}$ here. Then, 
\begin{equation}
\widehat{\mu} - \mu  =  \frac{1}{J} \sum_{j = 1}^{J} \frac{1}{N_j} \sum_{i = 1}^{N_j} \varepsilon_{ij}=\frac{1}{J} \sum_{j = 1}^{J}  \frac{\sigma_j}{N_j} \sum_{i = 1}^{N_j} e_{ij}. \label{mu_hat-mu_diff_var}
\end{equation}
From \eqref{mu_hat-mu_diff_var} we obtain $\mathbb{E}[\widehat{\mu} - \mu]=0$, and $\var[\widehat{\mu} - \mu]  = J^{-2} \sum_{j = 1}^{J}  N_j^{-1}\sigma_j^2\asymp J^{-2} \sum_{j = 1}^{J}  N_j^{-1}=N^{-1}J^{-1} \sum_{j = 1}^{J}  q_j^{-1}=O_p(N^{-1})$. Then $\widehat \mu-\mu=O_p(N^{-1/2})$. For $\widehat\sigma^2_j-\sigma^2_j$, we have
\begin{equation}\label{sigma_j^2_hat-sigma_j^2}
\begin{aligned}
\widehat{\sigma}_j^2 - \sigma_j^2 & = \frac{1}{N_j} \sum_{i = 1}^{N_j} \left(\left(Y_{i j}-\widehat{\mu}\right)^2-\sigma_j^2\right) = \frac{1}{N_j} \sum_{i = 1}^{N_j} \left( \varepsilon_{ij}^2 - \sigma_j^2 \right) + \left( \widehat{\mu} - \mu \right)^2- 2\left( \widehat{\mu} - \mu \right) \frac{1}{N_j} \sum_{i = 1}^{N_j} \varepsilon_{ij}\\
& = \frac{\sigma_j^2}{N_j} \sum_{i = 1}^{N_j} \left( e_{ij}^2 - 1\right) + \left( \widehat{\mu} - \mu \right)^2- 2\left( \widehat{\mu} - \mu \right) \frac{\sigma_j}{N_j} \sum_{i = 1}^{N_j} e_{ij}\\
& = O_p\left(\frac{1}{\sqrt{N_j}}\right)+O_p\left(\frac{1}{N}\right)+O_p\left(\frac{1}{\sqrt{N_j N}}\right)=O_p\left(\frac{1}{\sqrt{N_j}}\right),
\end{aligned}
\end{equation}
by CLT for $N_j^{-1/2} \sum_{i = 1}^{N_j} e_{ij} \overset{d}{\to} \mathcal{N}(0,1)$ and $N_j^{-1/2} \sum_{i = 1}^{N_j} ( e_{ij}^2 - 1) \overset{d}{\to} \mathcal{N}(0,\mathbb{E}[e^4]-1)$ as $N_j \to \infty$ and \eqref{mu_hat-mu_diff_var}.

We next prove some preliminary results such that
\begin{equation}
\sum_{j = 1}^{J} \frac{N_j}{\sigma_jN} \left\{ \frac{1}{N_j} \sum_{i = 1}^{N_j} \dot\pi_k (Z_{ij}) \phi \left( e_{ij}\right)\right\} = \mathbb{E} \left[\dot\pi_k(Z)\phi\left(e\right)\right] \sum_{j = 1}^{J} \frac{p_j}{\sigma_j}+O_p\left(\frac{1}{\sqrt{N}}\right),\label{c1k_different_var}
\end{equation}
\begin{align}
&\frac{1}{N} \sum_{j = 1}^J \frac{1}{N_j} \left\{\sum_{i_1 = 1}^{N_j}\sum_{i_2=1}^{N_j} \dot\pi_k\left(Z_{i_1 j}\right) \phi\left(e_{i_1 j}\right) e_{i_1 j}\left(e_{i_2 j}^2-1\right)\right\}\nonumber\\
=& \mathbb{E} \left[\dot\pi_k(Z)\phi\left(e\right)e\right]\frac{1}{N}\sum_{j=1}^{J}\sum_{i=1}^{N_j}(e_{ij}^2-1)+o_p\left(\frac{1}{\sqrt{N}}\right), \label{c2k_different_var}
\end{align} 
\begin{equation}
\frac{1}{N} \sum_{j = 1}^J \frac{1}{N_j}    \sum_{i_1 = 1}^{N_j}\sum_{i_2=1}^{N_j} \dot\pi_k\left(Z_{i_1 j}\right) \phi\left(e_{i_1 j}\right) e_{i_1 j} e_{i_2 j} = \mathbb{E} \left[\dot\pi_k(Z)\phi\left(e\right)e\right]\frac{1}{N}\sum_{j=1}^{J}\sum_{i=1}^{N_j}e_{ij}+o_p\left(\frac{1}{\sqrt{N}}\right), \label{c2k_different_var_minus}
\end{equation}
\begin{equation}
\frac{1}{N} \sum_{j = 1}^J \frac{1}{\sigma_j^2} \left\{\sum_{i = 1}^{N_j} \dot\pi_k\left(Z_{i j}\right) \phi\left(e_{ij}\right) e_{ij}\right\} = O_p(1),\label{DN2_Op1}
\end{equation}
as $\min\{N_1,\ldots,N_J\}\to\infty$, $J = o( N^{1/2})$. For \eqref{c1k_different_var}, straightforward calculation leads to
\begin{equation*}
\mathbb{E}\left[\sum_{j = 1}^{J} \frac{N_j}{\sigma_jN} \left\{ \frac{1}{N_j} \sum_{i = 1}^{N_j} \dot\pi_k (Z_{ij}) \phi \left( e_{ij}\right)\right\}\right] =\mathbb{E} \left[\dot\pi_k(Z)\phi\left(e\right)\right] \sum_{j = 1}^{J} \frac{p_j}{\sigma_j},
\end{equation*}
and
\begin{align*}
&\var\left[\sum_{j = 1}^{J} \frac{N_j}{\sigma_jN} \left\{ \frac{1}{N_j} \sum_{i = 1}^{N_j} \dot\pi_k (Z_{ij}) \phi \left( e_{ij}\right)\right\}\right] =\var\left[\dot\pi_k (Z) \phi \left( e\right)\right] \sum_{j = 1}^{J} \frac{p^2_j}{N_j\sigma^2_j}\\
=&\var\left[\dot\pi_k (Z) \phi \left( e\right)\right]\frac{1}{N^2} \sum_{j = 1}^{J} \frac{N_j}{\sigma^2_j}=O\left(\frac{1}{N}\right).
\end{align*}
Thus, \eqref{c1k_different_var} is verified. The derivation of \eqref{c2k_different_var} is analogous to that of \eqref{c1k_different_mean} if we first rewrite 
$N^{-1}\sum_{j=1}^{J}\sum_{i=1}^{N_j}(e_{ij}^2-1)$ on the right side as 
\begin{equation*}
\frac{1}{N}\sum_{j=1}^{J}\frac{1}{N_j}\sum_{i_1=1}^{N_j}\sum_{i_2=1}^{N_j}(e_{i_2j}^2-1),
\end{equation*}
and so is \eqref{c2k_different_var_minus}. \eqref{DN2_Op1} is obtained by 
\begin{equation*}
\mathbb{E}\left[\frac{1}{N}\sum_{j = 1}^J \frac{1}{\sigma_j^2} \{\sum_{i = 1}^{N_j} \dot\pi_k(Z_{i j}) \phi(e_{ij}) e_{ij}\}\right] =\mathbb{E}[\dot\pi_k(Z) \phi(e) e]\frac{1}{N}\sum_{j=1}^J\frac{N_j}{\sigma_j^2}=O(1),
\end{equation*}
and 
\begin{equation*}
\var\left[\frac{1}{N}\sum_{j = 1}^J \frac{1}{\sigma_j^2}\sum_{i = 1}^{N_j} \dot\pi_k(Z_{i j}) \phi(e_{ij}) e_{ij}\right]=O\left(\frac{1}{N}\right).
\end{equation*}

Now we start from \eqref{decomp}. For model \eqref{model_diff_var}, \eqref{decomp} and \eqref{main_term} still hold. The difference lies in the fact that the first term of \eqref{main_term} should be further expressed as
\begin{align*}
&\frac{1}{N} \sum_{j=1}^{J} \sum_{i=1}^{N_{j}} \dot\pi_k\left(Z_{i j}\right) \phi\left(\frac{\varepsilon_{i j}}{\sigma_j}\right)\left(\frac{\widehat{\varepsilon}_{i j}}{\widehat{\sigma}_j}-\frac{\varepsilon_{i j}}{\sigma_j}\right) \\
=&\frac{1}{N} \sum_{j=1}^{J} \sum_{i=1}^{N_{j}} \dot\pi_k\left(Z_{i j}\right) \phi\left(e_{ij}\right) \frac{\widehat{\varepsilon}_{i j}-\varepsilon_{i j}}{\widehat{\sigma}_j}\\
&-\frac{1}{N} \sum_{j=1}^{J} \sum_{i=1}^{N_{j}} \dot\pi_k\left(Z_{i j}\right) \phi\left(e_{ij}\right) \frac{\varepsilon_{i j} (\widehat{\sigma}_j^2-\sigma_j^2 )}{\widehat{\sigma}_j \sigma_j(\widehat{\sigma}_j+\sigma_j)}.\\
\equiv& D_{N1} - D_{N2}.
\end{align*}
For $D_{N1}$, under $H_0$, we have
\begin{align}
D_{N1} &= - \frac{\widehat{\mu} - \mu}{N} \sum_{j = 1}^{J} \frac{N_j}{\sigma_j} \left\{ \frac{1}{N_j} \sum_{i = 1}^{N_j} \dot\pi_k (Z_{ij}) \phi \left( e_{ij}\right)\right\} \left(1+o_p(1)\right) \nonumber\\
& = - \frac{1}{N} \left( \sum_{j = 1}^J \frac{N}{J N_j} \sum_{i = 1}^{N_j} \varepsilon_{ij}\right) \left( \sum_{j = 1}^{J} \frac{N_j}{\sigma_jN} \left\{ \frac{1}{N_j} \sum_{i = 1}^{N_j} \dot\pi_k (Z_{ij}) \phi \left( e_{ij}\right)\right\} \right) \left(1+o_p(1)\right) \nonumber\\
& = - \frac{1}{N} \left( \sum_{j = 1}^J \sum_{i = 1}^{N_j} \frac{\sigma_j e_{ij}}{q_j}\right) \left(c_{1k}\sum_{j = 1}^J \frac{p_j}{\sigma_j}  + O_p \left( \frac{1}{\sqrt{N}}\right)\right) \left(1+o_p(1)\right)\nonumber\\
& = - \frac{c_{1k}}{N}  \left[ \sum_{j = 1}^J \frac{ p_j}{\sigma_j} \right] \sum_{j = 1}^J \sum_{i = 1}^{N_j} \frac{\sigma_j e_{ij}}{q_j} + o_p \left( \frac{1}{\sqrt{N}}\right), \label{DN1}
\end{align}
where the second last line is followed by \eqref{c1k_different_var}, and $\mathbb{E} \left[\dot\pi_k(Z)\phi\left(e\right)\right]=c_{1k}$ under $H_0$. The last line is followed by $N^{-1} \sum_{j = 1}^J q_j^{-1}\sum_{i = 1}^{N_j} \varepsilon_{ij}=o_p(1)$ using LLN. For $D_{N2}$, under $H_0$, we have
\begin{align}
D_{N2}
=& \frac{1}{N} \sum_{j = 1}^J \frac{ (\widehat{\sigma}_j^2-\sigma_j^2)}{\widehat{\sigma}_j (\widehat{\sigma}_j+\sigma_j)} \left\{\sum_{i = 1}^{N_j} \dot\pi_k\left(Z_{i j}\right) \phi\left(e_{ij}\right) e_{ij}\right\} \nonumber\\
=& \frac{1}{N} \sum_{j = 1}^J \frac{1}{\widehat{\sigma}_j (\widehat{\sigma}_j+\sigma_j)} \left\{\sum_{i = 1}^{N_j} \dot\pi_k\left(Z_{i j}\right) \phi\left(e_{ij}\right) e_{ij}\right\}\left\{\frac{\sigma_j^2}{N_j} \sum_{i = 1}^{N_j} \left( e_{ij}^2 - 1\right) + \left( \widehat{\mu} - \mu \right)^2 \right. \nonumber \\
&\left.- 2\left( \widehat{\mu} - \mu \right) \frac{\sigma_j}{N_j} \sum_{i = 1}^{N_j} e_{ij}\right\}\nonumber\\
=& \frac{1}{N} \sum_{j = 1}^J \frac{\sigma_j^2}{N_j\widehat{\sigma}_j (\widehat{\sigma}_j+\sigma_j)} \left\{\sum_{i_1 = 1}^{N_j}\sum_{i_2=1}^{N_j} \dot\pi_k\left(Z_{i_1 j}\right) \phi\left(e_{i_1 j}\right) e_{i_1 j}\left(e_{i_2 j}^2-1\right)\right\}\nonumber\\
&+\left(\widehat{\mu} - \mu\right)^2\frac{1}{N} \sum_{j = 1}^J \frac{1}{\widehat{\sigma}_j (\widehat{\sigma}_j+\sigma_j)} \left\{\sum_{i = 1}^{N_j} \dot\pi_k\left(Z_{i j}\right) \phi\left(e_{ij}\right) e_{ij}\right\}\nonumber\\
&- 2 \left( \widehat{\mu} - \mu \right) \frac{1}{N} \sum_{j = 1}^J  \frac{\sigma_j}{N_j\widehat{\sigma}_j (\widehat{\sigma}_j+\sigma_j)}   \left\{\sum_{i_1 = 1}^{N_j}\sum_{i_2=1}^{N_j} \dot\pi_k\left(Z_{i_1 j}\right) \phi\left(e_{i_1 j}\right) e_{i_1 j} e_{i_2 j}\right\}\nonumber\\
=& \frac{1}{2N} \sum_{j = 1}^J \frac{1}{N_j} \left\{\sum_{i_1 = 1}^{N_j}\sum_{i_2=1}^{N_j} \dot\pi_k\left(Z_{i_1 j}\right) \phi\left(e_{i_1 j}\right) e_{i_1 j}\left(e_{i_2 j}^2-1\right)\right\}\left(1+o_p(1)\right)\nonumber\\
&+\left(\widehat{\mu} - \mu\right)^2\frac{1}{2N} \sum_{j = 1}^J \frac{1}{\sigma_j^2} \left\{\sum_{i = 1}^{N_j} \dot\pi_k\left(Z_{i j}\right) \phi\left(e_{ij}\right) e_{ij}\right\}\left(1+o_p(1)\right)\nonumber\\
&-\left( \widehat{\mu} - \mu \right) \frac{1}{N} \sum_{j = 1}^J \frac{1}{N_j \sigma_j} \left\{\sum_{i_1 = 1}^{N_j}\sum_{i_2=1}^{N_j} \dot\pi_k\left(Z_{i_1 j}\right) \phi\left(e_{i_1 j}\right) e_{i_1 j}e_{i_2 j}\right\} \nonumber \\
=&\frac{c_{2k}}{2}\frac{1}{N} \sum_{j = 1}^J \sum_{i = 1}^{N_j}  \left(e_{ij}^2 - 1\right) + o_p \left( \frac{1}{\sqrt{N}}\right), \label{DN2}
\end{align}
following \eqref{mu_hat-mu_diff_var}, \eqref{sigma_j^2_hat-sigma_j^2}, \eqref{c2k_different_var}, \eqref{c2k_different_var_minus} and \eqref{DN2_Op1}, as well as $\mathbb{E} \left[\dot\pi_k(Z)\phi\left(e\right)e\right]=c_{2k}$ under $H_0$.

It suffices to show $R_{N1}$ (in \eqref{RN1}) and $R_{N2}$ (in \eqref{RN2}) are $o_p(N^{-1/2})$ in this case. For $R_{N1}$, referring to the arguments in the proof of Theorem \ref{H0_different_mean}, here we have
\begin{equation*}
\vert R_{N 1}\vert \lesssim \frac{1}{N} \sum_{j=1}^{J} \sum_{i=1}^{N_{j}}\left(\frac{\widehat{\varepsilon}_{i j}-\varepsilon_{i j}}{\widehat{\sigma}_j}\right)^2+\frac{1}{N} \sum_{j=1}^{J} \sum_{i=1}^{N_{j}}\left(\frac{\varepsilon_{i j}(\widehat{\sigma}_j^2-\sigma_j^2)}{\widehat{\sigma}_j \sigma_j(\widehat{\sigma}_j+\sigma_j)}\right)^2 \equiv R_{N1}^A + R_{N1}^B.
\end{equation*}
For $R_{N1}^A$, we obtain
\begin{equation*}
R_{N1}^A = \frac{(\widehat{\mu} - \mu)^2}{N} \sum_{j = 1}^J \frac{N_j}{\widehat{\sigma}_j^2}  \asymp  \frac{1}{N} O_p \left( \frac{1}{N}\right) O_p (N)=O_p \left( \frac{1}{N}\right) = o_p \left( \frac{1}{\sqrt{N}}\right),
\end{equation*}
by \eqref{mu_hat-mu_diff_var} and
\begin{equation*}
\sum_{j = 1}^J \frac{N_j}{\widehat{\sigma}_j^2} = \sum_{j = 1}^J \frac{N_j}{\sigma_j^2}(1+o_p(1))=O_p(N).
\end{equation*}
For $R_{N1}^B$, we have
\begin{align*}
R_{N1}^B  \lesssim& \frac{1}{N} \sum_{j = 1}^{J} N_j \left( \widehat{\sigma}_j^2 - \sigma_j^2\right)^2 \left\{ \frac{1}{N_j} \sum_{i = 1}^{N_j} \varepsilon_{ij}^2 \right\}\\
=&\frac{1}{N} \sum_{j = 1}^{J}\sigma_j^2 \left( \frac{\sigma_j^2}{N_j} \sum_{i = 1}^{N_j} \left( e_{ij}^2 - 1\right) + \left( \widehat{\mu} - \mu \right)^2- 2\left( \widehat{\mu} - \mu \right) \frac{\sigma_j}{N_j} \sum_{i = 1}^{N_j} e_{ij}\right)^2 \left(\sum_{i = 1}^{N_j} e_{ij}^2\right) \\
\lesssim &\frac{1}{N} \sum_{j = 1}^{J}\frac{\sigma_j^6}{N_j^2}\left(  \sum_{i = 1}^{N_j} \left( e_{ij}^2 - 1\right) \right)^2 \left(\sum_{i = 1}^{N_j} e_{ij}^2\right)+\left( \widehat{\mu} - \mu \right)^4\left(\frac{1}{N} \sum_{j = 1}^{J}\sigma_j^2\sum_{i = 1}^{N_j} e_{ij}^2\right) \\
&+4\left( \widehat{\mu} - \mu \right)^2 \frac{1}{N} \sum_{j = 1}^{J}\frac{\sigma_j^4}{N_j^2} \left(  \sum_{i = 1}^{N_j} e_{ij} \right)^2 \left(\sum_{i = 1}^{N_j} e_{ij}^2\right)\\
\equiv& R_{N1}^{B,1}+\left( \widehat{\mu} - \mu \right)^4 R_{N1}^{B,2}+4 \left( \widehat{\mu} - \mu \right)^2 R_{N1}^{B,3}.
\end{align*}
For $R_{N1}^{B,1}$, we rewrite it as
\begin{equation*}
R_{N1}^{B,1}=\frac{1}{N}\sum_{j=1}^J\frac{\sigma_j^6}{N_j^2}\sum_{i_1 = 1}^{N_j} \sum_{i_2 = 1}^{N_j}\sum_{i_3 = 1}^{N_j}\left( e_{i_1 j}^2 - 1\right)\left( e_{i_2 j}^2 - 1\right) e_{i_3 j}^2.
\end{equation*}
Note that $\mathbb{E}[( e_{i_1 j}^2 - 1)( e_{i_2 j}^2 - 1) e_{i_3 j}^2]=\mathbb{E}[(e^2-1)^2e^2]\mathbbm{1}(i_1= i_2=i_3)+\mathbb{E}[(e^2-1)^2]\mathbbm{1}(i_1= i_2\ne i_3)$. Then,
\begin{align*}
\mathbb{E}\left[R_{N1}^{B,1}\right] &= \frac{1}{N}\sum_{j=1}^J\frac{\sigma_j^6}{N_j^2} \left(N_j\mathbb{E}\left[(e^2-1)^2e^2\right]+N_j(N_j-1)\mathbb{E}\left[(e^2-1)^2\right] \right)\\
&= O\left(\frac{J}{N}\right),
\end{align*}
which implies $R_{N1}^{B,1}=o_p(N^{-1/2})$. Similarly, for $R_{N1}^{B,3}$, by $\mathbb{E}[ e_{i_1 j} e_{i_2 j} e_{i_3 j}^2]=\mathbb{E}[e^4]\mathbbm{1}(i_1= i_2)+\mathbbm{1}(i_1= i_2\ne i_3)$, we have 
\begin{equation*}
\mathbb{E}\left[R_{N1}^{B,3}\right]=\frac{1}{N}\sum_{j=1}^J\frac{\sigma_j^4}{N_j^2} \left(N_j\mathbb{E}\left[e^4\right]+N_j(N_j-1)\right)=O\left(\frac{J}{N}\right),
\end{equation*}
which implies $R_{N1}^{B,3} = o_p(N^{-1/2})$. For $R_{N1}^{B,2}$, straightforward calculation leads to 
\begin{equation*}
\mathbb{E}\left[R_{N1}^{B,2}\right] = \mathbb{E}\left[\frac{1}{N} \sum_{j = 1}^{J}\sigma_j^2\sum_{i = 1}^{N_j} e_{ij}^2\right]=\frac{1}{N} \sum_{j = 1}^{J}N_j\sigma_j^2=O(1),
\end{equation*}
then $R_{N1}^{B,2}=O_p(1)$. Along with \eqref{mu_hat-mu_diff_var}, we obtain $R_{N1}^B = o_p ( N^{-1/2})$. Hence $R_{N1} = o_p ( N^{-1/2})$. On the other hand, for $R_{N2}$, we first note that 
\begin{align*}
\frac{1}{N} \sum_{j=1}^{J} \sum_{i=1}^{N_{j}}\left(\frac{\widehat{\varepsilon}_{i j}}{\widehat{\sigma}_j}-\frac{\varepsilon_{i j}}{\sigma_j}\right)^{4} & \lesssim \frac{1}{N} \sum_{j=1}^{J} \sum_{i=1}^{N_{j}}\left(\frac{\widehat{\varepsilon}_{i j}-\varepsilon_{ij}}{\widehat{\sigma}_j}\right)^{4}+\frac{1}{N} \sum_{j=1}^{J} \sum_{i=1}^{N_{j}}\left(\frac{\varepsilon_{i j}\left(\widehat{\sigma}_j^2-\sigma_j^2\right)}{\widehat{\sigma}_j \sigma_j(\widehat{\sigma}_j+\sigma_j)}\right)^{4} \\
& \asymp \left( \widehat{\mu} - \mu\right)^4 + \frac{1}{N} \sum_{j = 1}^J N_j \left( \widehat{\sigma}_j^2 - \sigma_j^2 \right)^4 \left\{ \frac{1}{N_j} \sum_{i = 1}^{N_j} \varepsilon_{ij}^4 \right\} \\
& \asymp O_p \left(\frac{1}{N^2} \right) +  o_p \left(\frac{1}{N} \right) =  o_p \left(\frac{1}{N} \right),
\end{align*}
since
\begin{align*}
&\frac{1}{N} \sum_{j = 1}^J N_j \left( \widehat{\sigma}_j^2 - \sigma_j^2 \right)^4 \left\{ \frac{1}{N_j} \sum_{i = 1}^{N_j} \varepsilon_{ij}^4 \right\} \\
\le& \left\{\max_{1\le j\le J}\left(\widehat\sigma_j^2-\sigma_j^2\right)^2\right\}\left\{\frac{1}{N} \sum_{j = 1}^J N_j \left( \widehat{\sigma}_j^2 - \sigma_j^2 \right)^2 \left\{ \frac{1}{N_j} \sum_{i = 1}^{N_j} \varepsilon_{ij}^4 \right\} \right\}\\
=&o_p\left(\frac{1}{N} \sum_{j = 1}^J N_j \left( \widehat{\sigma}_j^2 - \sigma_j^2 \right)^2 \left\{ \frac{1}{N_j} \sum_{i = 1}^{N_j} \varepsilon_{ij}^4 \right\}\right)\\
=&o_p\left(\frac{1}{\sqrt{N}}\right),
\end{align*}
where we use $\widehat\sigma^2_j-\sigma^2_j=o_p(1)$ from \eqref{sigma_j^2_hat-sigma_j^2}, and the last step can be obtained through similar arguments above. Besides, $N^{-1}\sum_{j=1}^{J} \sum_{i=1}^{N_{j}} \widetilde{\varepsilon}_{i j}^2/\widetilde{\sigma}_j^2=O_{p}(1)$ can be derived from the procedure in \eqref{RN2_Op1} analogously. Then we have
\begin{equation*}
\left\vert R_{N 2}\right\vert \lesssim\left[\frac{1}{N} \sum_{j=1}^{J} \sum_{i=1}^{N_{j}} \frac{\widetilde{\varepsilon}_{i j}^2}{\widetilde{\sigma}_j^2}\right]^{1 / 2}\left[\frac{1}{N} \sum_{j=1}^{J} \sum_{i=1}^{N_{j}}\left(\frac{\widehat{\varepsilon}_{i j}}{\widehat{\sigma}_j}-\frac{\varepsilon_{i j}}{\sigma_j}\right)^{4}\right]^{1 / 2}  = o_p \left( \frac{1}{\sqrt{N}}\right).
\end{equation*}
Therefore, \eqref{H0_different_var_decomp} is validated. This completes the proof of Theorem \ref{H0_different_var}. $\square$

\subsection{Proof of Theorem \ref{H1_different_var} and \ref{H1L_different_var}} 

The proof of Theorem \ref{H1_different_var} is analogous to that of Theorem \ref{H1_same}. Under $H_1$, in order to obtain \eqref{H1_different_var_decomp}, we just repeat the procedure in the proof of Theorem \ref{H0_different_var}, and replace $c_{1k}$ with $d_{1k}$ in \eqref{DN1}, $c_{2k}$ with $d_{2k}$ in \eqref{DN2}. From \eqref{H1_same_decomp} we know that
\begin{equation*}
\frac{1}{N}\sum_{j=1}^{J}\sum_{i=1}^{N_j}\pi_k\left(\widehat Z_{ij}\right) = \mathbb{E} \left[\pi_k (Z)\right] + o_p(1). 
\end{equation*}
Thus,
\begin{equation*}
\widetilde{\Psi}_K^2=\left(  \frac{1}{N}\sum_{j = 1}^J \sum_{i = 1}^{N_j} \bm\pi_K \left(\widehat Z_{ij}\right) \right)^\top \left(\sum_{j = 1}^J p_j \bm{\Omega}_K^{(j)} \right)^{-1} \left(  \frac{1}{N}\sum_{j = 1}^J \sum_{i = 1}^{N_j} \bm\pi_K \left(\widehat Z_{ij}\right) \right)   \overset{p}{\to}  \bm{a}_K^\top \left(\sum_{j = 1}^J p_j \bm{\Omega}_K^{(j)} \right)^{-1}\bm{a}_K.
\end{equation*}
Besides,
\begin{align}
&\sqrt{N} \left( \widetilde{\Psi}_K^2 - \bm{a}_K^\top \left(\sum_{j = 1}^J p_j \bm{\Omega}_K^{(j)} \right)^{-1}\bm{a}_K \right)  \nonumber\\
=& \sqrt{N} \left\{\left[ \bm{a}_K +  \frac{1}{N} \sum_{j = 1}^J \sum_{i = 1}^{N_j} \bm\pi_K\left(\widehat Z_{ij}\right)-\bm{a}_K\right]^\top \left(\sum_{j = 1}^J p_j \bm{\Omega}_K^{(j)} \right)^{-1} \left[ \bm{a}_K +  \frac{1}{N} \sum_{j = 1}^J \sum_{i = 1}^{N_j} \bm\pi_K\left(\widehat Z_{ij}\right)-\bm{a}_K\right] \right. \nonumber\\
&\left. \qquad- \bm{a}_K^\top \left(\sum_{j = 1}^J p_j \bm{\Omega}_K^{(j)} \right)^{-1} \bm{a}_K \right\} \nonumber\\
=& 2a^\top \left(\sum_{j = 1}^J p_j \bm{\Omega}_K^{(j)} \right)^{-1} \frac{1}{\sqrt{N}} \sum_{j = 1}^J \sum_{i = 1}^{N_j} \left( \bm\pi_K \left(\widehat Z_{ij}\right) - \bm{a}_K \right) +\left[\frac{1}{\sqrt{N}} \sum_{j = 1}^J \sum_{i = 1}^{N_j} \left( \bm\pi_K \left(\widehat Z_{ij}\right) - \bm{a}_K \right)\right]^\top\nonumber\\
&\left(\sum_{j = 1}^J p_j \bm{\Omega}_K^{(j)} \right)^{-1}\left[\frac{1}{\sqrt{N}} \sum_{j = 1}^J \sum_{i = 1}^{N_j} \left( \bm\pi_K \left(\widehat Z_{ij}\right) - \bm{a}_K \right)\right]. \label{Psi_tilde^2_derivation}
\end{align}
Under $H_1$, by CLT, we have
\begin{align*}
\frac{1}{\sqrt{N}}\sum_{j=1}^{J}\sum_{i=1}^{N_j}\left(\bm\pi_K\left(\widehat Z_{ij}\right)-\bm{a}_K\right) & =\sqrt{N}\left[\frac{1}{N}\sum_{j=1}^{J}\sum_{i=1}^{N_j}\pi_1\left(\widehat Z_{ij}\right)-a_1,\ldots,\frac{1}{N}\sum_{j=1}^{J}\sum_{i=1}^{N_j}\pi_K\left(\widehat Z_{ij}\right)-a_K\right]^\top \\ &\overset{d}{\to}  \sum_{j = 1}^J \sqrt{p_j} \bm{W}_j,
\end{align*}
where $\bm{W}_1, \ldots, \bm{W}_J$ are independently distributed as
\begin{equation*}
\bm{W}_j  \sim  \mathcal{N}_K \left( 0, \bm{\Lambda}_K^{(j)} \right),
\end{equation*}
with $\bm{\Lambda}_K^{(j)}=\left(\lambda_{kl}^{(j)}\right)_{K \times K}$ given by
\begin{align*}
\lambda_{kl}^{(j)} =& \mathbb{E} \left\{  \left(\pi_k (Z)-\mathbb{E}\left[ \pi_k (Z)\right]\right) -d_{1k} \left( \sum_{\ell = 1}^J \frac{p_{\ell}}{\sigma_{\ell}} \right) \frac{\sigma_j e}{q_j} - \frac{d_{2k}}{2} \left( e^2 - 1 \right)\right\}  \\
&\quad \left\{  \left(\pi_l (Z)-\mathbb{E}\left[ \pi_l (Z)\right]\right) - d_{1l}\left( \sum_{\ell = 1}^J \frac{ p_{\ell}}{\sigma_{\ell}} \right) \frac{\sigma_j e}{q_j} - \frac{d_{2l}}{2} \left( e^2 - 1 \right)\right\} \\
= &\mathbb{E} \left[ \pi_k (Z) \pi_l (Z) \right]-a_k a_l- \frac{ (d_{1k} d_{3l}+d_{1l} d_{3k}) \sigma_j}{q_j} \sum_{\ell = 1}^J \frac{p_{\ell}}{\sigma_{\ell}} + \frac{d_{1k}d_{1 l} \sigma_j^2}{q_j^2} \left( \sum_{\ell = 1}^J \frac{p_{\ell}}{\sigma_{\ell}}\right)^2 \\
&+ \frac{1}{2} \left[ a_k d_{2l} + a_l d_{2k} \right]-\frac{1}{2}\left[ d_{2k}d_{4l} + d_{2l}d_{4k}\right]+ \frac{ (d_{1k} d_{2l}+d_{1l} d_{2k}) \sigma_j}{2q_j} \sum_{\ell = 1}^J \frac{p_{\ell}}{\sigma_{\ell}}\mathbb{E} \left[e^3\right]\\
&+\frac{d_{2k} d_{2l} }{4} \left[\mathbb{E} \left[e^4\right]-1 \right].
\end{align*}
Therefore,
\begin{equation*}
\frac{1}{\sqrt{N}} \sum_{j = 1}^J \sum_{i = 1}^{N_j} \left( \bm\pi_K \left(\widehat Z_{ij}\right) - \bm{a}_K \right) \overset{d}{\to}  \mathcal{N}_K\left( \bm{0}, \sum_{j = 1}^J p_j \bm{\Lambda}_K^{(j)}\right),
\end{equation*}
and the asymptotic distribution of $\widetilde\Psi_K^2$ in \eqref{Psi_tilde_asym_normal} can be derived from \eqref{Psi_tilde^2_derivation} through similar arguments above. This completes the proof of Theorem \ref{H1_different_var}. 

The proof of Theorem \ref{H1L_different_var} can be referred to that of Theorem \ref{H1L_same} due to the similar arguments and techniques used. To obtain the results in Theorem \ref{H1L_different_var}, following the proof of Theorem \ref{H1_different_var}, it suffices to show that under $H_{1L}$, $\mathbb{E}[\pi_k (Z)]=\delta_N \Delta_k$,
$\mathbb{E}[\pi_k(Z)\pi_l(Z)]=\delta_{kl}+o(1)$, $d_{1k}=c_{1k}+o(1)$, $d_{2k}=c_{2k}+o(1)$, $d_{3k}=c_{1k}+o(1)$, $d_{4k}=c_{2k}+o(1)$, $\mathbb{E}[e^3]=o(1)$, and 
$\mathbb{E}[e^4]=3+o(1)$, which can be found in the proof of Theorem \ref{H1L_same}. $\square$

\subsection{Proof of Theorem \ref{thm_data driven}} 

We focus on the first case in Theorem \ref{thm_data driven}, and the latter two cases can be verified analogously. We primarily follow well-established steps in the literature, for example, the proof of Theorem 2 in \cite{ducharme2004goodness}, Lemma 1 in \cite{kraus2007data}, and Propositions 6 and 7 in \cite{duchesne2016estimating}. Note that $\{\widehat K=k\} \subseteq \{N\widehat\Psi^2_k-k \log N\ge N\widehat\Psi^2_{k^\prime}-k^\prime \log N\}$ for $k^\prime\ne k$, then we have
\begin{equation}
\mathbb{P}\left( \widehat K  = k \right) \le \mathbb{P}\left(N\widehat\Psi^2_k-k \log N\ge N\widehat\Psi^2_{k^\prime}-k^\prime \log N \right),\quad k^\prime\ne k. \label{K_hat_subadditive}
\end{equation}

Under $H_0$, in order to show $\mathbb{P}(\widehat K=1)\to 1$ as $N\to \infty$, we first show the contrary such that $\sum_{k=2}^{D}\mathbb{P}(\widehat K=k)\to 0$. In fact, from \eqref{K_hat_subadditive} we have
\begin{align*}
&\sum_{k = 2}^{D} \mathbb{P} \left( \widehat K  = k \right)\\
\le& \sum_{k = 2}^{D} \mathbb{P} \left( N\widehat\Psi^2_k  - k \log N \ge N\widehat\Psi^2_1  - \log N \right) \\
\le& \sum_{k = 2}^{D} \mathbb{P} \left( N\widehat\Psi^2_k  - k \log N \ge - \log N \right)\\
=& \sum_{k = 2}^{D} \mathbb{P} \left( N\widehat\Psi^2_k  \ge (k-1)\log N \right).
\end{align*}
From Corollary \ref{corollary1} we know that under $H_0$, $N\widehat\Psi^2_k\overset{d}{\to} \chi^2_k$ for each $k=1,\ldots,D$. Thus for $k=2,\ldots,D$, $\mathbb{P} ( N\widehat\Psi^2_k  \ge (k-1)\log N ) \to 0$ as $N\to \infty$. Therefore,
\begin{equation*}
\sum_{k = 2}^{D} \mathbb{P} \left( \widehat K  = k \right)\le\sum_{k = 2}^{D}\mathbb{P} \left( N\widehat\Psi^2_k  \ge (k-1)\log N \right) \to 0,
\end{equation*}
as $N\to \infty$, then $\mathbb{P} ( \widehat K  = 1 ) =1- \sum_{k = 2}^{D} \mathbb{P} ( \widehat K  = k ) \to 1$, and $N\widehat\Psi^2_{\widehat K}\overset{d}{\to} \chi^2_1$ follows.

Under $H_1^\prime$, \eqref{K_hat_subadditive} yields that
\begin{align*}
\mathbb{P}\left( \widehat K  = k \right) &\le \mathbb{P}\left(N\widehat\Psi^2_k-k \log N\ge N\widehat\Psi^2_{K_0}-K_0 \log N \right) \\
&= \mathbb{P}\left(\widehat\Psi^2_k-k \frac{\log N}{N}\ge \widehat\Psi^2_{K_0}-K_0 \frac{\log N}{N} \right). 
\end{align*}
From Theorem \ref{H1_same} we know that as $N\to \infty$,  $\widehat\Psi^2_k\overset{p}{\to} 0$ for $k=1,\ldots, K_0-1$, $\widehat\Psi^2_{K_0} \overset{p}{\to} \bm{a}_{K_0}^\top \bm{\Sigma}_{K_0}^{-1} \bm{a}_{K_0}>0$ and $\log N/N \to 0$. Thus $ \mathbb{P}( \widehat K  = k ) \to 0$ for $k=1,\ldots, K_0-1$, and 
\begin{equation*}
\lim_{N\to \infty} \mathbb{P}\left( \widehat K  \ge K_0 \right)=1-\lim_{N\to\infty} \sum_{k=1}^{K_0-1}  \mathbb{P}\left( \widehat K  = k \right)=1. 
\end{equation*}
Then, for any $x\in \mathbb{R}$, as $N\to\infty$,
\begin{align*}
\mathbb{P}\left(N\widehat\Psi^2_{\widehat K}\le x\right)&=\sum_{k=1}^D \mathbb{P}\left(N\widehat\Psi^2_k\le x,\widehat K=k \right)\\
&\le \sum_{k=1}^{K_0-1} \mathbb{P}\left( \widehat K  = k \right) + \sum_{k=K_0}^D \mathbb{P}\left(N\widehat\Psi^2_k\le x\right)\to 0,
\end{align*}
since $\widehat\Psi^2_k  \overset{p}{\to} \bm{a}_k^\top \bm{\Sigma}_k^{-1} \bm{a}_k>0$ for $k=K_0,\ldots,D$. Therefore, for any $x\in \mathbb{R}$,
\begin{equation*}
\lim_{N\to\infty}\mathbb{P}\left(N\widehat\Psi^2_{\widehat K}\le x\right) = 0,
\end{equation*}
which means the test is consistent against the alternatives given by \eqref{H1_data_driven}. $\square$

\subsection{Intuition behind the revised limiting null distribution \eqref{null_approx}} 

We primarily follow the arguments in \cite{kallenberg1999data}, \cite{janic2000data}, and \cite{kraus2007data}, and provide a heuristic derivation of the approximation $H(x)$ for the first case, as the latter two cases can be derived analogously. Note that
\begin{align} \label{eq_H(x)}
\mathbb{P} \left(N \widehat{\Psi}_{\widehat K}^2 \le x\right)=&\mathbb{P} \left(N \widehat{\Psi}_1^2 \le x, \widehat K = 1\right) + \mathbb{P} \left(N \widehat{\Psi}_2^2 \le x, \widehat K = 2\right)\nonumber\\
&+ \mathbb{P} \left(N \widehat{\Psi}_{\widehat K}^2 \le x,  \widehat K \ge 3\right).
\end{align}
The third term on the right-hand side of \eqref{eq_H(x)}, $\mathbb{P} (N \widehat{\Psi}_{\widehat K}^2 \le x,  \widehat K \ge 3)$, can be neglected under $H_0$ \citep{kallenberg1995data}. The event $\{\widehat K=1\}$ is approximated by $\{N\widehat\Psi^2_1-\log N \ge N\widehat\Psi^2_2-2\log N\}=\{N(\widehat\Psi^2_2-\widehat\Psi^2_1)\le \log N\}$, and $\{\widehat K=2\}$ is approximated by $\{N(\widehat\Psi^2_2-\widehat\Psi^2_1)> \log N\}$.

We first investigate the asymptotic distribution of $(N\widehat\Psi^2_1, N(\widehat\Psi^2_2-\widehat\Psi^2_1))^\top$. The limiting distributions of $N\widehat\Psi^2_1,N\widehat\Psi^2_2$ are functions of a bivariate normal vector $(R_1,R_2)^\top \sim \mathcal{N}_2(\bm{0},\bm{\Sigma}_2)$. We generally denote elements of $\bm{\Sigma}_2$ as $\begin{pmatrix} a & b\\ b & c \end{pmatrix}$ and $\rho=b/\sqrt{ac}$. The distribution of $(R_1,R_2)^\top$ can be constructed through two independent standard normal variables $G_1, G_2$ if we denote $\widetilde{R}_1= \sqrt{a}(\sqrt{1- \rho ^2}G_1+ \rho G_2)$ and $\widetilde{R}_2= \sqrt{c}G_2$, then $(\widetilde{R}_1, \widetilde{R}_2)^\top \overset{d}{=} (R_1,R_2)^\top$. Thus, $N\widehat\Psi^2_1\overset{d}{\to} R_1^2/a\overset{d}{=} \widetilde{R}_1^2/ a= (\sqrt{1-\rho^2}G_1+\rho G_2)^2$. For $N\widehat\Psi_2^2$, straightforward computations yield that
\begin{align*}
N\widehat\Psi_2^2&\overset{d}{\to} \left(\widetilde{R}_1, \widetilde{R}_2\right)\bm{\Sigma}_2^{-1}\left(\widetilde{R}_1, \widetilde{R}_2\right)^\top \\
&=\frac{1}{ac-b^2} \left(\widetilde{R}_1, \widetilde{R}_2\right)\begin{pmatrix} c & -b\\ -b & a\end{pmatrix}\left(\widetilde{R}_1, \widetilde{R}_2\right)^\top\\
&=\frac{1}{ac-b^2}\left(c\widetilde{R}_1^2-2b\widetilde{R}_1\widetilde{R}_2+a\widetilde{R}_2^2\right)\\
&=G_1^2+G_2^2.
\end{align*}
Then $N(\widehat\Psi_2^2-\widehat\Psi_1^2)\overset{d}{\to}(\rho G_1- \sqrt {1- \rho ^2}G_2)^2$. Since $(\sqrt{1-\rho^2}G_1+\rho G_2,\rho G_1-\sqrt {1- \rho ^2}G_2)^\top \sim \mathcal{N}_2(\bm{0},\mathbf{I}_2)$, we obtain that $N\widehat\Psi^2_1$, $N(\widehat\Psi^2_2- \widehat\Psi^2_1)$ are both asymptotically $\chi_1^2$ distributed and are asymptotically independent.

Now we calculate the approximation $H(x)$. We will treat $H(x)$ separately for $x\le \log N$, $\log N<x<2\log N$ and $x\geq2\log N$. For $x\le \log N$, note that
\begin{equation*}
\mathbb{P}\left(N\widehat\Psi^2_2\le x,\widehat K=2\right)\approx\mathbb{P}\left(N\widehat\Psi^2_2\le x,N\widehat\Psi^2_2-N\widehat\Psi^2_1>\log N\right)=0,    
\end{equation*}
by $N\widehat\Psi^2_1\geq0$. Thus, we have the approximation
\begin{align*}
&\mathbb{P} \left(N\widehat\Psi^2_1\le x, N\widehat\Psi^2_2- N\widehat\Psi^2_1\le \log N\right)\\
\approx&\left[ 2\Phi \left(\sqrt{x}\right) - 1\right] \left[ 2\Phi \left(\sqrt{\log N}\right)-1\right]\equiv H(x), \quad x\le \log N,    
\end{align*}
from the asymptotic $\chi_1^2$ distribution and asymptotic independence of $N\widehat\Psi^2_1$, $N(\widehat\Psi^2_2- \widehat\Psi^2_1)$. For $x\geq2\log N$, we have
\begin{align*}
&\mathbb{P} \left(N\widehat\Psi^2_2\le x,\widehat K=2\right)\\
\approx&\mathbb{P}\left(N\widehat\Psi^2_2\le x,N\widehat\Psi^2_2-N\widehat\Psi^2_1\ge\log N\right)\\
=&\mathbb{P}\left(N\widehat\Psi^2_2-N\widehat\Psi^2_1\ge\log N\right)-\mathbb{P}\left(N\widehat\Psi^2_2>x,N\widehat\Psi^2_2-N\widehat\Psi^2_1\ge\log N\right).
\end{align*}
As $N(\widehat\Psi^2_2-\widehat\Psi^2_1)$ is approximately $\chi_1^2$ distributed, we obtain
\begin{align*}
&\mathbb{P}\left(N\widehat\Psi^2_2-N\widehat\Psi^2_1\ge\log N\right)\\
\approx& 2\left(1-\Phi\left(\sqrt{\log N}\right)\right)\approx 2\frac{\phi\left(\sqrt{\log N}\right)}{\sqrt{\log N}}=\sqrt{\frac{2}{\pi N\log N}},
\end{align*}
by the fact that $t\phi(t)/(t+1) \le1-\Phi (t) \le \phi(t)/t$ for $t\to \infty$. Besides,
\begin{align*}
&\mathbb{P}\left(N\widehat\Psi^2_2>x,N\widehat\Psi^2_2-N\widehat\Psi^2_1\ge\log N\right)\\
\le&\mathbb{P}\left(N\widehat\Psi^2_2>x\right)\\
\le&\mathbb{P}\left(N\widehat\Psi^2_2>2\log N\right)\\
\approx&\exp\left\{-\frac{1}{2}\times 2\log N\right\}\\
=&\frac{1}{N},   
\end{align*}
where we use the distribution function of the limiting $\chi^2_2$ for $N\widehat\Psi^2_2$. Hence $\mathbb{P}(N\widehat\Psi^2_2-N\widehat\Psi^2_1\ge\log N)$ converges to 0 much slower than $\mathbb{P}(N\widehat\Psi^2_2>x,N\widehat\Psi^2_2-N\widehat\Psi^2_1\ge\log N)$, and thus the latter can be neglected. Therefore, for $x\geq2\log N$, we have the approximation
\begin{align*}
&\mathbb{P}\left(N\widehat\Psi^2_1\le x,N\widehat\Psi^2_2-N\widehat\Psi^2_1\le\log N\right)+\mathbb{P}\left(N\widehat\Psi^2_2-N\widehat\Psi^2_1\ge\log N\right)\\
\approx&\left[2\Phi\left(\sqrt{x}\right)-1\right]\left[2\Phi\left(\sqrt{\log N}\right)-1\right]+2\left[1-\Phi\left(\sqrt{\log N}\right)\right]\\
\equiv& H(x),\quad x\geq2\log N.
\end{align*}
For $\log N<x < 2\log N$, \cite{kallenberg1995data} suggested the linearization as follows
\begin{equation*}
H(x)=H(\log N)+\frac{x-\log N}{\log N}[H(2\log N)-H(\log N)],\quad\log N<x<2\log N.    
\end{equation*}
This completes the derivation of the approximation \eqref{null_approx} for the null distribution of the test statistic $N\widehat\Psi^2_{\widehat K}$. $\square$

\subsection{Implement issues in simulation studies} 

In our simulation study, for the proposed smooth tests, we adopt the orthonormal Legendre polynomials on $[0,1]$ for $\{\pi_k\}_{k = 1}^{\infty}$. The properties of orthonormal Legendre polynomials lead to the following expressions for constants $c_{1k}$ and $c_{2k}$:
\begin{equation*}
    \begin{aligned}
        c_{1k} &= \frac{\sqrt{2 k + 1}}{2^k} \, \sum_{j = 0}^{[k / 2]} (-1)^j \binom{k}{j} \binom{2(k - j)}{k} \int_0^1 (2 z - 1)^{k - 2j} \, \Phi^{-1}(z) \, \mathrm{d}z \\
        & = \sqrt{2 k + 1} \, \sum_{j = 0}^{[k / 2]} \left( \frac{-1}{4} \right)^j \binom{k}{j} \binom{2(k - j)}{k} \, \int_{-1/2}^{1/2} x^{k - 2j} \, \Phi^{-1} \left( x + \frac{1}{2}\right) \, \mathrm{d}z,
    \end{aligned}
\end{equation*}
and
\begin{equation*}
    c_{2k}  = \sqrt{2 k + 1} \, \sum_{j = 0}^{[k / 2]} \left( \frac{-1}{4} \right)^j \binom{k}{j} \binom{2(k - j)}{k} \, \int_{-1/2}^{1/2} x^{k - 2j} \, \left[\Phi^{-1} \left( x + \frac{1}{2}\right) \right]^2 \, \mathrm{d}z.
\end{equation*}
Note that $c_{2k}=0$ when $k$ is odd and $c_{1k}=0$ when $k$ is even. Therefore, in our experiments, we only compute the non-zero elements based on the above two formulae for $k=1,\ldots, K$ with $K\le 5$.

\subsection{Additional simulation results} \label{section_add_simu}

\begin{figure}[ht]
	\centering
	\includegraphics[scale=0.65]{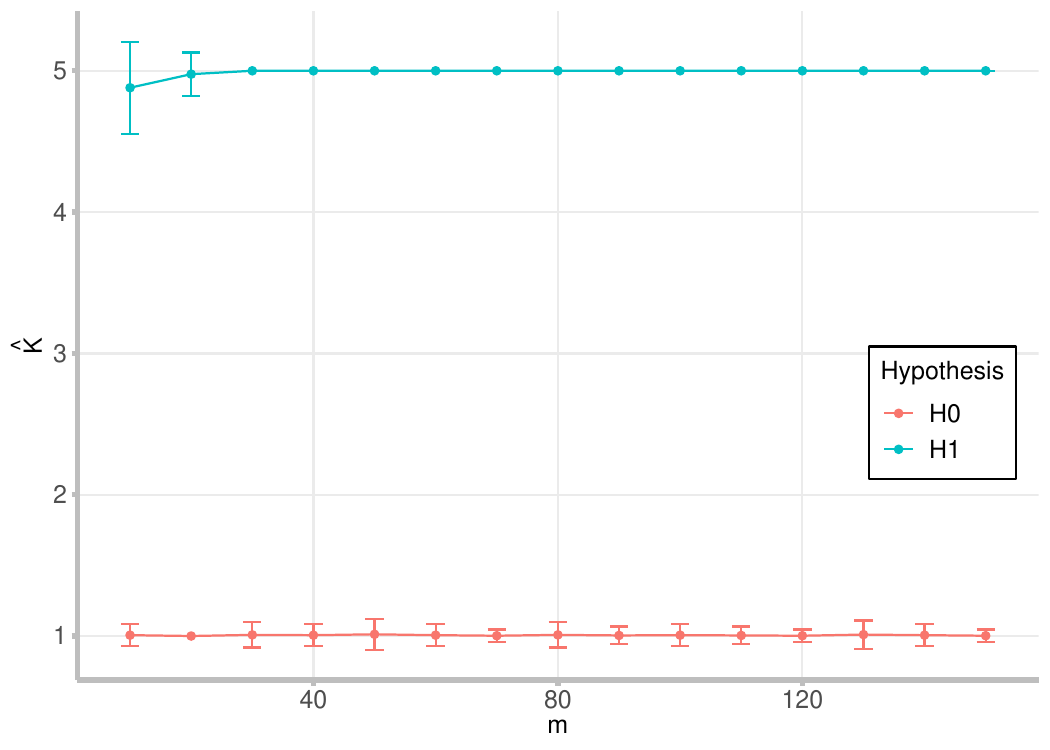}
	\caption{The sample means and error bars of $\widehat K$ in Experiment \uppercase\expandafter{\romannumeral3}$^\prime$}
	\label{Fig3.2'}
\end{figure}

The results for Experiment \uppercase\expandafter{\romannumeral3}$^\prime$ are reported in Tables \ref{rejection_rate_3.2'}, \ref{k_freq_3.2'} and Figure \ref{Fig3.2'}. They are broadly consistent with those obtained in Experiment \uppercase\expandafter{\romannumeral3}, with several additional findings. First, when $J=10$ and $m$ is relatively large, the total sample size $N$ exceeds the applicability range of the Shapiro--Wilk test (which requires $N \le 3000$), and hence the corresponding results are not reported in Table \ref{rejection_rate_3.2'}. In contrast, our proposed test is not subject to this limitation and remains applicable. Second, Table \ref{rejection_rate_3.2'} shows that the limiting $\chi^2_1$ distribution for the data-driven test provides valid control of the Type \uppercase\expandafter{\romannumeral1} error. Third, the empirical distribution of $\widehat{K}$ becomes more stable under both the null and alternative hypotheses, in some cases degenerating to a single value with probability one. These observations can be attributed to the increase in total sample size induced by a larger number of groups, which is in line with the theoretical results.

\begin{table}[ht]
\small
\centering
\caption{
Empirical rejection rates under $H_0$ in Experiment \uppercase\expandafter{\romannumeral3}$^\prime$. The empirical power under $H_1$ equals 1 for all sample sizes and all considered tests.
}
\begin{tabular}{c|ccccccc|cc}
\toprule
& \multicolumn{7}{c|}{Smooth tests} & \multirow{2}{*}{JB} & \multirow{2}{*}{KS} \\
\cmidrule(lr){2-8}
$m$  & $K = 1$ & $K = 2$ & $K = 3$ & $K = 4$ & $K = 5$ 
     & $\widehat K\&\chi^2_1$ & $\widehat K\&H(x)$ 
     \\
\midrule
10  & 0.028 & 0.046 & 0.062 & 0.036 & 0.054 & 0.060 & 0.048 & 0.050 & 0 \\
20  & 0.046 & 0.066 & 0.078 & 0.052 & 0.068 & 0.056 & 0.052 & 0.040 & 0 \\
30  & 0.050 & 0.056 & 0.044 & 0.040 & 0.048 & 0.044 & 0.042 & 0.054 & 0 \\
40  & 0.054 & 0.054 & 0.042 & 0.040 & 0.044 & 0.058 & 0.052 & 0.046 & 0 \\
50  & 0.052 & 0.062 & 0.052 & 0.056 & 0.054 & 0.048 & 0.042 & 0.066 & 0 \\
60  & 0.062 & 0.050 & 0.060 & 0.048 & 0.048 & 0.052 & 0.046 & 0.050 & 0 \\
70  & 0.068 & 0.052 & 0.052 & 0.054 & 0.044 & 0.058 & 0.054 & 0.040 & 0 \\
80  & 0.052 & 0.034 & 0.044 & 0.062 & 0.052 & 0.048 & 0.046 & 0.042 & 0 \\
90  & 0.048 & 0.048 & 0.034 & 0.050 & 0.058 & 0.042 & 0.040 & 0.044 & 0 \\
100 & 0.064 & 0.048 & 0.052 & 0.048 & 0.052 & 0.058 & 0.054 & 0.050 & 0 \\
110 & 0.044 & 0.048 & 0.058 & 0.058 & 0.032 & 0.060 & 0.056 & 0.048 & 0.002 \\
120 & 0.052 & 0.058 & 0.066 & 0.044 & 0.052 & 0.080 & 0.078 & 0.046 & 0 \\
130 & 0.046 & 0.050 & 0.038 & 0.058 & 0.062 & 0.058 & 0.052 & 0.046 & 0 \\
140 & 0.058 & 0.046 & 0.032 & 0.054 & 0.046 & 0.056 & 0.050 & 0.056 & 0 \\
150 & 0.046 & 0.040 & 0.060 & 0.050 & 0.054 & 0.038 & 0.038 & 0.042 & 0.002 \\
\bottomrule
\end{tabular}
\label{rejection_rate_3.2'}
\end{table}

\begin{table}[ht]
\small
\centering
\caption{Empirical frequency of $\widehat K$ in Experiment \uppercase\expandafter{\romannumeral3}$^\prime$}
\begin{tabular}{c|ccccc|ccccc}
\toprule
 & \multicolumn{5}{c|}{Under $H_0$} & \multicolumn{5}{c}{Under $H_1$} \\
\cmidrule(lr){2-6}\cmidrule(lr){7-11}
$m$ & $\widehat K=1$ & $\widehat K=2$ & $\widehat K=3$ & $\widehat K=4$ & $\widehat K=5$
    & $\widehat K=1$ & $\widehat K=2$ & $\widehat K=3$ & $\widehat K=4$ & $\widehat K=5$ \\
\midrule
10  & 0.994 & 0.006 & 0     & 0     & 0     & 0     & 0     & 0     & 0.120 & 0.880 \\
20  & 1     & 0     & 0     & 0     & 0     & 0     & 0     & 0     & 0.024 & 0.976 \\
30  & 0.992 & 0.008 & 0     & 0     & 0     & 0     & 0     & 0     & 0     & 1     \\
40  & 0.994 & 0.006 & 0     & 0     & 0     & 0     & 0     & 0     & 0     & 1     \\
50  & 0.988 & 0.012 & 0     & 0     & 0     & 0     & 0     & 0     & 0     & 1     \\
60  & 0.994 & 0.006 & 0     & 0     & 0     & 0     & 0     & 0     & 0     & 1     \\
70  & 0.998 & 0.002 & 0     & 0     & 0     & 0     & 0     & 0     & 0     & 1     \\
80  & 0.992 & 0.008 & 0     & 0     & 0     & 0     & 0     & 0     & 0     & 1     \\
90  & 0.996 & 0.004 & 0     & 0     & 0     & 0     & 0     & 0     & 0     & 1     \\
100 & 0.994 & 0.006 & 0     & 0     & 0     & 0     & 0     & 0     & 0     & 1     \\
110 & 0.996 & 0.004 & 0     & 0     & 0     & 0     & 0     & 0     & 0     & 1     \\
120 & 0.998 & 0.002 & 0     & 0     & 0     & 0     & 0     & 0     & 0     & 1     \\
130 & 0.990 & 0.010 & 0     & 0     & 0     & 0     & 0     & 0     & 0     & 1     \\
140 & 0.994 & 0.006 & 0     & 0     & 0     & 0     & 0     & 0     & 0     & 1     \\
150 & 0.998 & 0.002 & 0     & 0     & 0     & 0     & 0     & 0     & 0     & 1     \\
\bottomrule
\end{tabular}
\label{k_freq_3.2'}
\end{table}

\end{appendix}


\end{document}